\documentclass[aps,prx,twocolumn,letterpaper,superscriptaddress,showpacs]{revtex4-2}
\usepackage{amsmath, amsthm}
\usepackage{keyval}
\usepackage{graphicx,latexsym}
\usepackage{dcolumn}
\usepackage{bm}
\usepackage{color}
\usepackage{url}
\usepackage{CJK}
\usepackage{float}
\usepackage{hyperref}

\begin{document}
\title{Ground State Phase Diagram of the $t$-$t'$-$J$ model}

\begin{CJK*}{UTF8}{}
\author{Shengtao Jiang(\CJKfamily{gbsn}蒋晟韬)}
\altaffiliation{shengtaj@uci.edu}
\affiliation{Department of Physics and Astronomy, University of California, Irvine, California 92697, USA}

\author{Douglas J. Scalapino}
\altaffiliation{djs@physics.ucsb.edu}
\affiliation{Department of Physics, University of California, Santa Barbara, California 93106, USA}

\author{Steven R. White}
\altaffiliation{srwhite@uci.edu}
\affiliation{Department of Physics and Astronomy, University of California, Irvine, California 92697, USA}
\date{\today}

\begin{abstract}
We report results of large scale ground state density matrix renormalization group(DMRG) calculations on $t$-$t'$-$J$ cylinders with circumferences 6 and 8.  We determine a rough phase diagram which appears to approximate the 2D system. While for many properties, positive and negative $t'$ values ($t'/t = \pm 0.2$) appear to correspond to electron and hole doped cuprate systems, respectively, the behavior of superconductivity itself shows an inconsistency between the model and the materials. The $t'<0$ (hole doped) region shows antiferromagnetism limited to very low doping, stripes more generally, and the familiar Fermi-surface of the hole doped cuprates. However, we find $t'<0$ strongly suppresses superconductivity. The $t'>0$ (electron doped) region shows the expected circular Fermi pocket of holes around the $(\pi,\pi)$ point and a broad low-doped region of coexisting antiferromagnetism and $d$-wave pairing with a triplet $p$ component at wave-vector $(\pi,\pi)$ induced by the antiferromagnetism and $d$-wave pairing. The pairing for the electron low-doped system with $t'>0$ is strong and unambiguous in the DMRG simulations. At larger doping another broad region with stripes in addition to weaker $d$-wave pairing and striped $p$-wave pairing appears.  In a small doping region near $x=0.08$  for $t'\sim-0.2$, we find a new type of stripe involving unpaired holes located predominantly on chains spaced three lattice spacings apart. The undoped two-leg ladder regions in between mimic the short-ranged spin correlations seen in  two-leg Heisenberg ladders.
\end{abstract}
\maketitle
\end{CJK*}

\section{introduction}
There has been considerable recent progress in numerical simulations of the models associated with the cuprate superconductors---the 2D Hubbard and $t$-$J$ models, and their variants\cite{zheng2017,absence-qin,hubreview-kivleson,hubreview-qin,4leghub-hcjiang,4leghub-yfjiang,4leghub-huang,4legtj-dodaro,4leghub-chung,4legtj-hcjiang,stripe-vmc1,stripe-vmc2,Corboz2014,corboz2011stripes,hub-bench2015,WS2009,hubw4stp-corboz2019,himeda2002stripe,xu2020competing,jiang2021-hightcinsl,tohyama}.
Like the materials themselves, these models have been found to exhibit a rich variety of phenomena, ranging from uniform antiferromagnetism(AFM) and $d$-wave superconductivity(SC) to charge and spin stripes.
Some previously controversial issues have been mostly resolved, such as the existence of stripes. Striped states were first found as a Hartree Fock solution to the doped Hubbard model\cite{stripe-Zaanen,hatreefock-rice,machida1,machida2}, and in the late 1990's two of us used the density matrix renormalization group\cite{white1992dmrg,white1993dmrg2} and found stripes as the ground state of the $t$-$J$ model\cite{WS1998}. At that time this result was controversial, since other powerful simulation methods, such as variational Monte Carlo, could not confirm our results\cite{vmcnostripe1,vmcnostripe2}. In the last few years, with progress in a variety of methods combined with the use of several simulation methods together,  striped ground states have been confirmed not just in the $t$-$J$ model\cite{corboz2011stripes,Corboz2014,vmctjstripe,4legtj-dodaro}, but also the Hubbard model\cite{hub-bench2015,zheng2017,4leghub-hcjiang,hubstp-Ido,hubstp-tocchio,4leghub-huang,4leghub-yfjiang}. 

While $d$-wave singlet pairing has consistently been favored over other types of pairing in many approaches, it has been less clear in both the $t$-$J$ and Hubbard models whether the ground state is superconducting or not.  The role of stripes in competing against or enhancing pairing has also been difficult to determine. It has long been clear, however, that next nearest neighbor hopping $t'$ has a crucial influence on the pairing. With a $-t_{ij}$ sign convention for hopping, our early DMRG for the $t$-$J$ model found that a positive $t'$  stabilizes the commensurate $(\pi,\pi)$ antiferromagnetic correlations and enhances the $d$-wave pairing correlations, whereas a negative $t'$ seemed to disfavor these correlations\cite{tt'j-ws1999,WS2009}. Other work has suggested instead that a negative $t'$ is important in destabilizing stripes so that SC can occur\cite{4leghub-hcjiang}.

While DMRG has not changed fundamentally since the 1990's, there has been steady improvements in techniques, software, and computers since then.  Here we have used these developments to return to a study of the 
$t$-$t'$-$J$ model.
We report here a detailed description of the ground state phase diagram as a function of $t^\prime$ and doping $x$, based on $L\times 8$ cylinders with $L$ up to 50, with confirmation of the qualitative features using width 6 cylinders. 
Note that this model can be used to describe both the hole and electron doped cuprates:
for $t'/t<0$ it describes a hole doped system with electron filling $n_e=1-x$, while for $t'/t>0$, based on a particle-hole transformation, one has an electron doped system with $n_e=1+x$.

Two important results of our study are (1) the finding of a coexistent antiferromagnetic $d$-wave SC and induced $\pi$-triplet $p$-wave SC regime in the electron doped system and (2) the lack of long range SC order in the hole doped case.

This paper is organized as follows: in Sec.~\ref{sec:modelmethod} we introduce the $t$-$t'$-$J$ model, the DMRG methods used, and the main observables that we study. In Sec.~\ref{sec:phasediagram} we show the $t'-x$ ground state phase diagram for $J/t=0.4$ obtained from the $8\times L$ cylinder calculations. Here the spin, charge and $d$-wave pairing strength for $t'/t=\pm0.2$ are shown as the doping $x$ varies slowly along the length of the cylinder. Combined with similar calculations with doping $x$ fixed and $t'$ slowly varying along the length of the cylinder, these results(details shown in Sec.~\ref{sec:scan}) are used to determine the $t'-x$ ground state phase diagram.

The resulting phase diagram exhibits four distinct regions. In Sec.~\ref{sec:dpip} we examine the lightly($x\lesssim0.14$) electron doped region in which there is coexisting AFM and strong $d$-wave pairing order. In addition there is necessarily also a $p$-wave triplet pairing component with center of mass momentum $(\pi,\pi)$. This order parameter does not depend upon a $p$-wave pairing interaction, but is dynamically generated by coexisting AFM and $d$-wave SC order parameters\cite{psaltakis1983,zhang1997so5,aperis2008,rowe2012spin,almeida2017induced}. While it has the same form as the generator of infinitesimal rotations between AFM and SC order parameters in the $SO(5)$ theory\cite{zhang1997so5},it appears here as an additional order parameter. Its strength relative to the AFM and d-wave order will be discussed. In the more heavily doped electron region discussed in Sec.~\ref{sec:stp} we find stripes with local AFM, and weaker $d$-wave and $p$-wave triplet pairing.
In Sec.~\ref{sec:csandw3} we discuss the $t'<0$ part of the phase diagram where we find conventional stripe and an W3 stripe region but negligible pairing. In Sec.~\ref{sec:cuprates} we discuss the relationship of the sign of $t'$ to the electron and hole doped cuprates and compare our results to experiments. Sec.~\ref{sec:sum} contains our conclusions. 

\section{model and method}
\label{sec:modelmethod}
We study the $t$-$t^\prime$-$J$ model on a square lattice, with Hamiltonian
\begin{equation}
\begin{aligned}
    \label{eq:tj}
    H=&\sum_{\langle \langle ij \rangle \rangle\sigma}-t_{ij}(c^\dag_{i\sigma}c_{j\sigma}+h.c.)\\
    &+J\sum_{\langle ij \rangle}(\Vec{S_i}\cdot \Vec{S_j}-\frac{1}{4}n_i^{tot} n_j^{tot})-\mu\sum_i n_i^{tot}
\end{aligned}
\end{equation}
here $n_i^{tot}=n_{i\uparrow}+n_{i\downarrow}$ is the total particle density on site $i$. $\langle \langle ij \rangle \rangle$ includes both nearest neighbor and next nearest neighbor pairs of sites while $\langle ij \rangle$ includes only nearest neighbor pairs of sites. Doubly occupied states are specifically excluded in the Hilbert space. 
For all calculations we set $t$=1 and $J$=0.4.
A chemical potential $\mu$ is used to control the doping level in some of the calculations; in others the number of particles is fixed.
We study cylinders, with open boundary conditions in the $x$ direction and periodic boundary conditions in the $y$ direction. We study width-6 and width-8 cylinders with lengths up to 50. Our primary focus is on the width-8 systems.  Behavior in width 6 is similar and provides an indication that our width 8 results are relevant for 2D. 

We use finite system DMRG using the ITensor library\cite{itensor}.
For this size system, keeping about 3000-4000 states is sufficient to measure local properties, provided that the calculation is not stuck in a metastable state.  To control this issue, we perform a variety of simulations with different starting states and temporary pinning fields, comparing energies and convergence of different states, to gain an understanding of the low energy states and their orders.  Some of the details of this process in subtle cases is discussed below.  In many cases, such as a conventional striped state, starting in a product state with the holes near where they end up is all that is necessary, but one should try different fillings and spacings of the stripes.  For example, 8 holes in a striped state might make either two 4-hole stripes or four 2-hole stripes.  In such a case we would try both possibilities and compare energies.

We focus on local measurements of the density, the  magnetization, and pairing. The hole density and magnetization are measured using $S_z=\frac{1}{2}(n_{i\uparrow}-n_{i\downarrow})$ and $n_{hole}=1-n_{i\uparrow}-n_{i\downarrow}$. To detect the superconducting order and structure in the grand canonical ensemble, we use the singlet (s)  and triplet (t) link pairing operators
\begin{equation}
\label{eq:sin}
    \Delta^\dag_{s,t}(l)=
    \frac{1}{\sqrt{2}}(c^\dag_{l_1,\uparrow}c^\dag_{l_2,\downarrow}\pm
    c^\dag_{l_2,\uparrow}c^\dag_{l_1,\downarrow})
\end{equation}
where the $+$ and $-$ are for singlet and triplet, respectively, and $l_1$ and $l_2$ are the two sites of  link $l$.

The expectation value $\langle\Delta^\dag+\Delta\rangle$ gives the local pairing strength.
For a $d$-wave superconductor, $\langle\Delta^\dag_s+\Delta_s\rangle$ switches sign between bonds in the $x$ and $y$ directions\cite{dwavesc}.

\section{Phase diagram}
\label{sec:phasediagram}
We begin by presenting the approximate phase diagram of the model in the doping ($x$)--next nearest neighbor hopping ($t'$) plane, the detailed features of which are explained later. A key distinction is the difference between positive and negative $t^\prime$.  The $t$-$J$ model cannot be doped above half-filling, so it does not appear that one can simulate electron-doped cuprates.  However, a particle-hole transformation of the single band Hubbard model  maps electron doping to hole doping, but with a change in the sign of $t^\prime$. One can then view the $t$-$J$ model as a low energy description of this particle-hole transformed Hubbard model.  We discuss this in more detail in Sec.~\ref{sec:cuprates}.  The key point is that
we can view $t^\prime \approx -0.2$ as applicable to the hole-doped cuprates, while $t^\prime \approx 0.2$ is applicable to the electron doped cuprates.  We will refer to the regions of the phase diagram using this terminology.

\begin{center}
\begin{figure}[t]
	\vspace{-0.0cm}
    \includegraphics[width=1.0\columnwidth,clip=true]{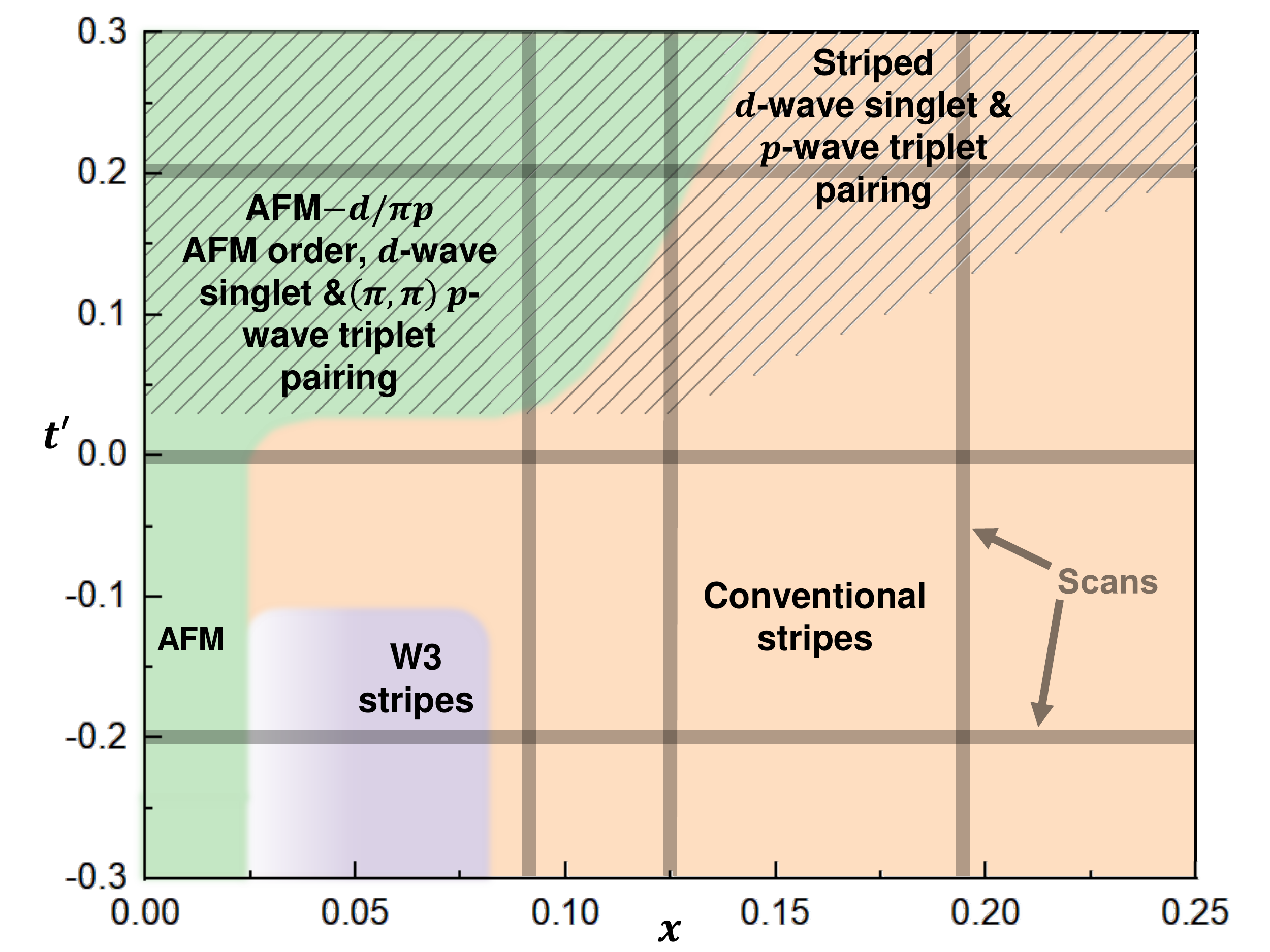}
	\vspace{-0.5cm}
	\caption{
		\label{fig:phase2d}
          Approximate phase diagram in the $x-t'$ plane, where $x$ is the doping and $t'$ is the next-nearest neighboring hopping, for width-8 cylinders. The six gray lines indicate the location of parameter-varying ``scan'' calculations on long cylinders which were the main tool to determine the phase diagram. The green indicates commensurate AFM order; the beige represents conventional stripes, which modulate $\pi$-phase shifted domain walls in the AFM order. 
          The slanted lines indicate $d$-wave pairing order. The simultaneous presence of $d$-wave pairing and AFM correlations induces  weaker momentum-($\pi$,$\pi$) $p$-wave order.}
	\vspace{-0.2cm}
\end{figure}
\end{center}

Our most useful tool in determining the phase diagrams are scan calculations, where in a long cylinder we slowly vary one parameter of the model, either $t'$ or the chemical potential $\mu$, to scan a whole line of the phase diagram.  These lines are shown in gray in Fig.~\ref{fig:phase2d}. These scans are detailed in the next section.  

Much of the phase diagram is taken up by a phase with conventional
stripes. These stripes are lines of increased hole density two or three sites wide which act as domain walls to $\pi$-phase shifted AFM (or at least significant local AFM correlations). Although the holes in these stripes correlate into pairs, the pairs tend to lack phase coherence, and pairing correlations are weak.  Significantly, negative $t'$ is found to decrease the pairing correlations. This phase makes up most of the $t'<0$ side of the diagram.  

%%%%%%%%%%%FIGURE%%%%%%%%%%%%%%%%%%%%%%%%%%%%%%%%%%%
%%%%%%%%%%%%%%%%%%%%%%%%%%%%%%%%%%%%%%%%%%%%%%%%%%%%%

\vspace{-1.0cm}
There is also a novel type of striped phase in a small region of the phase diagram with $t'<0$.  In this W3 (width-3) striped phase, the stripes are predominately one site wide with exactly two rows of mostly undoped sites between them acting as a spin-ladder. The holes within stripes are unpaired with a spacing of about 4 between holes.  The Heisenberg two-leg ladder is spin-gapped, with very short range spin correlations\cite{white1994resonating}, and the ``ladders'' in the W3 phase behave similarly. The stripes in W3 do not induce a $\pi$ phase shift to the spins on either side, probably due to the combination of low doping within the stripe and the short range spin correlations. Instead of acting as a domain walls, they decouple the spin ladders. The W3 phase appears to have substantial decoupling between the undoped ladders and the doped chains, and the transverse period is locked at 3 lattice spacings. The W3 phase (like a $t$-$J$ chain\cite{xiang92}) does not exhibit pairing. 
For $t'<0$, commensurate AFM order is present only very close to zero doping; stripes break up the AFM order very quickly on doping.

For $t'$ positive above a small threshold, one enters a very different low-doped phase.  This phase has three types of order simultaneously.  The two dominant forms of order are AFM and $d$-wave superconductivity, which have also been found in recent studies of the Hubbard model with positive $t'$\cite{hubw4stp-corboz2019}. There are no signs of stripes at low doing, and the magnetic order is commensurate at $Q=(\pi,\pi)$. The $d$-wave order is robust; unlike in previous studies at $t'=0$ where determining whether the system is superconducting or not requires careful scaling, here its presence is very clear.
These two dominant orders, $d$-wave pairing and AFM, combine to form a weaker triplet $p$-wave order at wavevector $Q=(\pi,\pi)$\cite{psaltakis1983,senechal2005,foley2019coexistence,kyung2000,pitriplet-kato,pitriplet-machida,pitriplet-kato,pitriplet-murakami}. This order comes about because the AFM order breaks SU(2) symmetry, so that singlet and triplet pairing are no longer distinct, and the nonzero wavevector reflects the wavevector of the AFM order. There is no separate attractive conventional interaction in the $p$ channel; this derivative order is purely a consequence of the other two orders. 
%%%%%%%%%%%FIGURE%%%%%%%%%%%%%%%%%%%%%%%%%%%%%%%%%%%
\begin{figure}[t]
    \begin{center}
	\vspace{-0.0cm}
    \includegraphics[width=1.0\columnwidth,clip=true]{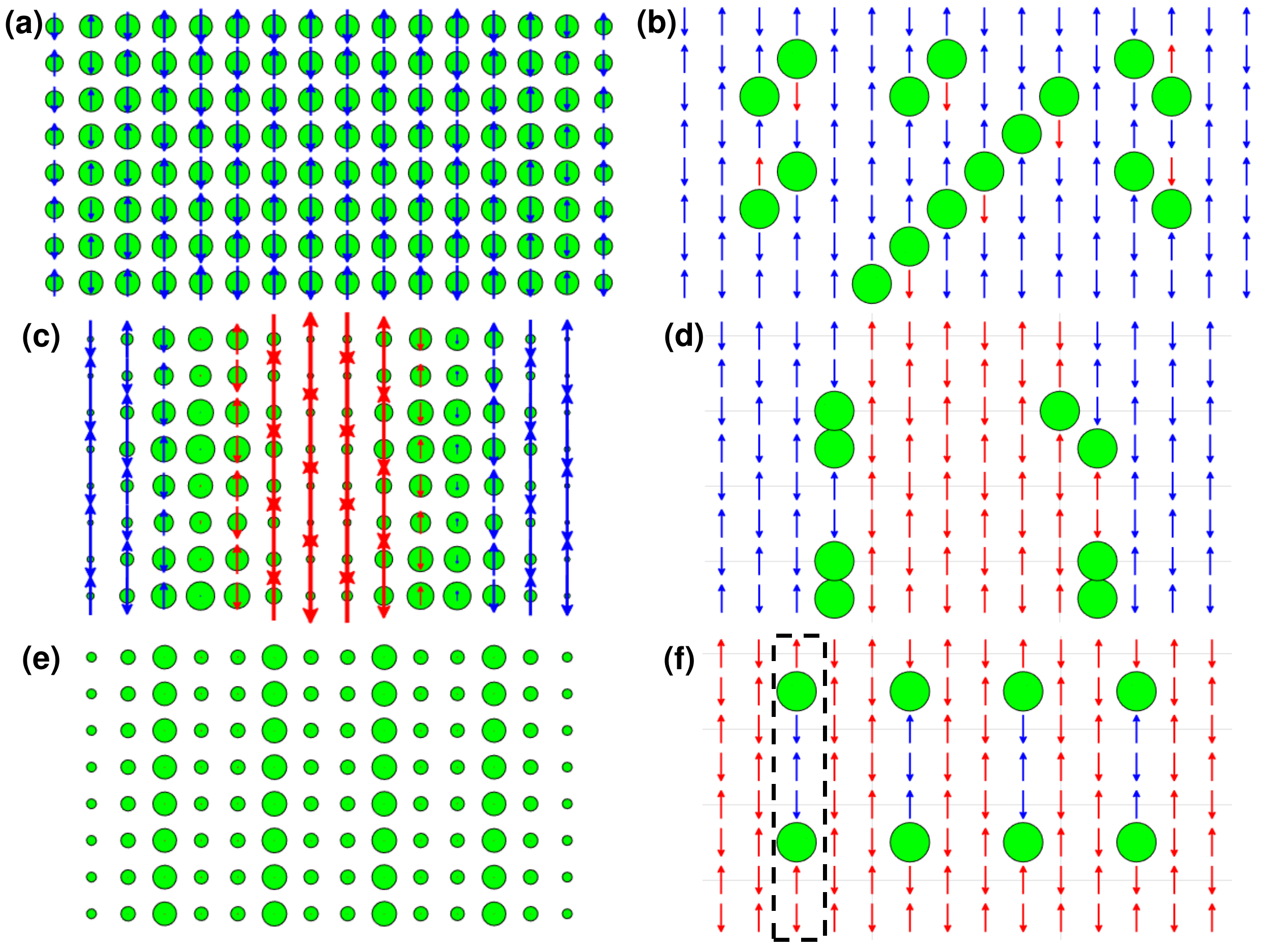}
	\vspace{-0.7cm}
	\caption{
    		\label{fig:snapshot}
    		Local properties (left) and a high-probability product state (right) of systems in different phases. The left panels show conventional local measurements for ground states in three different phases. The length of the arrows,  the diameter of the circles, and the width of the lines represent $\langle S_z\rangle$, local doping, and link singlet pairing magnitudes, respectively. The spins are colored to indicate different AFM domains. The right panels show particular product states which  occurred with maximum probability in a particular search within the corresponding state (see text). (a) and (b): electron low doped phase with $t^\prime=0.2, x=0.12$, with simultaneous pairing and AFM order. The highest probability configuration of a pair of holes is diagonal-next-nearest neighbor. (c) and (d): Conventional stripe phase at $t^\prime=0, x=0.07$, where half-filled stripes form, and pairing is visible within the stripes, but lacking phase coherence.  (e) and (f): W3 striped phase at $t^\prime=-0.2, x=0.07$, with holes unpaired within the stripes. 
    		}
	\vspace{-0.2cm}
	\end{center}
\end{figure}
%%%%%%%%%%%%%%%%%%%%%%%%%%%%%%%%%%%%%%%%%%%%%%%%%%%%%

In the higher-doped $t'>0$ region, a superconducting striped phase appears. This phase looks locally much like the lower doped phase, but with stripes.  The stripes look like conventional stripes in most respects, but they exhibit significant $d$-wave pairing, unlike the $t'<0$ striped phase. The stripes act as domain walls to the AFM order, and locally one sees derivative $\pi p$ triplet order as well, modulated by the stripes. The pairing overall is somewhat weaker in the higher doped phase, probably because of the stripes.  But while the stripes somewhat compete with superconductivity, the main driver  against pairing appears to be negative $t'$ itself, rather than $t'$ acting through stripe formation.  A likely mechanism for this is that positive $t'$ directly increases the mobility of pairs\cite{jeckelmann1998}, making it easy for them to phase cohere and to avoid becoming locked into stripes. 

%%%%%%%%%%%FIGURE%%%%%%%%%%%%%%%%%%%%%%%%%%%%%%%%%%%
\begin{center}
\begin{figure*}[t]
	\vspace{0cm}
    \includegraphics[width=1.7\columnwidth,clip=true]{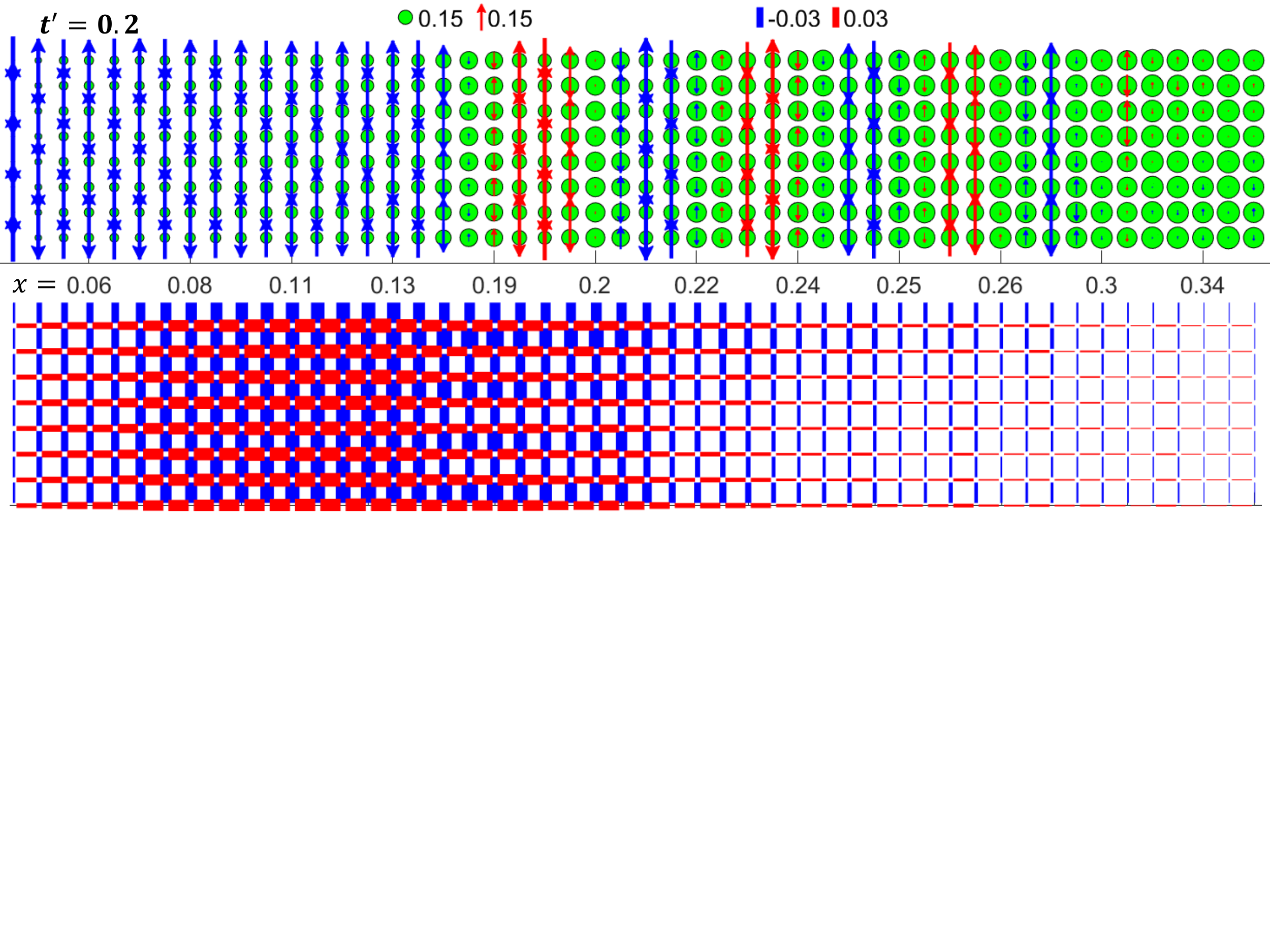}
	\vspace{-5.3cm}
    \caption{
	\label{fig:dpva0.2}
	    A doping-varying scan on a 50$\times$8 cylinder with $t^\prime$=0.2, appropriate for electron-doped cuprates. Spin, charge are shown in the upper plot in the same way as in Fig.~\ref{fig:snapshot}. The lower plot shows the $d$-wave pairing with its sign/amplitude indicated by the color/thickness of bonds.  The numbers on the middle axis indicate the approximate local doping.  No pair field is applied. In the underdoped region ($x\lesssim$0.13) the system exhibits AFM with strong $d$-wave pairing. In the overdoped region ($x>$0.13) stripes are present, with  pairing persisting to about $x \approx 0.25-0.3$.}
	\vspace{0.2cm}
\end{figure*}
\end{center}
%%%%%%%%%%%%%%%%%%%%%%%%%%%%%%%%%%%%%%%%%%%%%%%%%%%%
%%%%%%%%%%%FIGURE%%%%%%%%%%%%%%%%%%%%%%%%%%%%%%%%%%%
\begin{center}
\begin{figure*}[ht]
	\vspace{0.3cm}
    \includegraphics[width=1.7\columnwidth,clip=true]{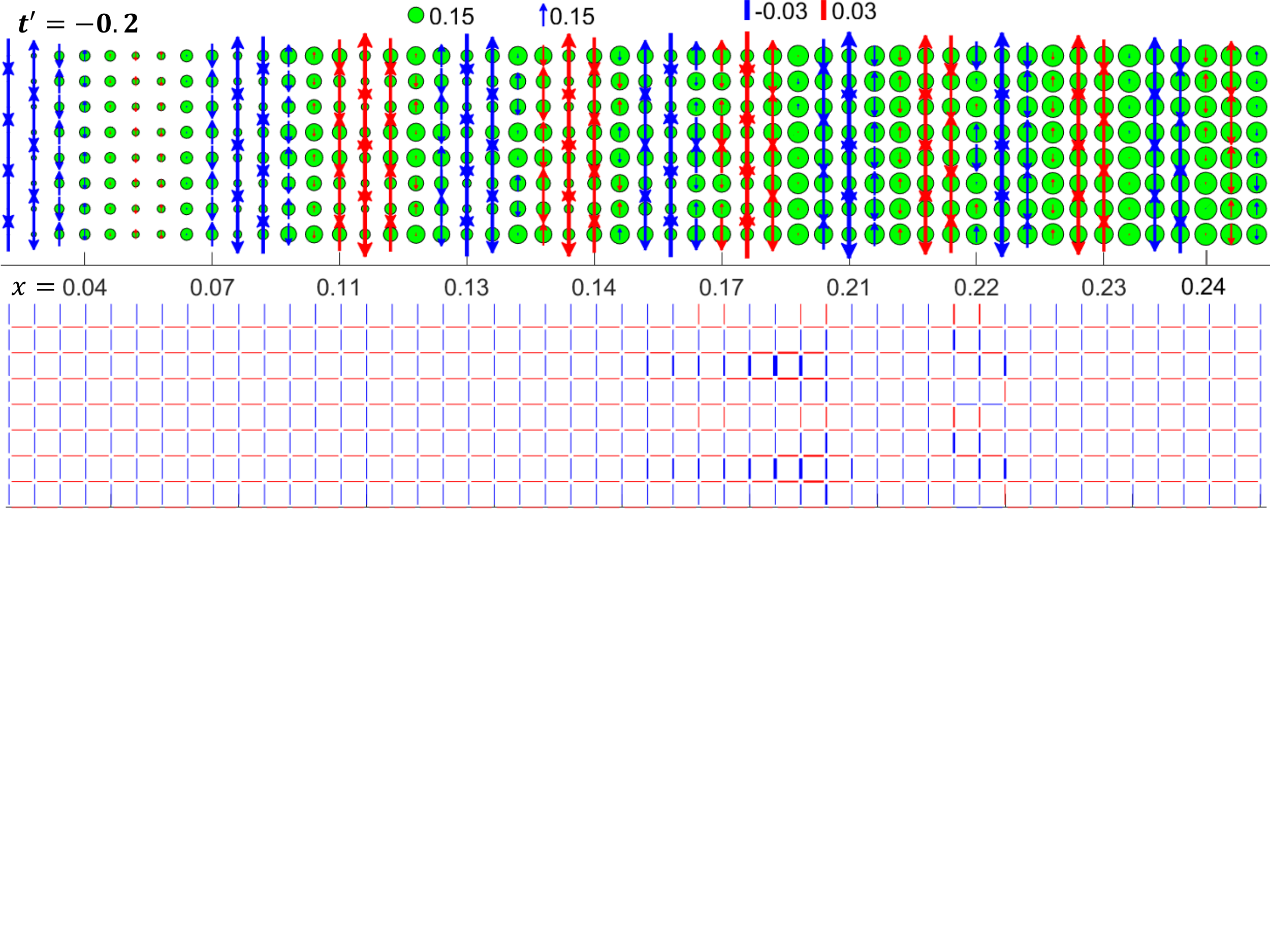}
	\vspace{-5.3cm}
    \caption{
		\label{fig:dpva-0.2}
         A doping-varying scan on a 50$\times$8 cylinder with $t^\prime$=-0.2, appropriate for hole-doped cuprates. A staggered magnetic pinning field of 0.03 is applied on the left edge. A global $d$-wave pair field of 0.005 is applied to measure the pairing response. The system exhibits stripes across the whole doping range shown here with minimal pairing response to the applied pairing field.}
	\vspace{0.0cm}
\end{figure*}
\end{center}
%%%%%%%%%%%%%%%%%%%%%%%%%%%%%%%%%%%%%%%%%%%%%%%%%%%%%

\vspace{-1.5cm}
In Fig.~\ref{fig:snapshot} we show non-scan simulations of three of the phases, emphasizing their differences.  An alternate view is given by the high-probability product state plots shown in the right panels.  These product states are determined from the ground state by a limited search for the most probable product state. One way to search for a probable spin configuration is to sequentially go through the sites of the lattice, and at each site, pick the most probable spin state.  After each spin is picked, the wavefunction is projected to reflect this, just as a physical measurement of spin $i$ (say finding the up state) would leave the wavefunction projected into the associated up-$i$ space. However, this approach fails in the presence of holes, since at low doping there are far fewer holes than spins, and the holes end up appearing at the end of the search path, i.e. mostly on the right side of the system, which is a low probability configuration. Instead, here we will search for the hole positions separately, finding the most probable position for a hole over all sites at each step, using the hole density of the projected wavefunction.  After the holes are found, then we perform a determination of the spin configuration with a fixed path through the rest of the sites, optimizing each spin and projecting.

This gives a view of the states that is hard to see in local measurements or correlation functions.  In particular, in the $d$-wave phase in panel (a)-(b), one sees the holes grouped in pairs, with the most probably configuration of a pair being diagonally next-nearest neighbor, as found in earlier work\cite{white1997hole}. In panel (b) one sees an apparent diagonal stripe, but there is no sign of this in the ordinary measurements of panel (a).  It may be that this is only slightly more likely than many other different configurations.

In the conventional striped state shown in panels (c)-(d),
pairs of holes appear as the most probable state, but the state has only short ranged pairing correlations (not shown) and the small hole probability  between the stripes suggests that the binding of the pairs to the stripes is suppressing superconductivity. In the W3 striped state shown in panels (e) and (f), the most probable state has the holes at their maximum separation within the stripes, with domain walls visible in the vertical direction across the holes, instead of the horizontal.   This configuration is consistent with the idea that the stripes here imitate the 1D $t-J$ chain \cite{xiang92}, and that one can view the holes as holons living in a squeezed Heisenberg chain of spins.  

%%%%%%%%%%%FIGURE%%%%%%%%%%%%%%%%%%%%%%%%%%%%%%%%%%%
\begin{center}
\begin{figure*}[ht]
	\vspace{0.0cm}
    \includegraphics[width=1.7\columnwidth,clip=true]{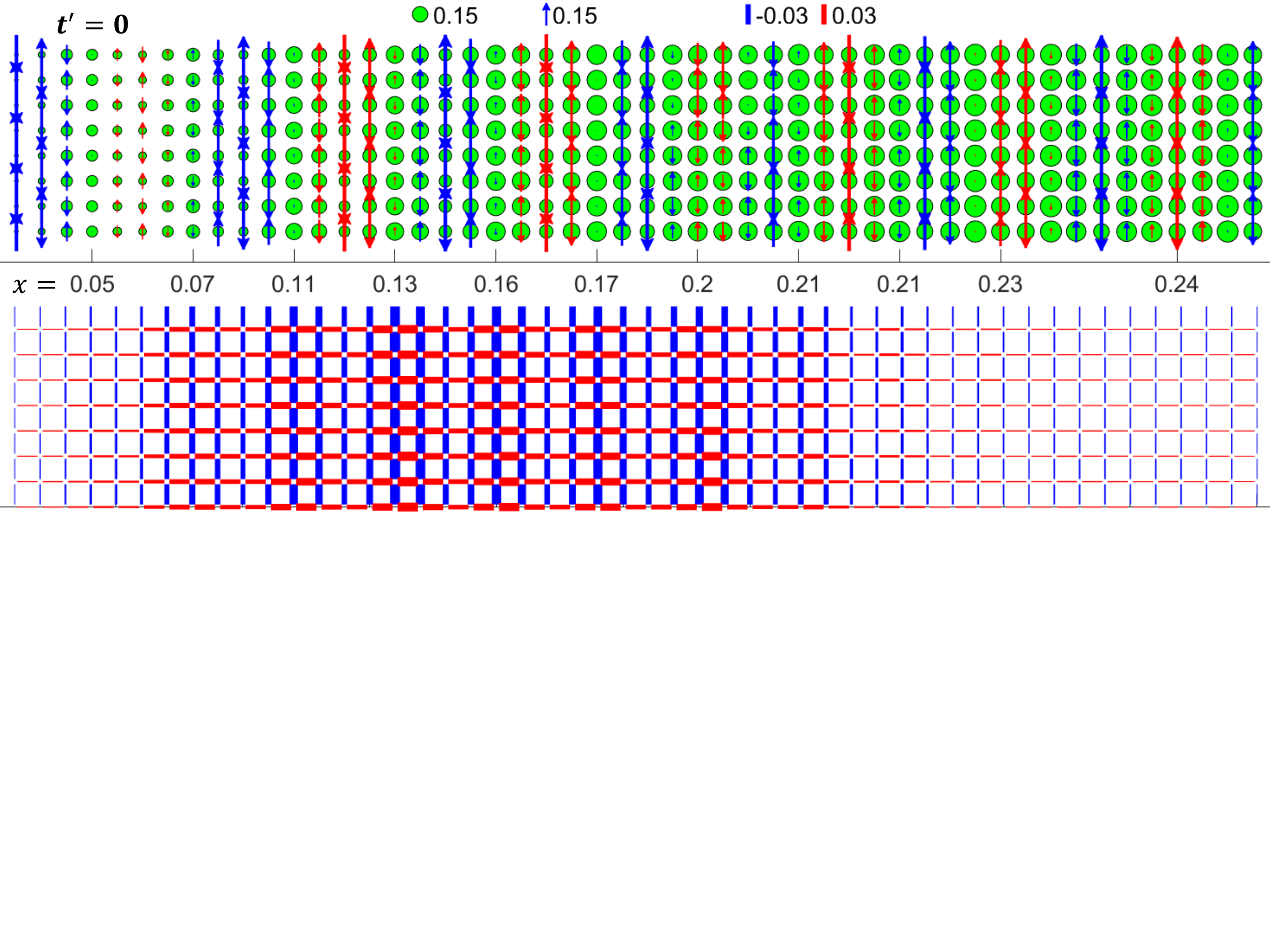}
	\vspace{-5.4cm}
	\caption{
		\label{fig:dpva0}
          A doping-varying scan on a 50$\times$8 cylinder with $t^\prime$=0.  A staggered magnetic pinning field of 0.05 is applied on the both edges. A global $d$-wave pairfield of 0.005 is applied to measure the pairing response. The system is striped and pairing response peaks near $x$=0.15, but without the applied field (not shown) the system shows no local pairing and the pair-pair correlations die rapidly with separation. }
	\vspace{0.4cm}
\end{figure*}
\end{center}
%%%%%%%%%%%%%%%%%%%%%%%%%%%%%%%%%%%%%%%%%%%%%%%%%%%%%
%%%%%%%%%%%FIGURE%%%%%%%%%%%%%%%%%%%%%%%%%%%%%%%%%%%
\begin{figure*}[ht]
\begin{center}
	\vspace{0.0cm}
    \includegraphics[width=1.7\columnwidth,clip=true]{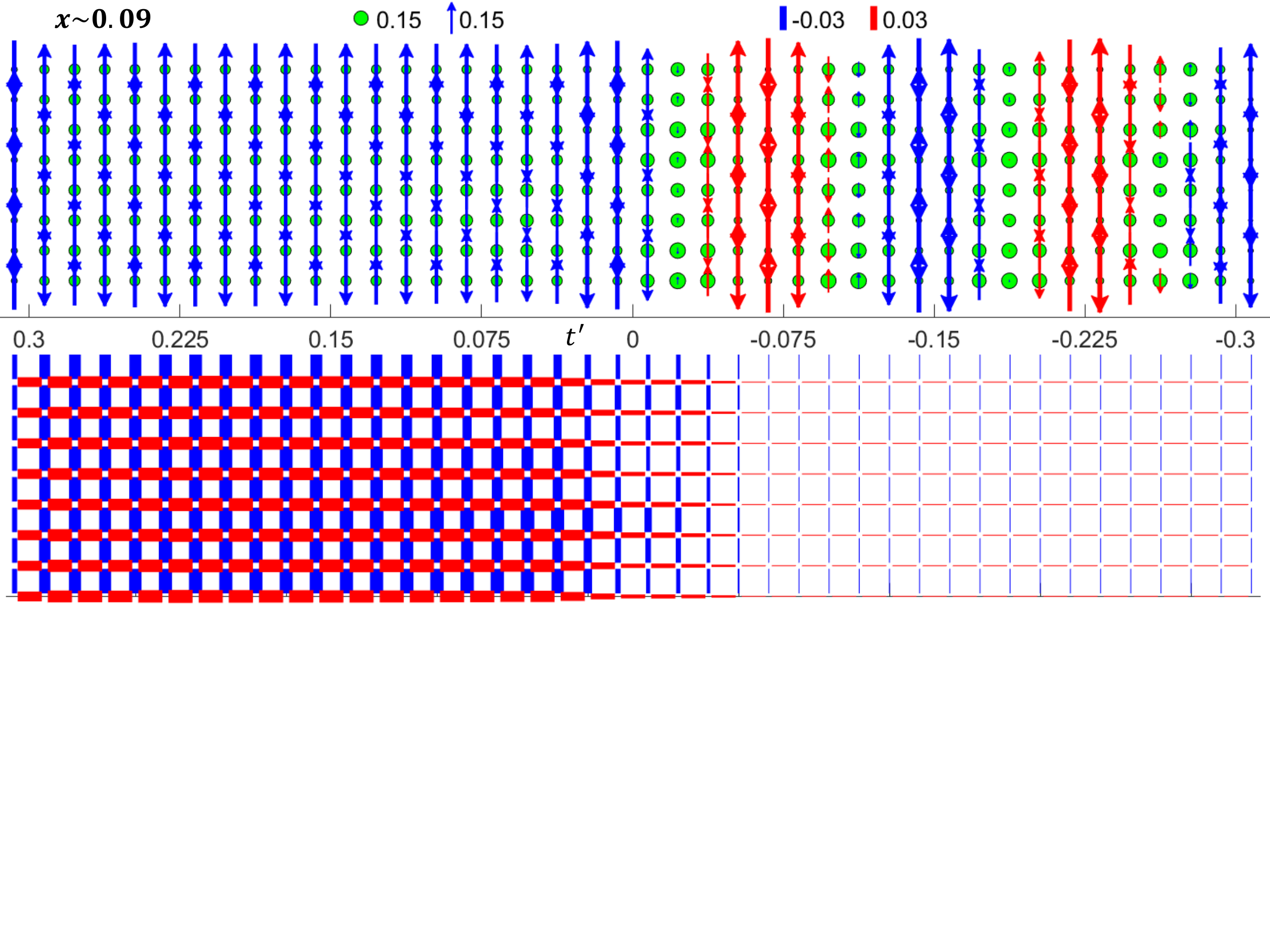}
	\vspace{-4.2cm}
	\caption{
		\label{fig:tpva0.09}
          A $t'$-varying scan on a 42$\times$8 cylinder with doping $x \sim 0.09$.A staggered magnetic pinning field of 0.03 is applied on both edges. No pair field is applied. For $t^\prime>0$ we see the AFM-$d\pi p$ phase, while in the region with negative $t^\prime$ we find a conventional stripe phase and pairing is suppressed.}
	\vspace{-0.2cm}
\end{center}
\end{figure*}
%%%%%%%%%%%%%%%%%%%%%%%%%%%%%%%%%%%%%%%%%%%%%%%%%%%%%

\vspace{-1.0cm}
\section{Scans varying doping and $t'$}
\label{sec:scan}
We now discuss the scan calculations which were used in constructing the phase diagram. As shown in Fig.~\ref{fig:dpva0.2}, in one set of scans, $t'$ was fixed and the doping $x$ was slowly varied along the length of the cylinder while in another set $x$ was fixed and $t'$ was varied.  
A linear variation of the chemical potential with the length $l_x$ down the cylinder gave an essentially linear increase of the doping x. However, for the $t'$=0.2 scan, it was necessary to vary the chemical potential slowly in the low doping region where AFM, $d$-wave and $\pi-p$-wave triplet pairing appeared.
In the $t^\prime$ varying case, the chemical potential also needed to be adjusted to keep the doping approximately constant across the system\footnote{The chemical potential is of form $\mu(l_x)=\mu_0+a\sqrt{1+(b|2l_x-L_x|/L_x)^2}$ with $a$ and $b$ to be adjusted and different for left and right half. This form of chemical potential varies slower and connects smoothly in the middle.}.

A key feature of the scan calculations is to reduce the problem of metastable states.  For fixed parameters (i.e. a non-scan calculation) one may happen to be near a phase boundary, and determining which side one is on may involve small energy differences. In contrast, using a scan one is likely to pass through the phase boundaries, and the system will automatically adjust the location of the boundaries to account for the energies. The calculations are stabilized by the parts of the cylinder where the system is well within one phase or another.

%%%%%%%%%%%FIGURE%%%%%%%%%%%%%%%%%%%%%%%%%%%%%%%%%%%
\begin{figure*}[ht]
\begin{center}
	\vspace{-0.0cm}
    \includegraphics[width=1.7\columnwidth,clip=true]{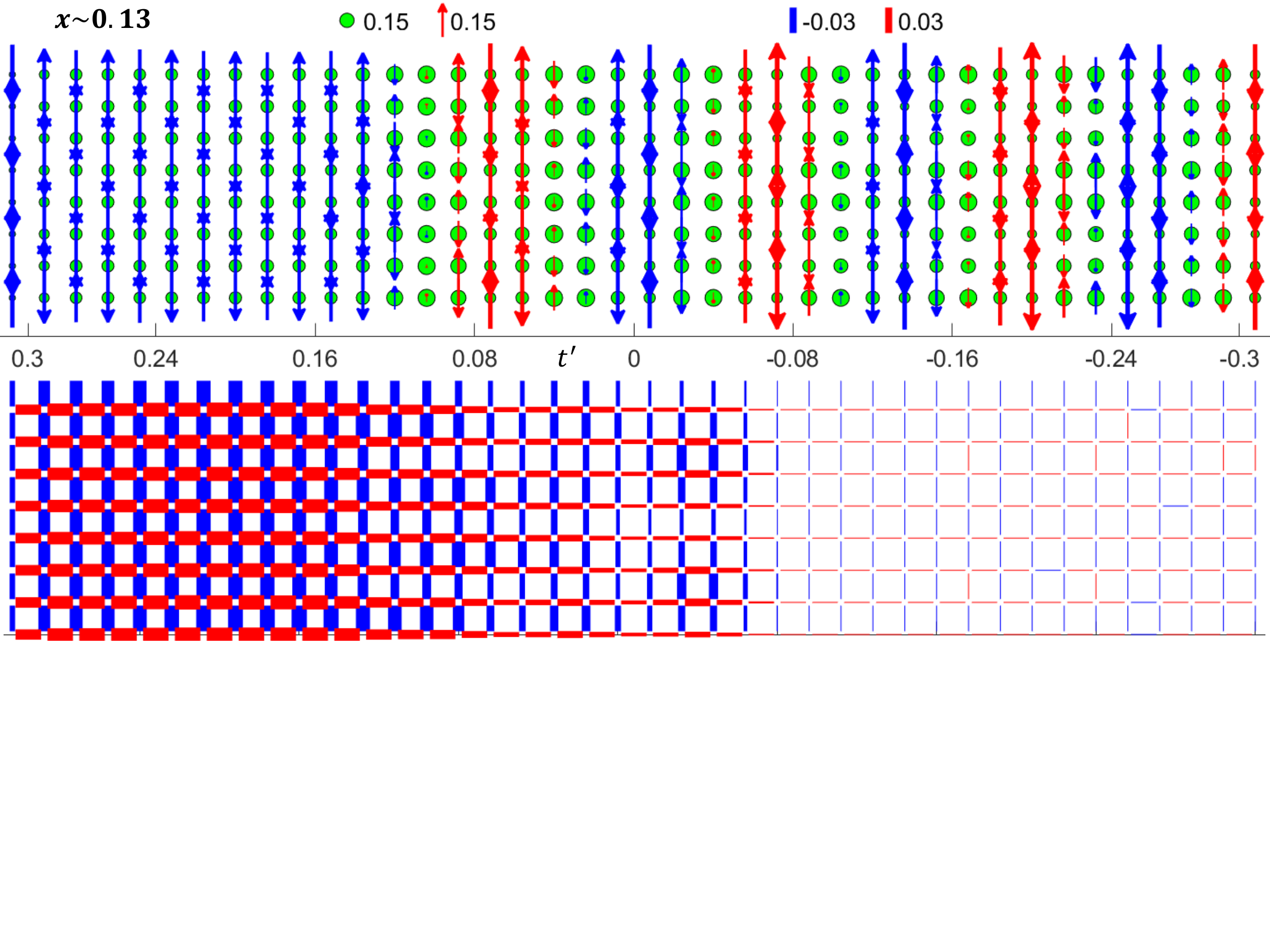}
	\vspace{-3.8cm}
	\caption{
		\label{fig:tpva0.13}
          A $t'$-varying scan on a 40$\times$8 cylinder with doping $x \sim 0.13$. No pair field is applied. A staggered magnetic pinning field of 0.03 is applied on both edges.  For $t^\prime>0$ we see the AFM-$d\pi p$ phase which becomes striped for smaller $t'$. In the negative $t^\prime$ region the stripes continue but without pairing.}
	\vspace{-0.2cm}
\end{center}
\end{figure*}
%%%%%%%%%%%%%%%%%%%%%%%%%%%%%%%%%%%%%%%%%%%%%%%%%%%%%
%%%%%%%%%%%FIGURE%%%%%%%%%%%%%%%%%%%%%%%%%%%%%%%%%%%
\begin{figure*}[ht]
    \begin{center}
	\vspace{-0.3cm}
    \includegraphics[width=1.7\columnwidth,clip=true]{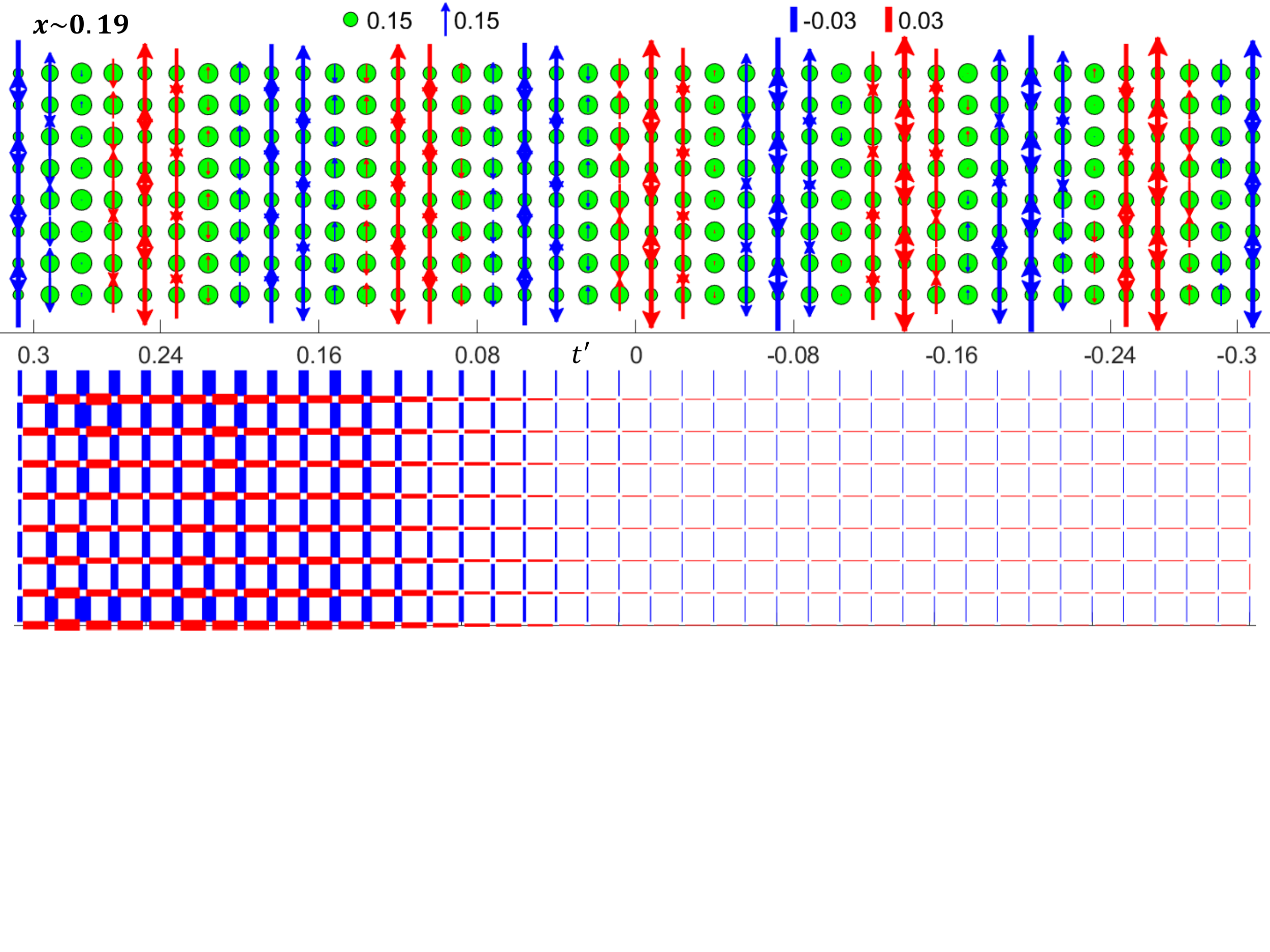}
	\vspace{-4.0cm}
	\caption{
    		\label{fig:tpva0.19}
          A $t'$-varying scan on a 40$\times$8 cylinder with doping $x \sim 0.19$. No pair field is applied. A staggered magnetic pinning field of 0.03 is applied on both edges.  There are stripes across the whole system, but pairing only for larger positive $t^\prime$.}
	\vspace{-0.2cm}
	\end{center}
\end{figure*}
%%%%%%%%%%%%%%%%%%%%%%%%%%%%%%%%%%%%%%%%%%%%%%%%%%%%%

In non-scan calculations where one is looking for a particular order, it is common to ``pin the edges'' with a corresponding field applied to the edge sites. This is not so clear-cut a procedure for a scan going through different phases, but, in fact, often in a DMRG calculation pinning fields are not necessary. Instead, DMRG can self-pin in a large 2D system. This aspect of DMRG calculations is well-known among DMRG experts, but less so by others, so we give a detailed explanation of this effect in Appendix A.  The gist is that DMRG tends to break a continuous symmetry and develop an order parameter just as a real experimental sample does.  The broken symmetry goes away in the limit of large bond dimension, but for a range of moderate bond dimensions the system develops an order parameter similar in magnitude to that of the 2D thermodynamic limit. The required bond dimension to eliminate the broken symmetry increases rapidly with system size.  Attempts to converge beyond this broken symmetry plateau can be counter-productive, since in the symmetric phase the order can only be seen through correlation functions, and one can miss unexpected orders. In addition, correlation functions are inherently much less accurate than local measurements\cite{White_and_Chernyshev}. In our scans, we use this effect to our advantage: in systems where there does appear to be robust $d$-wave superconductivity(SC), we do not pin it with an external field but rather we allow the system to self-pin. In systems where $d$-wave SC is suppressed, we apply a weak global pairing field.  In this case, we may get some false positive signals of 
SC, but its absence is a clear sign that a superconducting state is not a low energy state. In some cases we also apply a magnetic pinning field on one or both edges to reduce edge effects\cite{2ddmrg-miles}. It should be pointed out that such fields have almost no effect on the magnetic order in the bulk which appear as long as we are in the ``broken symmetry plateau''.

Figure~\ref{fig:dpva0.2} shows an $x$-varying scan with a fixed $t^\prime=0.2$ corresponding to electron-doped cuprates. 
In the underdoped region we find coexisting uniform AFM, strong $d$-wave singlet pairing and $(\pi,\pi)~p$-wave triplet pairing (detailed discussion later in Sec.~\ref{sec:dpip}). 
As one increases doping away from half-filling, the pairing increases rapidly and $|\langle S_z \rangle|$ decreases slowly.  
When the doping is further increased beyond $x\sim0.14$, conventional-looking stripes emerge. The transition to stripes is rather sharp. The stripes still have robust pairing, but the magnitude of the order parameter is reduced.  The point of optimal doping, where pairing is maximum, is near $x=0.14$, in the uniform phase. Within the striped phase, pairing decreases with higher doping. It eventually disappears in a smooth way near $x \sim 0.25-0.3$

Figure~\ref{fig:dpva-0.2} shows a similar $x$-varying scan but with  $t^\prime=-0.2$, corresponding to hole-doped cuprates.
Other than a small region showing signs of the W3 striped phase around $x\sim0.06$, the whole scan shows conventional stripes.
As one increases doping, the magnitude of the density oscillations first increases until around $x=0.2$, and then decreases.
Pairing is almost completely suppressed despite a global $d$-wave pairfield of 0.005. 
Pairing remains weak even if the pairfield is made rather strong, say, $~0.03$.  In terms of the pairing response and the applicability of the $t$-$t^\prime$-$J$ model to the cuprates, Figure~\ref{fig:dpva0.2} and Fig.~\ref{fig:dpva-0.2} indicate a clear contradiction:  pairing is much stronger in the hole doped cuprates than we find in the $t$-$t'$-$J$ model.

Figure~\ref{fig:dpva0} shows an $x$-varying scan with $t^\prime=0$. While this case does not directly map to the cuprates, it has been studied often because of its simplicity.
We find a hole density and spin pattern similar to the $t^\prime=-0.2$ scan. 
Around $x\sim0.07$ there is again some signs of a W3 striped phase, but separate calculations with fixed doping find this W3 stripe is meta-stable and higher in energy ($o(0.001t)$ per site) than the conventional striped phase at $t^\prime=0$.
Despite a similar striped structure, the pairing response with $t^\prime=0$ is much stronger than at $t^\prime=-0.2$. 
Under a global $d$-wave pairfield of 0.005, the pairing peaks around $x=0.15$ with a value $\langle \Delta^\dag+\Delta\rangle \sim 0.06$.
If the pairfield is turned off, pairing decays slowly. Further calculations (not shown) indicate that at $t^\prime=0$, the paired state is not the ground state.  In the global phase diagram, the boundary line where pairing appears is at slightly positive $t^\prime$. 
This closeness of the boundary helps explain why in previous studies at $t^\prime=0$, it has been very difficult to determine if the ground state is superconducting.

We now show another three scans where we keep $x$ approximately constant  and vary $t^\prime$ from 0.3 to -0.3.
Figure~\ref{fig:tpva0.09} shows a low doping case, $x\sim0.09$.  The contrast between positive and negative $t^\prime$ is striking. 
For $t^\prime>0$ we find the AFM-$d/\pi p$ phase with uniform AFM and strong pairing.
For $t^\prime<0$ we find conventional stripes and a rapid and strong suppression of the pairing response. 

Figure~\ref{fig:tpva0.13} shows a medium doping $x\sim 0.13$. 
Here the boundary of uniform density versus stripe order has shifted to $t^\prime\sim 0.1$. The striped state for $t'>0$ has pairing, although it is weaker than in the AFM-$d/\pi p$ phase. 

Figure~\ref{fig:tpva0.19} shows a high doping case with $x\sim 0.19$.
In this case there are stripes for the whole range of  $t^\prime$. However, pairing order is only present for $t^\prime \gtrsim 0.1$.

\section{Electron low-doped phase with coexisting uniform AFM, $d$-wave singlet and $(\pi,\pi)~p$-wave triplet pairing (AFM-$d/\pi p$)}
\label{sec:dpip}
%%%%%%%%%%%FIGURE%%%%%%%%%%%%%%%%%%%%%%%%%%%%%%%%%%%
\begin{center}
\begin{figure*}[t]
	\vspace{-0.0cm}
	\hspace*{-0.4cm}
    \includegraphics[width=1.5\columnwidth,clip=true]{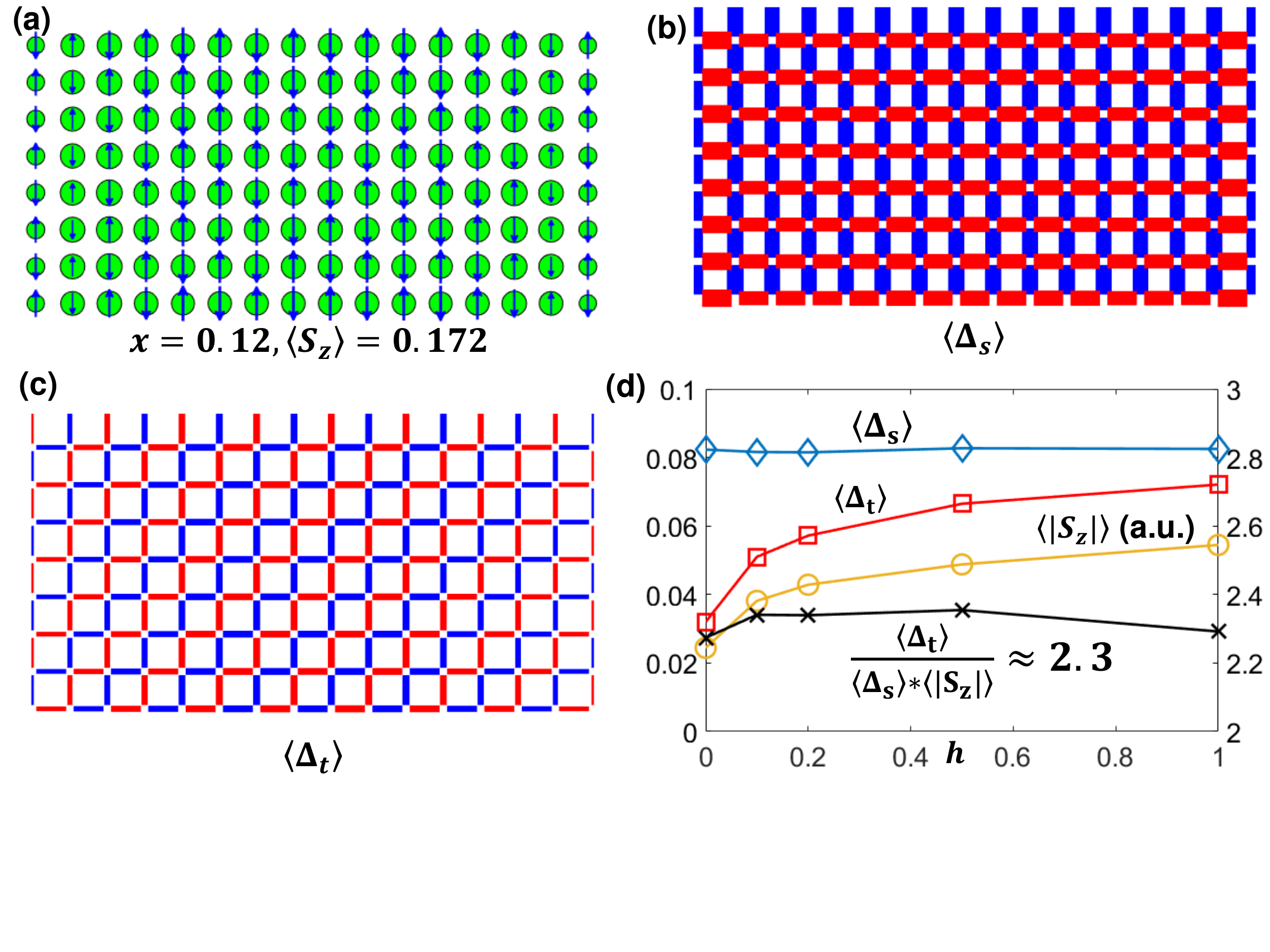}
	\vspace{-1.9cm}
	\caption{
		\label{fig:afm}
         The AFM $d/\pi p$ phase at fixed $t^\prime$=0.2 and $x=$0.12. (a) and (b) show spin, charge, and $d$-wave singlet pairing as in previous plots. In (c), we show triplet link pairing, where we find $(\pi,\pi)~p$-wave order. For (d), we plot singlet and triplet pairing, as well as the spin expectation value (left-axis scale), for systems which have had a global staggered magnetic field $h$ applied; each value of $h$ is a different simulation. The singlet pairing is nearly independent of $h$  while the triplet pairing and magnetization both increase with $h$, but the indicated ratio (black crosses, right axis) is nearly constant. 
         }
	\vspace{-0.2cm}
\end{figure*}
\end{center}
%%%%%%%%%%%%%%%%%%%%%%%%%%%%%%%%%%%%%%%%%%%%%%%%%%%%%

\vspace{-1.0cm}
We now consider the individual phases in detail, starting with the phase with three order parameters at low doping and $t^\prime>0$. 
This region corresponds to electron-doped cuprates.
There are two dominant orders: uniform AFM and a strong $d$-wave singlet pairing. As a result of these two orders there is also a $(\pi,\pi)~p$-wave triplet pairing.  We call this phase AFM-$d/\pi p$, and it exists in a broad region defined roughly by $t^\prime>0$ and $x<0.14$.
Details of this phase for a non-scan calculation at
 $x=0.12$ and $t^\prime=0.2$ are shown in Fig.~\ref{fig:afm}. In parts (a) and (b) we show the doping, spin, and singlet pairing expectation values.
All these quantities are uniform across the system. 
No applied pairing field was used.
The presence of nonzero pairing order helps the density become more uniform, counteracting any oscillations due to the open boundaries. The magnitude of the pairing order is 
$\langle \Delta^{ \dag}_{s}+\Delta_{s} \rangle =0.081$, and the difference in sign between vertical and horizontal bonds signifies $d$-wave order. 
We judge the size of the order parameter to be quite large; in particular, if one does apply a pairing field, one cannot readily make it much larger. Also, the simulations are clear and unambiguous; there do not appear to be any other competing states.
Starting from a non-pairing initial product state,  the system spontaneously breaks particle-conservation symmetry and produces the pairing order.
As mentioned before,  Appendix~\ref{sec:broksym} has a detailed discussion for spontaneously broken symmetry in DMRG.

As shown in Fig.~\ref{fig:afm}(c) we also observe a smaller $(\pi,\pi)~p$-wave triplet pairing in addition to the strong $d$-wave singlet pairing. It is uniform in amplitude and has $(\pi,\pi)~p_x-p_y$ form:
\begin{equation}
    \langle \Delta_{t}(l_x,l_y) \rangle =e^{i\pi(l_x+l_y)}[\langle \Delta_{t}(l_x,l_y,x) \rangle
    -\langle \Delta_{t}(l_x,l_y,y) \rangle]
\end{equation}
with $\Delta_{t}(l_x,l_y,x/y)$ being a triplet pairing on a horizontal/vertical link with  left/lower site($l_x,l_y$).
The overall phase of triplet pairing is determined by the overall phase of the AFM order and $d$-wave pairing. This triplet order is a consequence of the other two orders, not a competing order. 
As mentioned before, The existence of AFM order breaks SU(2) spin symmetry, so that singlet and triplet pairing are no longer distinct, making the $d$-wave pairing have a triplet component.  The nonzero wavevector reflects the wavevector of the AFM order.
The magnitude of the triplet pairing is roughly proportional to the singlet pairing: $ \langle \Delta_{t} \rangle / \langle \Delta_{s} \rangle \approx 0.4$, if no magnetic field is applied, and this ratio is mostly $t'$ independent.

To further study the interplay of AFM, singlet and triplet pairing, we applied a global staggered magnetic field to the system which directly enhances the AFM order.
Figure~\ref{fig:afm}(d) shows that under this field, both magnetization and triplet pairing are enhanced while singlet pairing is mostly unchanged.  It is interesting that there is no competition between strong AFM order and $d$-wave pairing; they happily coexist, but as a consequence of increased AFM order the triplet order gets larger. 
Defining $A(x)$ through the following relation between these three quantities:
\begin{equation}
    \langle \Delta_{t} \rangle=A(x)
    \langle \Delta_{s} \rangle  \langle S_z \rangle,
\end{equation}
we find that $A(x)$ varies slowly with doping (2.3 for $x=0.12$, 2.0 for $x=0.065$).
We further verify that this relation holds when a global $d$-wave singlet pairfield is applied.
This relation is qualitatively consistent with several studies where there is coexistence of AFM, $d$-wave singlet pairing and $(\pi,\pi)~p-$wave triplet pairing\cite{foley2019coexistence,almeida2017induced,rowe2012spin}.
This further implies that this $(\pi,\pi) p$-wave triplet pairing is purely parasitic and relies on the existence of both AFM and $d$-wave singlet pairing.
    
\section{Higher electron doping phase: stripes with $d$-wave singlet and striped $p$-wave pairing}
\label{sec:stp}
%%%%%%%%%%%FIGURE%%%%%%%%%%%%%%%%%%%%%%%%%%%%%%%%%%%
\begin{center}
\begin{figure}[ht]
	\vspace{-0.0cm}
    \includegraphics[width=1.0\columnwidth,clip=true]{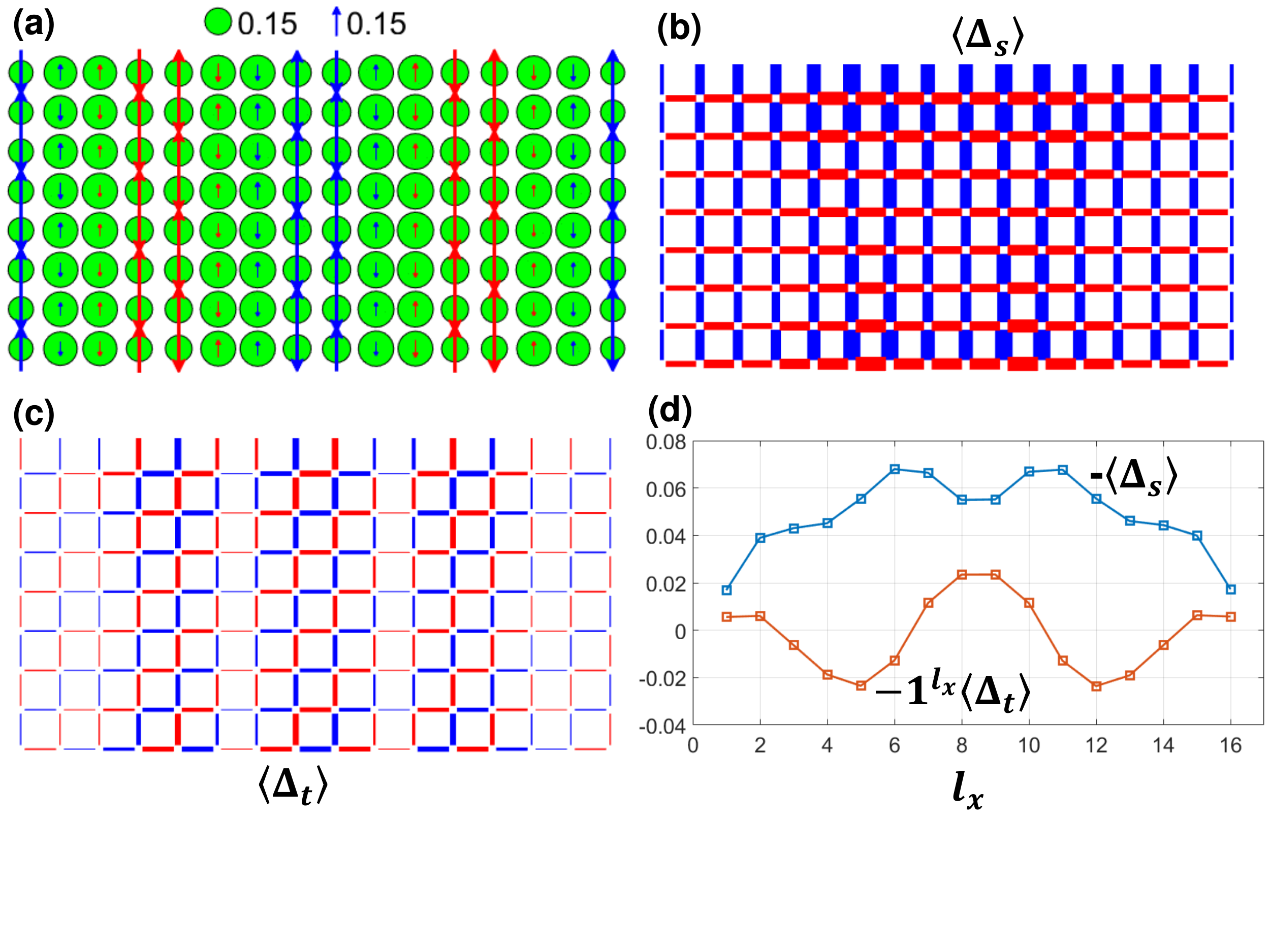}
	\vspace{-1.6cm}
	\caption{
		\label{fig:stp}
          Higher doped positive $t'$ phase with stripes and pairing.  The doping is $x=0.20$ and $t^\prime=0.2$. (a), (b), and (c) are as in the previous figure.  In (c), we see that the $p$ pairing is modulated by the domain walls in the antiferromagnetism. This is apparent in (d), which shows the singlet pairing and the triplet pairing multiplied by $-1^{l_x}$ versus $l_x$. The singlet pairing has small modulations with peaks at the stripes,  while the triplet pairing  oscillates  with nodes at the stripes.
         }
	\vspace{-0.2cm}
\end{figure}
\end{center}
%%%%%%%%%%%%%%%%%%%%%%%%%%%%%%%%%%%%%%%%%%%%%%%%%%%%%

\vspace{-1.0cm}
In the electron overdoped, $t^\prime>0$ region of the phase diagram we observe a striped phase with roughly uniform $d$-wave singlet pairing and modulated $p$-wave triplet pairing.
In Fig.~\ref{fig:stp}(a-c) we show local expectation values for a point in this phase at $x=0.20$ and  $t^\prime=0.2$. 
The striped phase looks like stripes at $t^\prime<0$ if one only looks at the charge and spin. 
Unlike that case, here we have clear $d$-wave pairing, although not as strong as at lower doping.
In this case, starting in a product state, the system can get stuck in an unpaired state, but applying a pairfield for a few sweeps allows it to go to the lower energy (by about $o(0.001t)$ per site) paired state with a singlet pairing order$\langle \Delta^\dag_{s}+\Delta_{s} \rangle$=0.044. (In a width 6 system, the unpaired state is not metastable; starting from a product state, the DMRG sweeps readily find the paired state. More comparisons with width 6 systems are made in Appendix ~\ref{sec:w6}.)
The pairing is only slightly larger on the domain walls compared to the region in-between them, as shown in  Fig.~\ref{fig:stp}(d). 

Because there are local AFM regions between the stripes, one would expect also a triplet pairing component to appear. 
Figure~\ref{fig:stp}(c) shows the $p$-wave triplet pairing for this system. Interestingly, the $p$-wave triplet pairing $\langle \Delta_{t} \rangle$ which is modulated by stripes shows a wave-like amplitude as one can see more clearly in Fig.~\ref{fig:stp}(d). 
In contrast to the $\Delta_{s}$ which is only slightly bigger at the domain walls, the $\Delta_{t}$ order has nodes at the domain walls, reflecting it parasitic dependcen on the AFM order.

\section{Conventional stripe phase and low-doped width-3(W3) stripe phase}
\label{sec:csandw3}
While the striped phase described in the previous section has a ground state with pairing, in a broader parameter region which includes the whole $t'<0$ part except for extreme low doping, stripes still form but the ground state has no pairing. This phase is the conventional striped phase.

The hole and spin pattern of conventional stripes without pairing is very similar to stripes with pairing. One small difference, which one can see 
by comparing Fig.~\ref{fig:dpva0.2} and Fig.~\ref{fig:dpva-0.2}, is that the stripes at $t^\prime=0.2$ are homogeneous in hole density and spin along the stripes, while at $t^\prime=-0.2$ there is a small modulation along each stripe. 
The difference in pairing is much more significant: the conventional striped state is non-superconducting.
A state with pairing is nearby for $t^\prime$ near 0, and it can be seen as a metastable state in DMRG, but its energy rises as $t^\prime$ is made more negative and it is no longer metastable.
For $t^\prime=-0.2$, even  a strong global pairfield  triggers only a weak pairing response and the pair-pair correlation function shows exponential decay.

%%%%%%%%%%%FIGURE%%%%%%%%%%%%%%%%%%%%%%%%%%%%%%%%%%%
\begin{figure}[t]
    \begin{center}
	\vspace{-0.0cm}
    \includegraphics[width=1.0\columnwidth,clip=true]{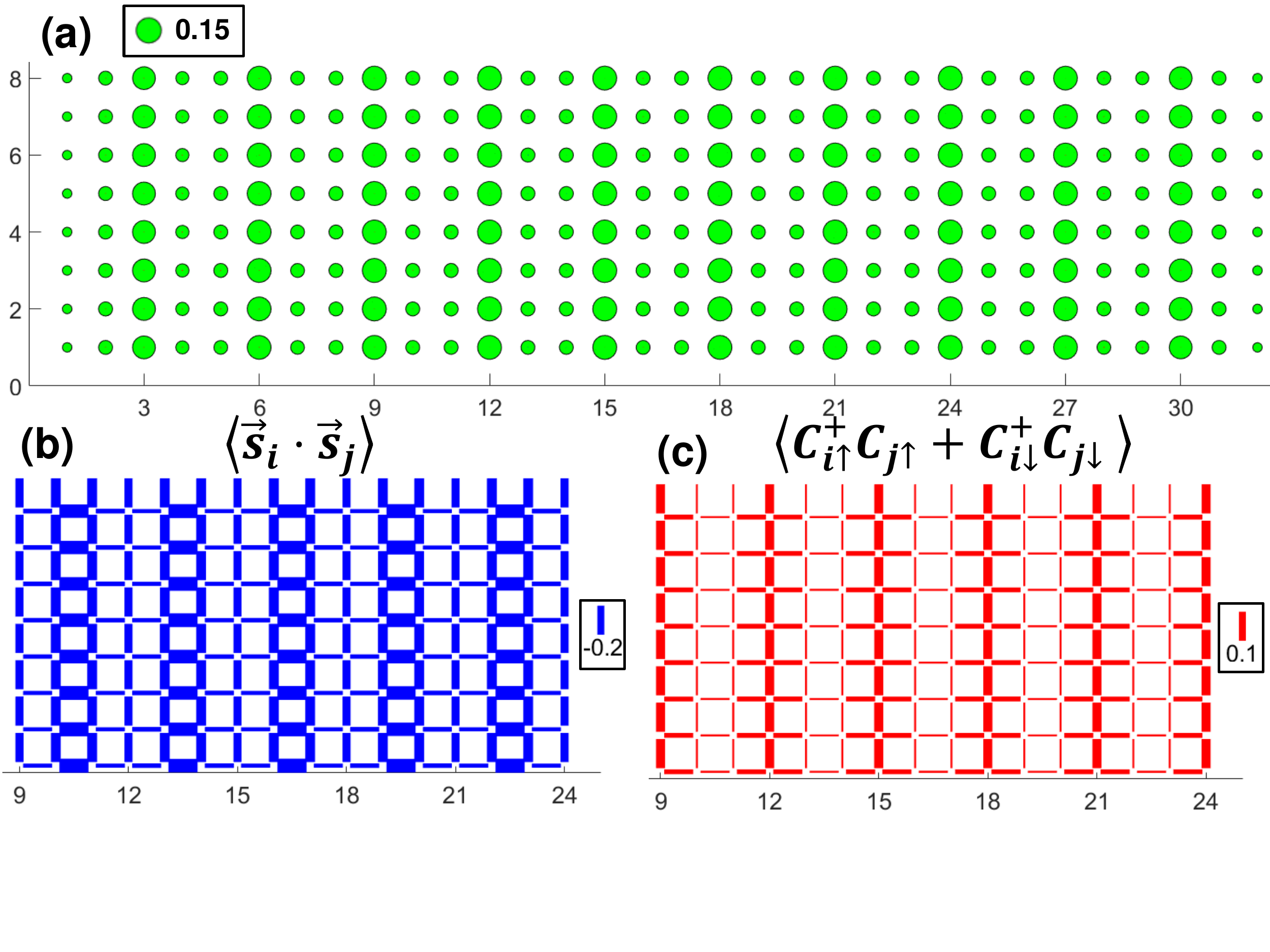}
	\vspace{-1.3cm}
	\caption{
    		\label{fig:w3stp}
    		W3 striped phase on a 32 $\times$ 8 cylinder at $x=0.08,~t'=-0.2$. In (a) only the local doping is shown; the local spin measurements are zero. The stripes are strongly associated with single columns. In (b)  we show the nearest neighbor spin-spin correlations, which are much stronger in the undoped width-two ``ladders''.  Longer distance spin-spin correlations decay rapidly (not shown). In (c) we show the link hopping, which is very strong along the stripes but also exhibits limited hops onto the ladders.  The results together suggest significant decoupling between the chains and ladders.
    		}
	\vspace{-0.2cm}
	\end{center}
\end{figure}
%%%%%%%%%%%%%%%%%%%%%%%%%%%%%%%%%%%%%%%%%%%%%%%%%%%%%
%%%%%%%%%%%FIGURE%%%%%%%%%%%%%%%%%%%%%%%%%%%%%%%%%%%
\begin{figure*}[ht]
    \begin{center}
	\vspace{-0.0cm}
    \includegraphics[width=1.7\columnwidth,clip=true]{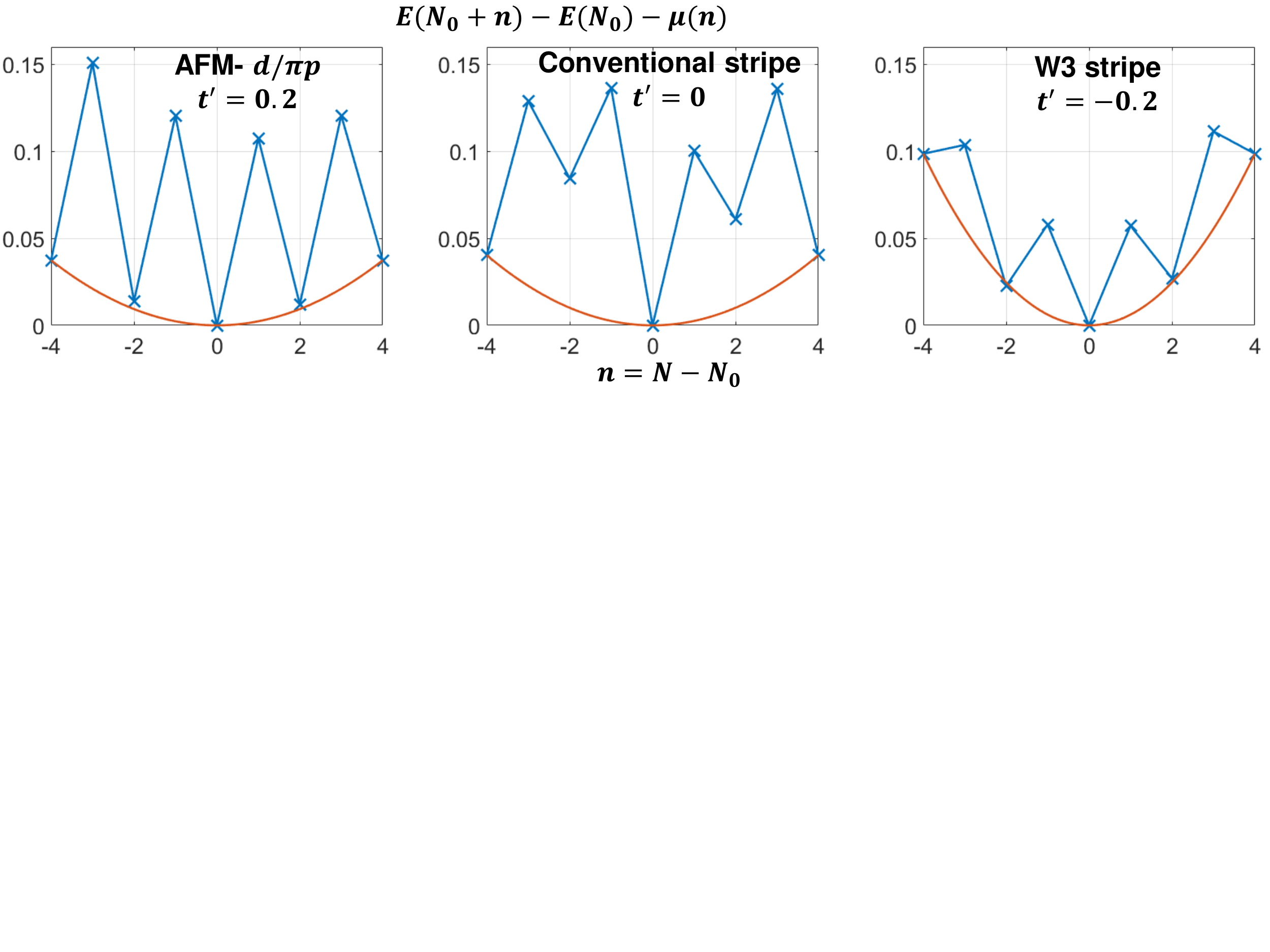}
	\vspace{-6.7cm}
	\caption{
    		\label{fig:gap}
    		Energy versus number of particles, from which one can read off various gaps. The overall curvature is the result of using a finite length cylinder ($32\times6$); the orange curve shows a quadratic fit to the points touching it. The chemical potential has been set to make the slope at the midpoint zero. Left: AFM-$d/\pi p$ phase with $t'$=0.2 at $x\sim 0.08$.  Middle: conventional striped phase with $t'$=0 at $x\sim 0.08$.   Right: W3 striped phase with $t'=-0.2$ at $x \sim 0.07$. Here the W3 stripe runs horizontally on width 6, displayed in Appendix B.
    		}
	\vspace{-0.2cm}
	\end{center}
\end{figure*}
%%%%%%%%%%%%%%%%%%%%%%%%%%%%%%%%%%%%%%%%%%%%%%%%%%%%%

The width-3 (W3) striped phase is distinct from the conventional striped phase, although both occur for $t^\prime<0$. 
Figure~\ref{fig:w3stp} shows non-scan results for the W3 phase, at a doping of 0.08, at $t^\prime=-0.2$. The key to understanding the W3 phase is a Heisenberg two-leg ladder.  A two leg ladder has a spin gap of about $J/2$, and we can think of this gap not just as the raising of excited state energies, but also the lowering of the ground state energy, making width two undoped ladder regions favored. The two-leg Heisenberg spin ladder has short-range spin correlations, with a correlation length of about 3.19\cite{white1994resonating}. This is in contrast to, say, a three leg ladder which is gapless with power law spin correlations. We do not find a W4 phase similar to the W3 phase but with 3-leg ladders; the two-leg ladder W3 configuration is the only such phase found. The stripes themselves resemble $t$-$J$ chains, with one hole per about four lattice spacings; holes are unbound, and there is strong hopping along the chain.  There is also transverse hopping onto the ladders but this seems predominantly a single hop away from the chains. In the spin-squeezed picture of the $t$-$J$ chain\cite{xiang92}, the holes act as mobile domain walls in a Heisenberg chain; thus, for example, instantaneous singlet spin correlations are present across each hole. There is no attraction between holes.  All these feature seem to also describe the stripes in the W3 phase.  The W3 phase seems even farther from superconducting than the conventional striped phase:  there is no sign of paired holes(see also Fig.~\ref{fig:snapshot}(f)).

In the W3 phase the spin correlations are short ranged in both directions.  The low doping of each stripe and the weak transverse hopping make them unable to create $\pi$ phase shifts in the local AFM correlations. There are negligible spin correlations between the ladders. Within the ladders, vertical-separation spin-spin correlations are also short ranged, with a much more rapid decay compared to the conventional striped phase. 

The W3 stripe shown in Fig.~\ref{fig:w3stp} is at its ideal filling for a width 8 cylinder: two holes per three columns or a doping of $x=\frac{2}{24}=0.0833$. If we decrease the doping the phase does not change in a smooth way on width 8. Two holes per stripe and two-leg-ladder undoped regions are both favored in a quantized way. Decreasing the doping on width 8 causes defects(see Fig.~\ref{fig:w3defects} in Appendix): limited regions which have wider spacing so that most of the cylinder can maintain a spacing of 3 between stripes.  If the two-leg Heisenberg ladder picture is correct for the W3 phase, then the spacing of 3 would hold on any width cylinder. However, there is nothing in this picture that says the spacing of holes along a stripe must be exactly 4. We do not expect an odd number of holes per stripe, as that would require a spin excitation. But one might have different spacings on a much wider cylinder: for example, one might find four-hole stripes not just on a width-16 cylinder, but also, say, 14 and 18.    

Another system where we can see the W3 phase is a width 6 cylinder, where the stripes run along the length of the cylinder. This is shown in Appendix B. In this case two stripes and two ladders just fit. In the width 6 cylinder, we do find the spacing of holes on each stripe can be varied away from 4 slightly by adjusting the doping, consistent with the discussion above for wide cylinders.

\section{Energy gaps}
\label{sec:gap}
%%%%%%%%%%%FIGURE%%%%%%%%%%%%%%%%%%%%%%%%%%%%%%%%%%%
\begin{center}
\begin{figure*}[t]
	\vspace{-0.3cm}
    \includegraphics[width=1.5\columnwidth,clip=true]{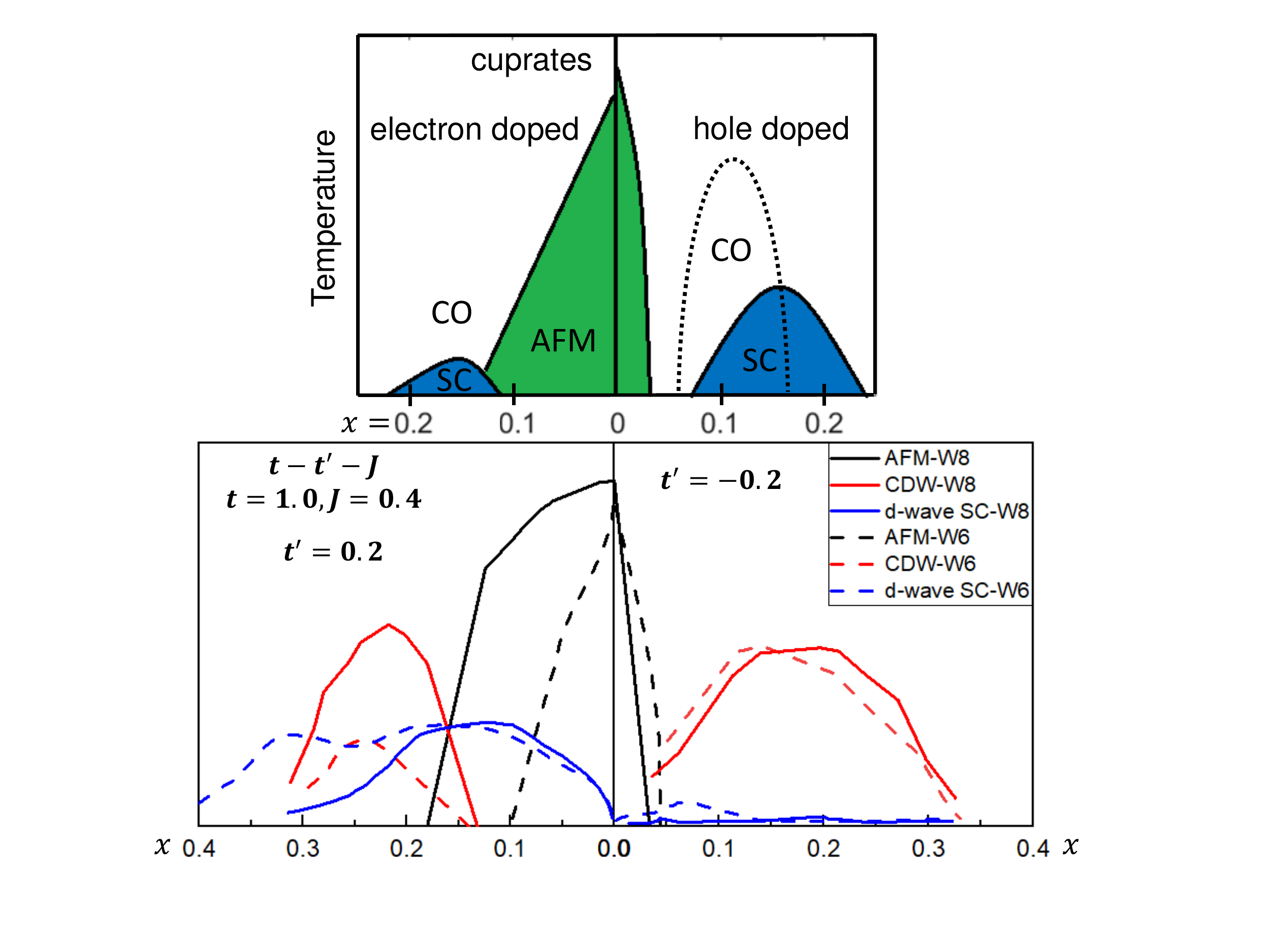}
	\vspace{-1.0cm}
    \caption{
	\label{fig:phase1d}
      Upper panel: Experimental phase diagram of a typical cuprate superconductor, following\cite{da2015charge}, with antiferromagnetic(AFM), charge ordered(CO) and superconducting(SC) phases. The vertical axis is temperature and horizontal axis indicates electron doping (left side) and hole doping (right side).  \quad Lower panel:  Antiferromagnetic, superconducting  and charge-density wave order parameters at zero temperature in the $t$-$t^\prime$-$J$ model from DMRG calculations. Solid lines are for width-8(W8) cylinders and dashed lines are for width-6(W6) cylinders.}  
	\vspace{-0.2cm}
\end{figure*}
\end{center}
%%%%%%%%%%%%%%%%%%%%%%%%%%%%%%%%%%%%%%%%%%%%%%%%%%%%%

\vspace{-1.0cm}
We can get further insight into the nature of the phases by studying their energy gaps associated with adding or removing particles. A generic formula for this sort of energy gap is
\begin{equation}
    \Delta E_n= [E(N_0+n)+E(N_0-n)-2E(N_0)]/2,
\end{equation}
for adding or removing $n$ particles at a time, where $E(N)$ is the ground state energy with $N$ particles. This formula exhibits finite size effects due to the finite size of the system.  The finite length manifests as an overall curvature of $E(N)$, which can be viewed as a shift in the chemical potential with $N$. In an infinite system, the chemical potential would not shift when adding a finite number of particles. 

Rather than extrapolations in system size,  we find it more convenient to plot $E(N)$ directly, and fit the lower envelope of points to a quadratic function. This is shown in Fig.~\ref{fig:gap}. The gaps are then measured by how many points rise above the quadratic fit.  These calculations were done on a width 6 system for higher accuracy. For the W3 stripe calculations on width 6, two stripes  run in the horizontal direction, as shown in Fig.~\ref{fig:extrapattern}(b) of Appendix B. Changing the number of particles changes the filling of these two stripes.

In the AFM $d-\pi/p$ phase, we see that an odd number of particles is higher in energy.  This is because in this superconducting phase, an odd-$N$ system has an extra quasiparticle, and we interpret the associated gap as the superconducting gap. Here, this is about 0.12. There is no sign of gaps associated with higher numbers of particles.  In contrast, in the conventional stripe phase, we see two gaps involved.   Systems with odd $N$ exhibit the highest energies, corresponding to broken pairs.  However, we also see that in the even-$N$ sector, multiples of four are lower in energy than non-multiples of four. This is because the stripes in this system have four holes, composed of two pairs, which are bound.  A non-multiple of four must have an isolated pair(see Fig.~\ref{fig:extrapattern}(a) in appendix), with an energy higher by about 0.05. 

In the right panel showing a W3 striped phase, we see a  smaller single particle gap compared to the previous two cases. This state is unpaired but the energy is still sensitive to having half-integer total spin.  One would expect an extra mainly living in a stripe, since a $t$-$J$ chain is gapless but a Heisenberg ladder has a large spin gap.  It is not clear whether the finite gap seen is a consequence of the finite length, the even circumference of the cylinder, or some other effect.

\section{comparison to cuprates}
\label{sec:cuprates}
\begin{center}
\begin{figure*}[t]
	\vspace{-0.0cm}
    \includegraphics[width=1.7\columnwidth,clip=true]{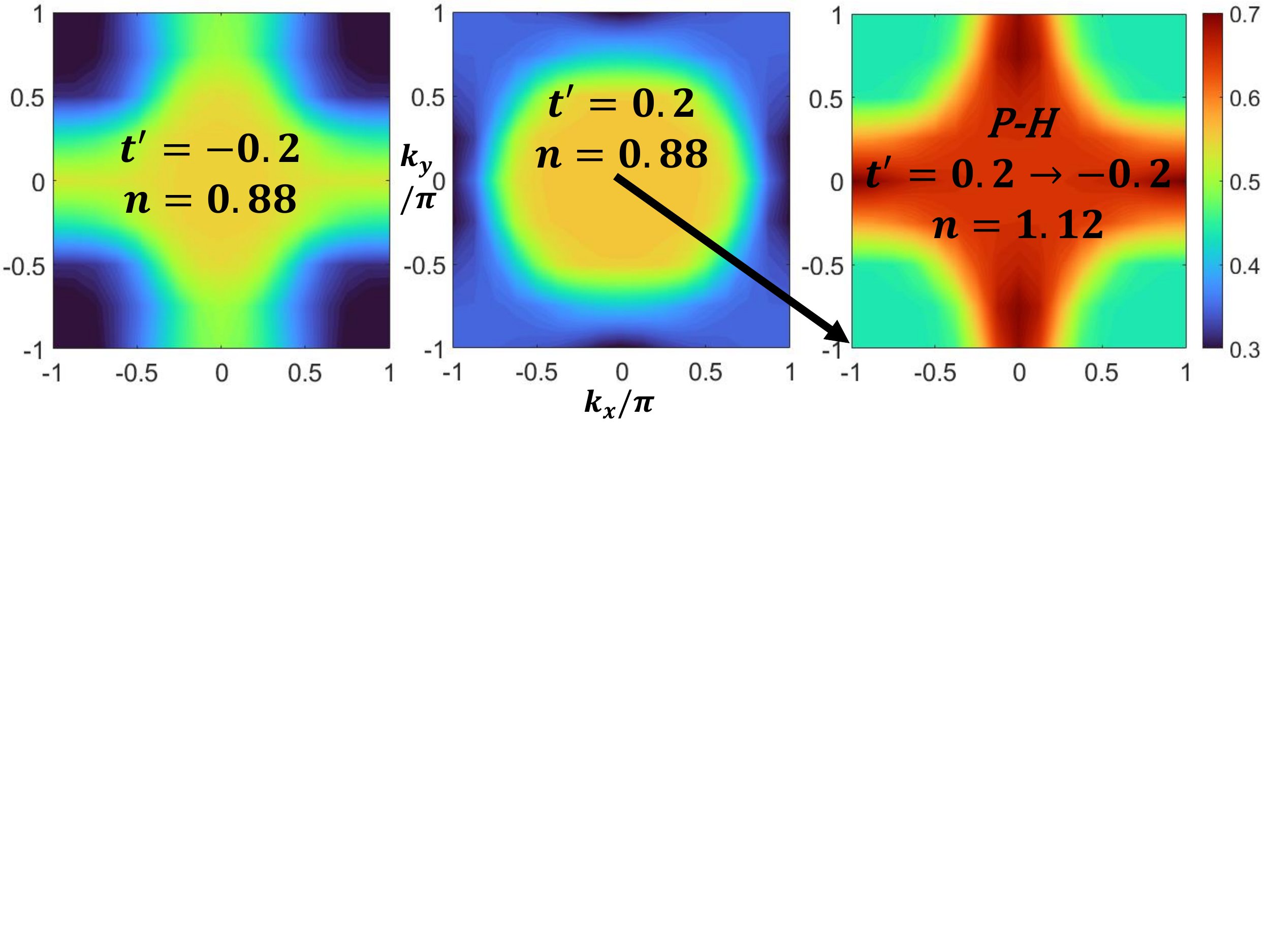}
	\vspace{-6.5cm}
	\caption{
		\label{fig:nk}
        Momentum space occupancy $n(\vec{k})$ for a single spin in the width-8 cylinder. The left figure shows $n(\vec{k})$ for the fermions in the $t$-$t'$-$J$ model with $t'=-0.2$ and $n$=0.88 fermions per site. This corresponds to the momentum distribution for the electrons in a hole doped system. The center figure shows the momentum distribution of the fermions in the $t$-$t'$-$J$ model with n=0.88 fermions per site and $t'=0.2$. Under a particle-hole transformation, which includes a $(\pi,\pi)$ shift of the origin, one obtains the figure on the right. Here $n(\vec{k})$ represents the momentum occupation of the electrons for an electron doped system with $t'=-0.2$ and $n$=1.12 electrons per site. }
	\vspace{-0.2cm}
\end{figure*}
\end{center}

To test the applicability of our model to cuprates, we first look at the momentum distribution function $n(\vec{k})$ by measuring the single-particle Green's function in real space and Fourier transforming it. The results are shown in in Fig.~\ref{fig:nk}.
Both cases are for a fermion doping of $x\approx0.125$ of the $t$-$t'$-$J$ model. 
In the left figure with $t'=-0.2$, $x$ represents the hole doping and $n(\vec{k})$ is the momentum occupation of the electrons for a hole doped system with $n=1-x \sim 0.875$ electrons per site. This Fermi surface is similar to what is seen in the hole-doped cuprates.
In the right part of Fig.~\ref{fig:nk} with $t' = 0.2$, the fermions occupy a circular region centered at the origin. Under a particle-hole transformation, $\vec{k}$ is shifted by $(\pi,\pi)$ and these fermions represent holes in a region centered about $(\pi,\pi)$. In this case, the system is electron doped with $n=1+x \sim 1.125$ electrons per site, and has a Fermi surface that is  similar to what is seen in the electron-doped cuprates.
%%%%%%%%%%%FIGURE%%%%%%%%%%%%%%%%%%%%%%%%%%%%%%%%%%%
%%%%%%%%%%%%%%%%%%%%%%%%%%%%%%%%%%%%%%%%%%%%%%%%%%%%%

\vspace{-0.8cm}
If we collect the antiferromagnetic(AFM), charge ordered(CO) and superconductivity(SC) pairing from the various scans with $t'=0.2$ and $t'=-0.2$, we can construct a zero-temperature order parameter diagram as shown in the lower panel of Fig.~\ref{fig:phase1d}. This can be compared to the nominal cuprates phase diagram in the upper panel taken from \cite{da2015charge}, where here the vertical axis is temperature. 
We see several similarities: a much broader AFM dome on the electron doped side than the hole doped side and a charge ordered region at intermediate doping on both sides.
However, contrary to the cuprate phase diagram, the SC pairing is significantly suppressed on the hole doped side with $t'=-0.2$, whereas in the hole doped cuprates there is a broad SC dome. Moreover, on the electron doped side with $t'=0.2$, we find that the $t$-$t'$-$J$ model exhibits a broad range of doping over which there is coexisting AFM, $d$-wave SC and $\pi$-triplet-$p$-wave SC order, contrary to what is observed in the cuprate phase diagram. 

Thus we conclude that one must go beyond the $t-t'-J$ model to understand superconductivity in the cuprates.  
This immediately suggests an important question:  would a Hubbard model with $t'$ do better?  In renormalizing away the two particle states to go from the Hubbard to the $t-t'-J$ model, there are terms that are of the same order as $J$ that are  omitted\cite{tt'j-sasha}. It could be that these terms are important for representing the physics of the cuprates. Alternatively, it may be that other interactions are needed to represent superconductivity properly.

\section{Summary}
\label{sec:sum}
In summary, we have carried out large scale ground state DMRG calculations on $t$-$t'$-$J$ cylinders with width six and eight which approximate the behavior of 2D systems. We have established an approximate phase diagram for this model.  

On the positive $t'$ side, which corresponds to electron doped cuprates, at low doping we find an AFM-$d/\pi p$ phase with coexisting uniform AFM and strong $d$-wave singlet pairing. As a result of these two orders, there also exists $(\pi,\pi),p$-wave triplet pairing. Pairing in this electron low-doped region is strong and unambiguous. At higher doping there is a striped phase with relatively weaker $d$-wave singlet pairing, as well as triplet pairing with an amplitude modulated by the stripes.

On the negative $t'$ side, which corresponds to hole doped cuprates, there is a broad striped phase. States with pairing go from being meta-stable and only slightly higher in energy near $t'=0$ to significantly suppressed with $t'=-0.2$.  At low doping, near $x=0.08$, we find a novel width-3 stripe phase that has chains of unpaired holes separated by two-leg spin ladders. The hole chains behave like 1D $t$-$J$ chains while spins on two-leg ladders mimic the short-ranged spin correlations seen in two-leg Heisenberg ladders. For $t'<0$, AFM order only exists for a very narrow doping range near half-filling.

Despite the fact that this $t$-$t'$-$J$ model manages to capture several aspects of the electron and hole-doped cuprates, including the broad AFM dome on the electron side and a much more narrow one on the hole side, as well as charge order on both sides, the superconducting properties exhibit significant discrepancies with respect to the cuprates. The hole doped cuprates exhibit strong superconductivity while the corresponding region of the model does not. In contrast, for the electron doped region of  the model we find strong superconductivity over a broader doping range than in the cuprates, and for a substantial range of doping this pairing coexists with AFM and triplet $p$-wave superconductivity. 

\textit{Note:} As this paper was being finished, S. Gong, W. Zhu and D. N. Sheng posted a preprint where DMRG was used to study superconducting, magnetic, and charge
correlations on width 4 and 6 $t$-$t'$-$J$ cylinders\cite{gong2021robust}. Their main conclusion was that for $t'>0$ there appears to be robust d-wave superconductivity.
Although their phase diagram differs in several respects from ours, we believe their correlation functions are quite consistent with ours. In particular, they find AFM correlations in  the AFM-$d/\pi p$ region of the phase diagram which decay exponentially but with a rapidly increasing correlation length going from width 4 to 6.   In our results, we also find the AFM correlations have a similar decay on width 6, but by width 8 the correlations always extend beyond the length of the systems we study, manifesting as long-range order with a broken symmetry ground state. 

\section*{acknowledgements}
We thank Steven A. Kivelson, A. L. Chernyshev and Judit Romhanyi for stimulating discussions. 
SJ and SRW were supported by the NSF under DMR-1812558.
DJS was supported by the Scientific Discovery through Advanced Computing (SciDAC) program funded by the U.S. Department of Energy. 

\appendix
\section{Spontaneously broken symmetry in DMRG calculations}
\label{sec:broksym}

In the limit of a bond dimension of $m=1$, a matrix product state(MPS) is a simple product state, and 
DMRG represents a simple sort of mean field theory.  The resulting state breaks symmetry, 
like mean-field techniques in general. As the bond dimension increases, the MPS approaches
the result of an exact diagonalization.  Thus it is useful to think of DMRG as interpolating in
some sense between mean field theory and exact diagonalization.  In 1D systems, one is often
very close to the exact diagonalization side of this, so one can usually ignore questions
of broken symmetry.  For 2D systems, it is important to be aware of how DMRG may
or may not break symmetries.

How DMRG can break symmetry depends on what quantities are conserved in the calculation.
Consider the antiferromagnetic Heisenberg model on a square lattice. Say we use $U(1)$ symmetry,
i.e. we conserve total $S^z$. To get the ground state, we choose the $S^z=0$ sector,
and for $m=1$, DMRG gives 
one of the two N\'eel states oriented in the spin $z$ direction. A N\'eel state oriented in
another direction does not have definite $S^z$. Now suppose we increase $m$, but not
too much.  The system then tries to reduce the energy subject to having low entanglement,since $m$ controls the entanglement that can be captured when one splits the system in two.

\begin{figure}[t]
    \begin{center}
    \hspace*{0.8cm}
    \includegraphics[width=1.3\columnwidth,clip=true]{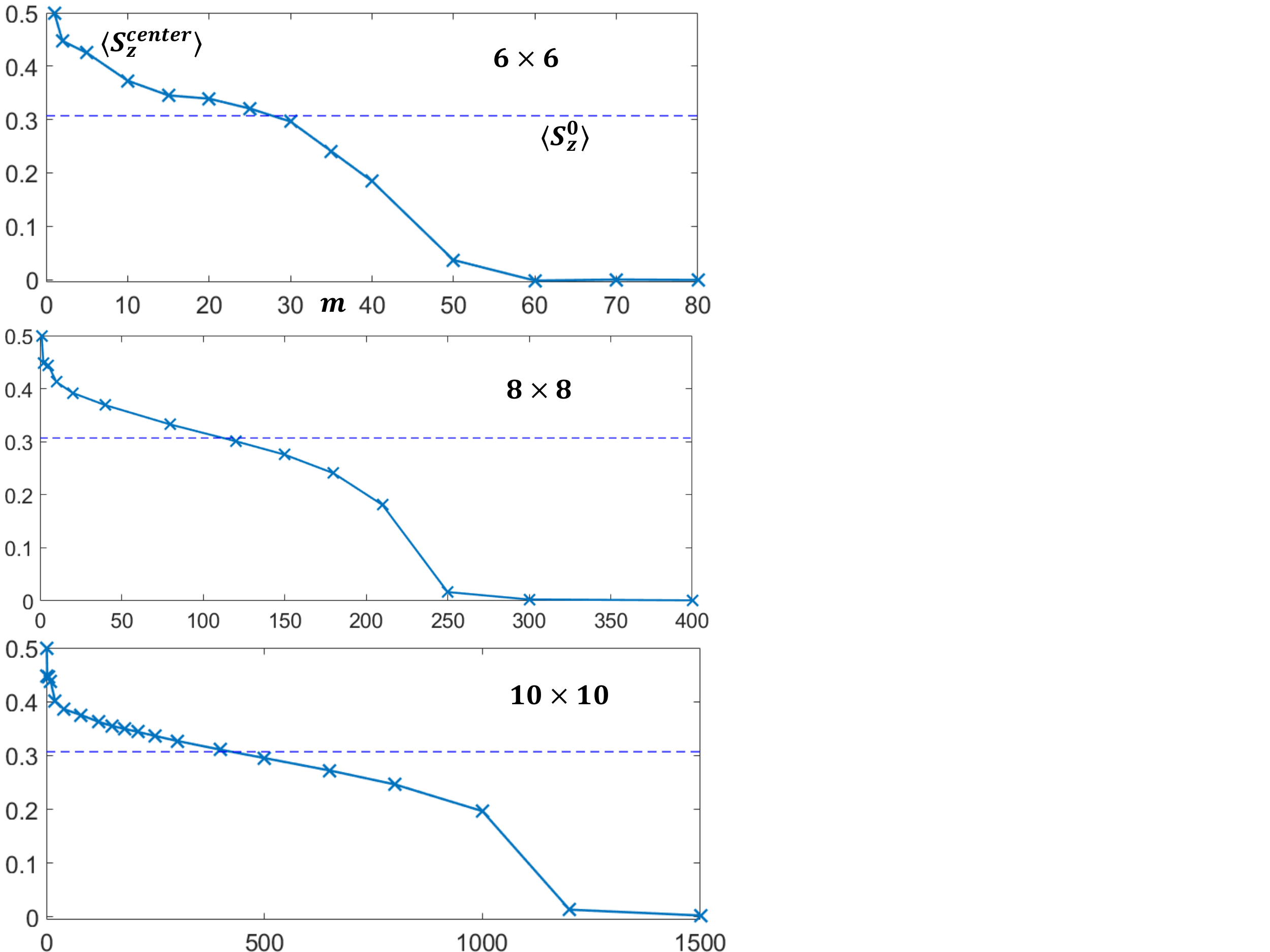}
	\vspace{-0.5cm}
	\caption{
    		\label{fig:broksym}
    		For the Heisenberg model, well-converged spin moment $\langle S_z^{center} \rangle$ for the center spin as a function of bond dimension $m$ in $6\times6$, $8\times8$ and $10\times10$ systems, using $S_z$ conservation ($U(1)$ symmetry). The blue dashed line is $\langle S_z^0 \rangle=0.307 $, the known broken symmetry magnetization for an infinite system\cite{sandvik97}. At bond dimension 1, DMRG mimics a simple mean field theory, and gives a perfect N\'eel order parameter of 0.5.  At very large bond dimension, DMRG gives the exact result of zero (overall spin singlet) for any finite system with an even number of sites.  In between, we see a plateau around the 2D broken symmetry value which sharpens with increasing system size.
    		}
	\vspace{-0.2cm}
	\end{center}
\end{figure}

The easiest way to decrease the energy is to build in local spin fluctuations, but to stay
close to one N\'eel state. Any superposition of N\'eel states would produce larger entanglement
with little benefit in energy.  These effects increase with system size.
Thus, a $U(1)$ DMRG calculation will show a broken spin orientation symmetry 
along the $z$ direction. As one increases $m$, at first the local spin fluctuations
become better and better, and one may approach a plateau near 
$|\langle S^z_0 \rangle|$ or $ \tilde M \approx 0.307$, where $M$ is the bulk magnetization.
The accuracy of this plateau depends on system size. As one increases $m$ farther,
eventually the energy optimization notices that different directions of the order can be
mixed, and one tends to $\langle S^z_i \rangle \to 0$, the exact diagonalization result 
for an even number of sites, satisfying the Lieb-Mattis theorem.\cite{lieb1962}.
Figure~\ref{fig:broksym} shows $\langle S_z^{center} \rangle$ at a center site for 6x6,  8x8 and 10x10 clusters as a function of the number of states(or bond dimension $m$).  One can see that  $\langle S_z^{center} \rangle$ has a plateau around $\langle S_z^0 \rangle$ before approaching the exact solution of a spin-singlet. The bond dimension required to form the overall spin-singlet increases rapidly with the width of the system.

If one does not use $U(1)$ symmetry, the system could form a N\'eel state in any direction.
Using $U(1)$ speeds up the calculation considerably, so we avoid leaving it out except in special cases, such as applying a spin-twist boundary condition.  Some
DMRG programs conserve $SU(2)$, but not ITensor, the library used here.  
In that case, the ground state calculation is 
constrained to the singlet sector and it cannot break spin symmetry.  Keeping $SU(2)$
means that one keeps more states for the same computational cost, but this is offset
to a varying extent, in terms of how low the energy is, by having to construct a 
superposition of N\'eel states. Thus either $SU(2)$ or $U(1)$ calculations are more effective
depending on what one is studying.
%%%%%%%%%%%FIGURE%%%%%%%%%%%%%%%%%%%%%%%%%%%%%%%%%%%
%%%%%%%%%%%%%%%%%%%%%%%%%%%%%%%%%%%%%%%%%%%%%%%%%%%%%

If symmetry is not broken, then determining the order parameter requires measuring
a correlation function. Correlation functions tend to converge slowly with the 
bond dimension in DMRG, compared
to local quantities. Correlation functions also produce the square of the order parameter,
which, if the order parameter is small, causes further inaccuracy. It can be much 
more effective to measure order parameters by strongly pinning the edges of a cylinder and
measuring the local order parameter in the center\cite{White_and_Chernyshev}.
This can be very accurate if one adjust the aspect ratio of the cylinder to cancel
leading finite size effects\cite{White_and_Chernyshev}.
However, if one is not trying for a quantitative determination of the 2D order parameter, one gets good results from measuring the local order due to the self-pinning of limited symmetry DMRG. 

For a potentially superconducting system, one may or may not conserve the number of
fermions modulo 2.  One cannot turn off the number conservation completely, since
this breaks the fermion parity and interferes with putting in the right statistics.
We refer to these two choices as conserving or not conserving particle number. If
the system is superconducting, a mean field theory such as BCS would break symmetry
to give a definite phase, making a superposition of different numbers of particles.
If one conserves particle number, one cannot have a definite phase. In this case,
in order to measure pairing, one would have to use correlation functions. Since the order parameters tend to be small, measuring its square with DMRG is inaccurate, both because of the small quantity being measured and the inaccuracy of correlation functions.  
If one turns off number conservation, then one can have a phase.  However, usually,
for efficiency, we use only real wavefunctions. In this case the allowed broken symmetry
reduces from $U(1)$ (the phase) to $Z_2$ ($+$ or $-$).  The gain in numerical
efficiency by being near a broken symmetry state is less clear for a $d$-wave superconducting 
state than the magnetic  case, because of the more complicated nature of the local
order. At $m=1$, an MPS cannot represent a singlet living on two adjacent sites. 
However, the advantages associated with measuring local quantities rather than correlation
functions makes the non-particle-conserving approach highly preferred.

In contrast to the magnetic case, a system which is superconducting may not break number 
symmetry at low bond dimension, particularly if the pairing is relatively weak. The
system may get stuck in a particle-conserving state, even if non-particle conserving states
are allowed and are lower in energy for a fixed bond dimension.  However, one can turn on a pairing/proximity-effect field $\Delta + \Delta^\dagger$
for a few sweeps to get it unstuck, and then see if the local pairing grows in subsequent
sweeps, approaching an approximate plateau. Alternatively, one can apply   pairing fields
on the edges and measure the response in the center.  If the pairing is quite strong,
then one is very likely to see a robust spontaneous broken-symmetry pairing without
applying any temporary or edge fields.

\section{Magnetic order in the AFM-$d/\pi p$ phase}
%%%%%%%%%%%FIGURE%%%%%%%%%%%%%%%%%%%%%%%%%%%%%%%%%%%
\begin{figure}[t]
    \begin{center}
    \includegraphics[width=1.35\columnwidth,clip=true]{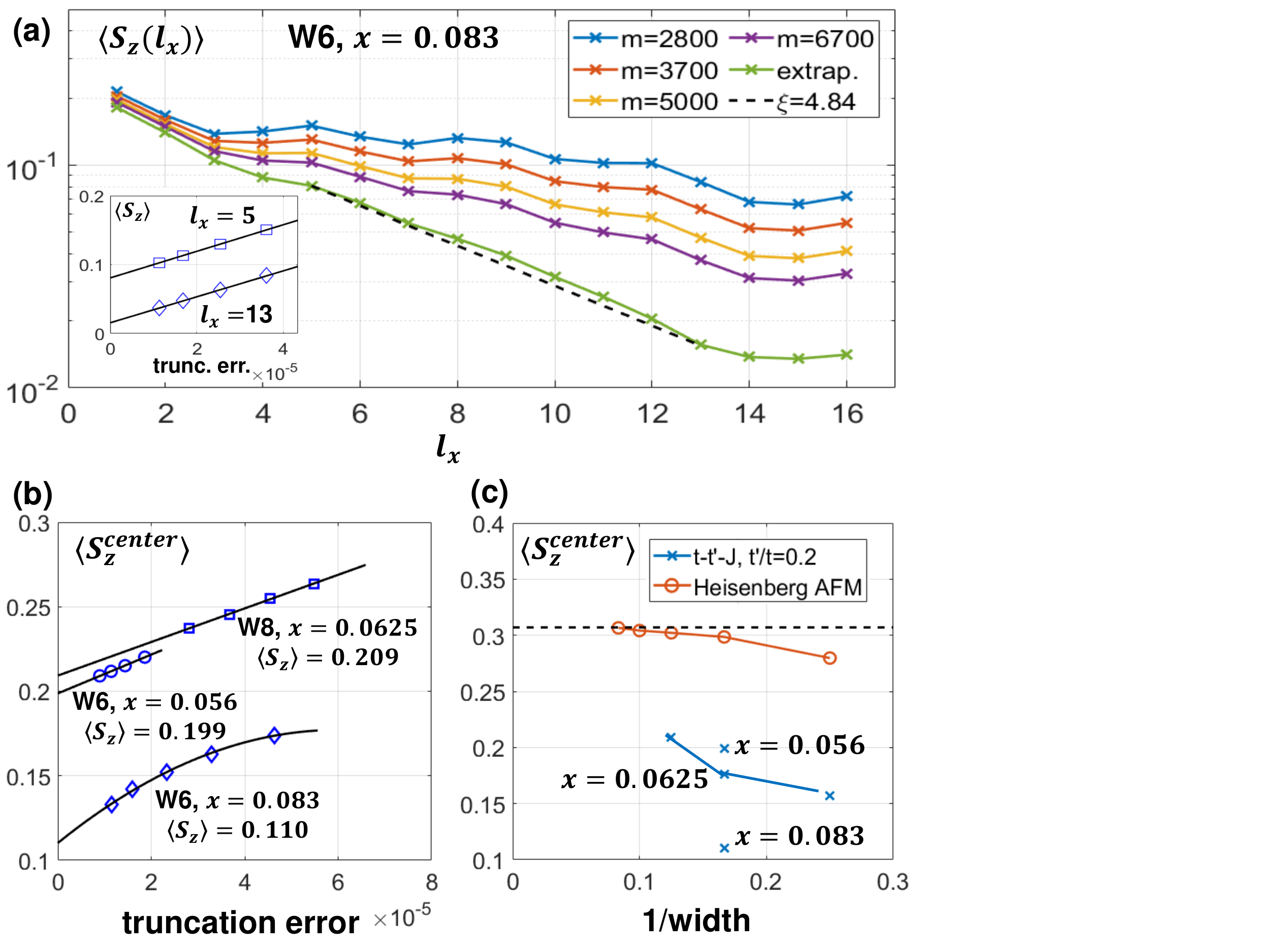}
	\caption{
    		\label{fig:magorder} 
    		(a): Decay of local magnetic order as one moves away from the left edge where a weak staggered magnetic field of 0.03 is applied for different bond dimension $m$. $\langle S_z(l_x)\rangle$ is extrapolated with truncation error(shown in inset) on a point by point basis.
    		The decay of magnetic order shows good agreement with correlation length $\xi=4.84$~\cite{gong2021robust}.
    		(b): For a length/width=2 cylinder with strong staggered magnetic pinning field of 0.2 on both edges\cite{White_and_Chernyshev}, local magnetic order in the center extrapolated with truncation error. All three cases are in the AFM-$d/\pi p$ phase with $t'=0.2$.
    		(c): Using the technique described in (b), local magnetic order in the center for cylinders of different widths for $t$-$t'$-$J$ model and the Heisenberg model.  For the Heisenberg model, the dashed line shows the precisely known order parameter 0.307 from\cite{sandvik97}.
    		}
	\vspace{-0.2cm}
	\end{center}
\end{figure}
%%%%%%%%%%%%%%%%%%%%%%%%%%%%%%%%%%%%%%%%%%%%%%%%%%%%%

There are several types of approaches one could take to try to infer the broken symmetries of the 2D system from our finite-bond dimension DMRG results on finite sized cylinders.  For the most part, our results in the main text seem clear-cut, with broken symmetries appearing in fairly large cylinders with small changes as the bond dimension is increased.  However, it is clearly very useful to try to consider particular points in the phase diagram for a more careful finite-sized analysis.  In this appendix we consider the antiferromagnetic order in the low-doped $t'>0$ system. In the recent paper\cite{gong2021robust}, the authors found that in this regime on width 6 cylinders strong $d$-wave pairing was indicated, but the authors did not clearly conclude that there was AF order.

The DMRG studies in \cite{gong2021robust} used full SU(2) symmetry, and kept more states than we have been able to.  However, in DMRG calculations, correlation functions, which were used in \cite{gong2021robust}, are inherently less accurate than local quantities, which we used. The spin-spin correlations decayed exponentially in\cite{gong2021robust}, but the mere fact of exponential decay does not tell much about possible long-range order in 2D.  For example, it is known that an even-leg Heisenberg ladder always has exponential decay of correlations, but in 2D the model has long range order. The crucial question is what is the decay length, and how does it scale with width?

%%%%%%%%%%%FIGURE%%%%%%%%%%%%%%%%%%%%%%%%%%%%%%%%%%%
\begin{center}
\begin{figure*}[ht]
	\vspace{0cm}
    \includegraphics[width=1.7\columnwidth,clip=true]{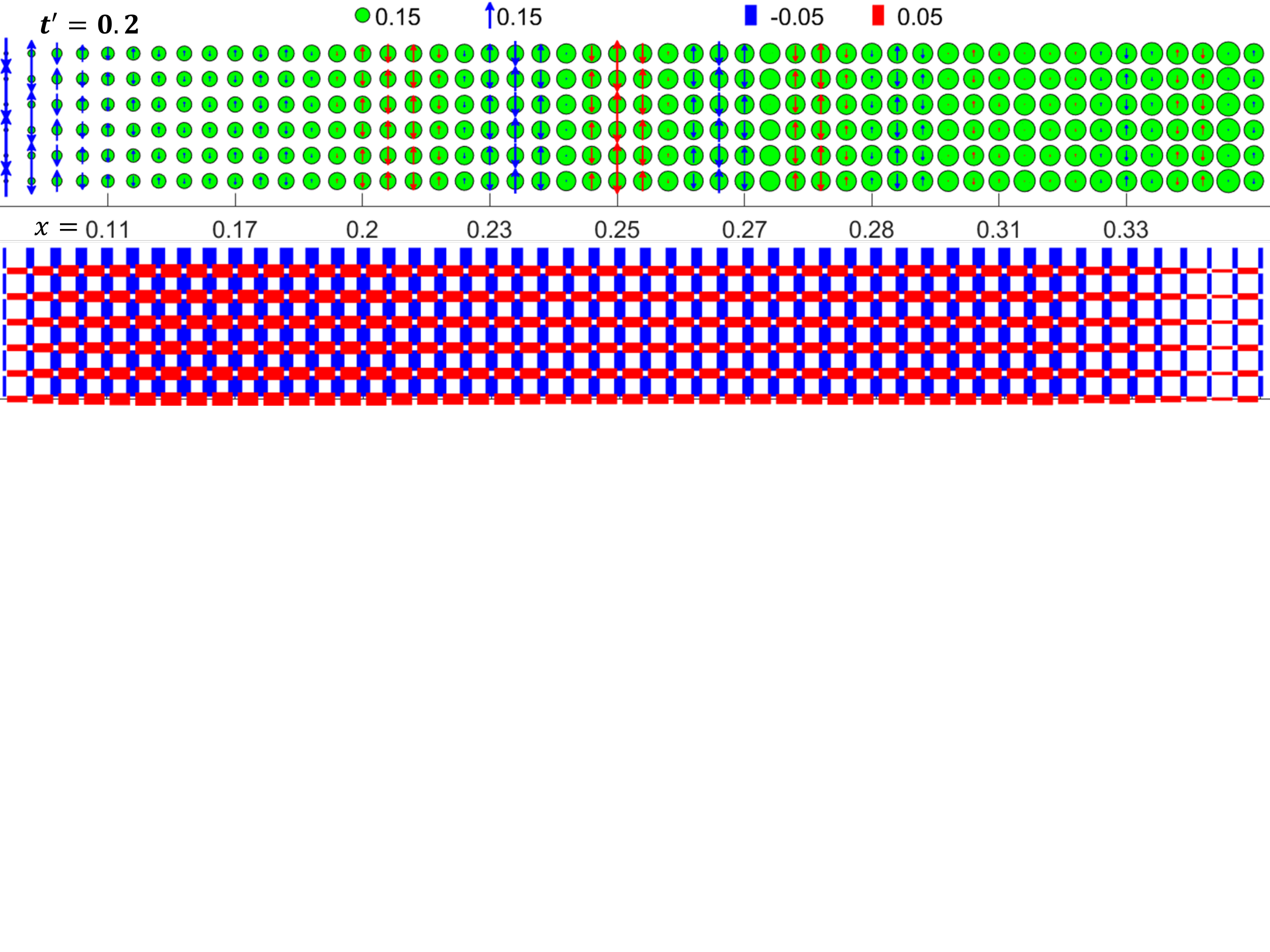}
	\vspace{-6.5cm}
    \caption{
	\label{fig:w6dpva0.2}
	    A doping-varying $50\times6$ cylinder with $t'=0.2$ which simulates electron-doped cuprates. The numbers on middle axis indicate the averaged local doping. A staggered magnetic pinning field of 0.03 is applied on the left edge. No pair field is applied. At low doping there is AFM order which continuously transition to stripe and then uniform pattern at higher doping.}
	\vspace{0.0cm}
\end{figure*}
\end{center}
%%%%%%%%%%%%%%%%%%%%%%%%%%%%%%%%%%%%%%%%%%%%%%%%%%%%%
%%%%%%%%%%%FIGURE%%%%%%%%%%%%%%%%%%%%%%%%%%%%%%%%%%%
\begin{center}
\begin{figure*}[ht]
	\vspace{-0.2cm}
    \includegraphics[width=1.7\columnwidth,clip=true]{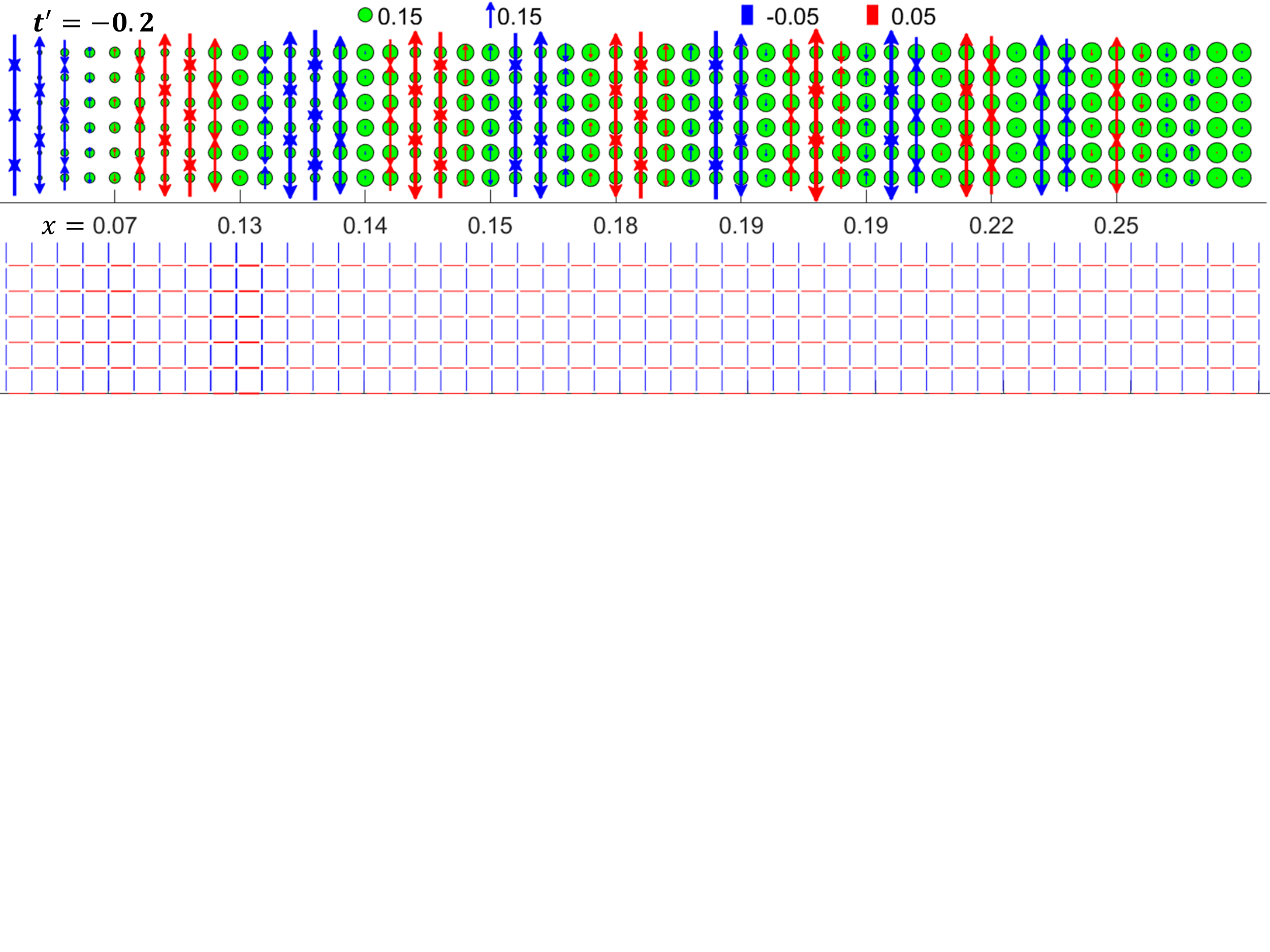}
	\vspace{-6.7cm}
    \caption{
	\label{fig:w6dpva-0.2}
	    A doping-varying $50\times6$ cylinder with $t'=-0.2$ which simulates hole-doped cuprates. A staggered magnetic pinning field of 0.03 is applied on the left edge. A global pair field of 0.005 is applied to measure the pairing response. There's stripe pattern across all doping with pairing suppressed}
	\vspace{0.4cm}
\end{figure*}
\end{center}
%%%%%%%%%%%%%%%%%%%%%%%%%%%%%%%%%%%%%%%%%%%%%%%%%%%%%

\vspace{-1.8cm}
Spin-spin correlation lengths can also be determined through the static response to a pinning field, allowing a local measurement.  In Fig.~\ref{fig:magorder}(a), we compare
measurements of the spin-spin correlation length of the width 6 system at a doping of $x=0.083$ and $t'=0.22, J=1/3, J'=0.016$.  The correlation length in \cite{gong2021robust} for this system was determined to be $\xi=4.84$.  By extrapolating the local response to an edge field, we find a completely consistent exponential decay with the same correlation length. The extrapolations in truncation error for each local measurement are very well behaved, as shown in inset of Fig.~\ref{fig:magorder}(a). Does $\xi=4.84$ on a width 6 cylinder correspond to long range 2D order?  Correlation lengths were shorter on width 4, but this information did not make it clear in \cite{gong2021robust} that there was long range AF order in 2D. 

Instead, we use the aspect ratio method discussed in reference\cite{White_and_Chernyshev}.  In this case strong pinning is applied to the ends of the cylinder and the order parameter is measured in the center.  The aspect ratio can be chosen to  eliminate the leading finite size correction, making the approach to the 2D order parameter very rapid. (In longer cylinders, the system looks more 1D-like, giving a smaller order parameter; in shorter cylinders, the strong pinning prevails, giving an overestimate.) For the Heisenberg square lattice, Fig.~\ref{fig:magorder}(c) shows how effective this approach is for an aspect ratio of 2, which is a little larger than the ideal aspect ratio near 1.9. With the slightly larger aspect ratio, the approach is from below as the width is increased. We apply this approach to the low-doped $t$-$J$ model at $x=0.0625$, using $8\times 4$, $12\times 6$, and $16\times8$ cylinders.
For width 6, the target doping corresponded to a non-integer number of pairs, so  a linear interpolation was made between two adjacent integral dopings.  The finite size behavior is not as flat as for the Heisenberg case, and the interpolation is only a rough treatment, but nevertheless the results provide solid evidence for a nonzero  AF order parameter, roughly between 0.2 and 0.25. 

\section{Comparisons with width-6 cylinders}
\label{sec:w6}
%%%%%%%%%%%FIGURE%%%%%%%%%%%%%%%%%%%%%%%%%%%%%%%%%%%
\begin{center}
\begin{figure*}[ht]
	\vspace{0.4cm}
    \includegraphics[width=1.7\columnwidth,clip=true]{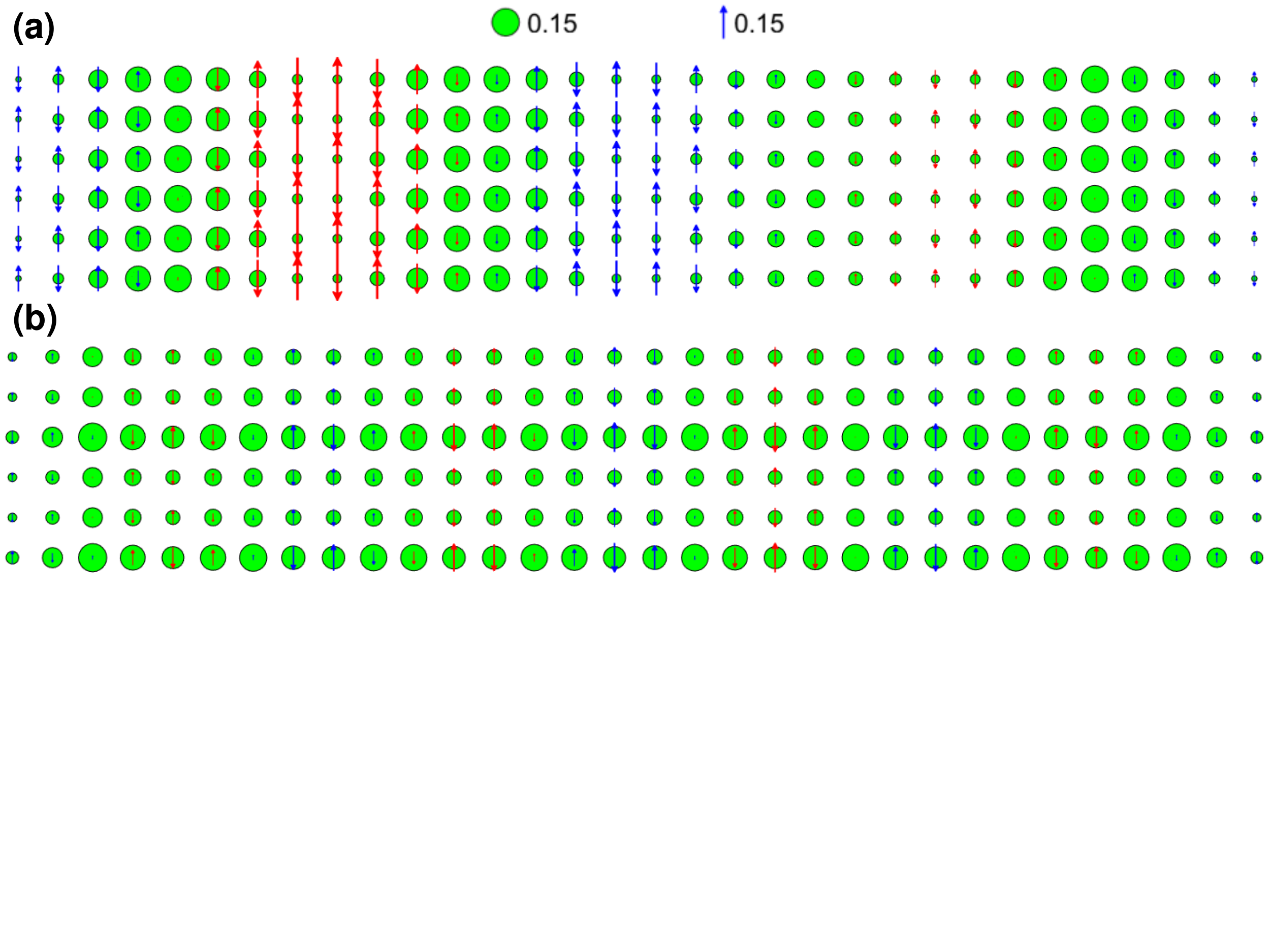}
	\vspace{-4.5cm}
    \caption{
	\label{fig:extrapattern}
	    (a) A conventional striped phase with two holes removed from the third stripe from the left.  (b)A W3 phase with the stripes running horizontally in a width-6 cylinder at a doing of $x=0.083$.
	    }
	\vspace{0.4cm}
\end{figure*}
\end{center}
%%%%%%%%%%%%%%%%%%%%%%%%%%%%%%%%%%%%%%%%%%%%%%%%%%%%%
%%%%%%%%%%%FIGURE%%%%%%%%%%%%%%%%%%%%%%%%%%%%%%%%%%%
\begin{figure}[ht]
    \begin{center}
    \hspace*{1.1cm}
    \includegraphics[width=1.5\columnwidth,clip=true]{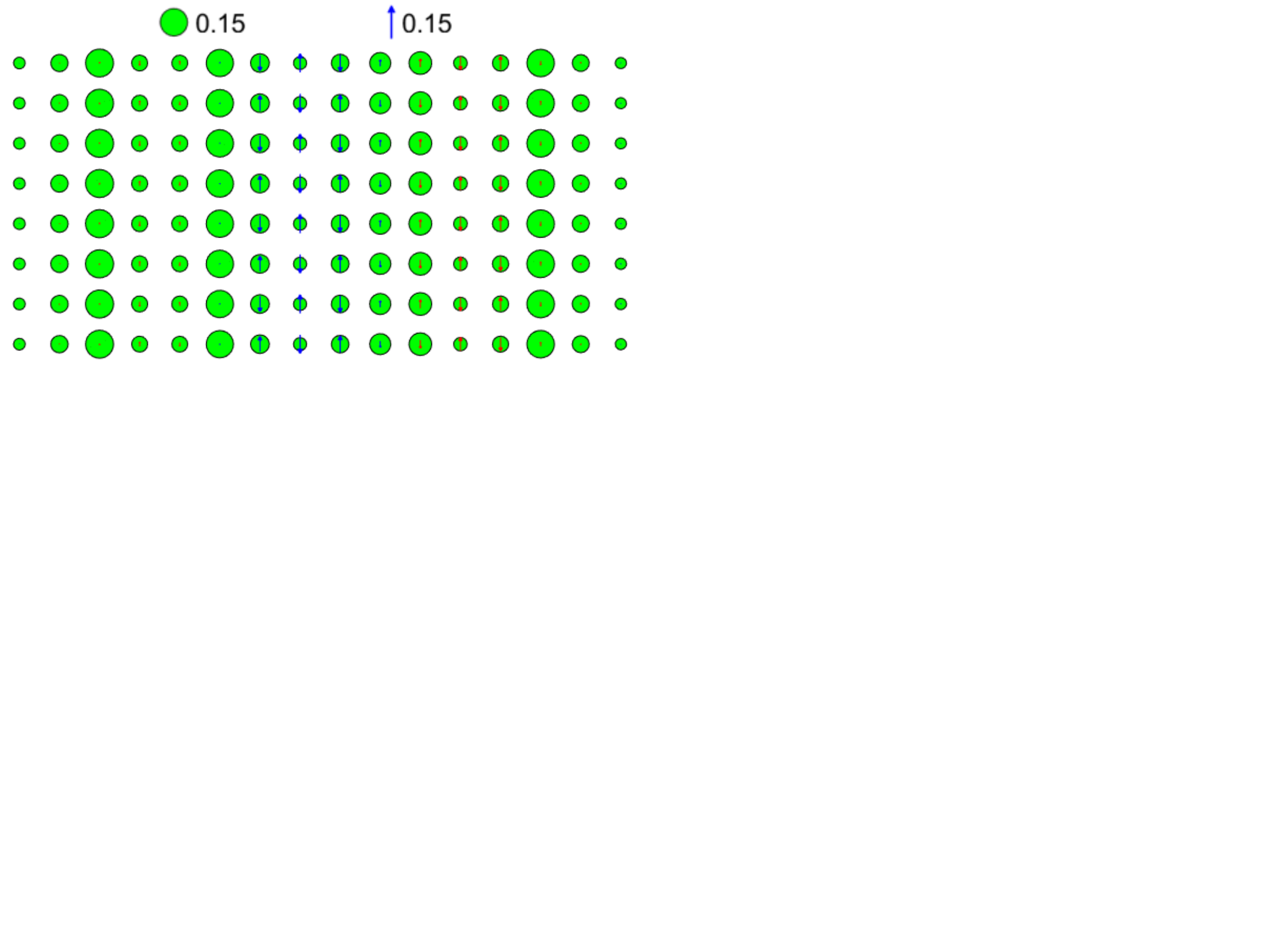}
	\vspace{-6.6cm}
	\caption{
    		\label{fig:w3defects}
    		A W3 striped phase with a defect. 
    		}
	\vspace{-0.2cm}
	\end{center}
\end{figure}
%%%%%%%%%%%%%%%%%%%%%%%%%%%%%%%%%%%%%%%%%%%%%%%%%%%%%

\vspace{-0.5cm}
In this Appendix we report results for two doping-varying scans in width-6 cylinders which support the qualitative features of the ground state phase diagram in Fig.~\ref{fig:phase1d}.  We also show some details on the higher energy states which determine the energy gaps in Sec.~\ref{sec:gap}. 

Figure~\ref{fig:w6dpva0.2} shows a doping-varying scan with fixed $t'=0.2$, corresponding to the electron-doped cuprates. Similar to what is seen on width-8 cylinders, there is a coexisting AFM-$d/\pi p$ phase at low doping, and a striped phase with pairing at higher doping. 
Quantitatively there are several differences. The AFM-$d/\pi p$ phase has a narrower doping range with weaker AFM order compared to the  width-8 system. The transition between the   AFM-$d/\pi p$ phase and the striped phase is less sharp in the width-6 system. 
Pairing in the width-6 system exists over a  wider doping range than in the width-8 system.

Figure~\ref{fig:w6dpva-0.2} shows a doping-varying scan with fixed $t'=-0.2$. This is similar to the width-8 results other than the lack of the W3 striped phase at low doping. A single vertical W3 stripe with two holes on width 6 has a substantially different doping per unit length than on width 8, where we do see the W3 stripes. 

In determining the energies shown in Fig.~\ref{fig:gap}, some of the states have unusual configurations, such as defects in the stripe pattern.  For example, when one removes two holes from a striped calculation, the missing holes are likely to be removed from one particular stripe.  Figure~\ref{fig:extrapattern}(a) shows an example of this, with two holes removed from the third stripe from the left. The state is higher in energy because of the binding of pairs into stripes.  We expect there to be other similar states at nearly the same energy, and a more precise calculation would show superpositions of these configurations. 

Figure~\ref{fig:extrapattern}(b) shows a W3  phase with the stripes running horizontally in a width-6 cylinder. This configuration allows one to vary the number of holes to get the energies in Fig.~\ref{fig:gap}. On this width, there are some noticeable spin moments and a slightly charge variation along the stripes due to the open boundaries.

Figure~\ref{fig:w3defects} shows a W3 striped phase on width 8, where changing the doping has produced a  defect. Each of the four stripes shown has two holes, but the overall length of the cylinder is too long for a perfect W3 configuration. In the figure,  the third stripe from the left has increased its width and the ladder in between it and the second stripe has increased to width 3.  The width-3 region shows longer spin-spin correlations, as expected.

\bibliography{ref}

\begin{thebibliography}{59}%
\makeatletter
\providecommand \@ifxundefined [1]{%
 \@ifx{#1\undefined}
}%
\providecommand \@ifnum [1]{%
 \ifnum #1\expandafter \@firstoftwo
 \else \expandafter \@secondoftwo
 \fi
}%
\providecommand \@ifx [1]{%
 \ifx #1\expandafter \@firstoftwo
 \else \expandafter \@secondoftwo
 \fi
}%
\providecommand \natexlab [1]{#1}%
\providecommand \enquote  [1]{``#1''}%
\providecommand \bibnamefont  [1]{#1}%
\providecommand \bibfnamefont [1]{#1}%
\providecommand \citenamefont [1]{#1}%
\providecommand \href@noop [0]{\@secondoftwo}%
\providecommand \href [0]{\begingroup \@sanitize@url \@href}%
\providecommand \@href[1]{\@@startlink{#1}\@@href}%
\providecommand \@@href[1]{\endgroup#1\@@endlink}%
\providecommand \@sanitize@url [0]{\catcode `\\12\catcode `\$12\catcode
  `\&12\catcode `\#12\catcode `\^12\catcode `\_12\catcode `\%12\relax}%
\providecommand \@@startlink[1]{}%
\providecommand \@@endlink[0]{}%
\providecommand \url  [0]{\begingroup\@sanitize@url \@url }%
\providecommand \@url [1]{\endgroup\@href {#1}{\urlprefix }}%
\providecommand \urlprefix  [0]{URL }%
\providecommand \Eprint [0]{\href }%
\providecommand \doibase [0]{https://doi.org/}%
\providecommand \selectlanguage [0]{\@gobble}%
\providecommand \bibinfo  [0]{\@secondoftwo}%
\providecommand \bibfield  [0]{\@secondoftwo}%
\providecommand \translation [1]{[#1]}%
\providecommand \BibitemOpen [0]{}%
\providecommand \bibitemStop [0]{}%
\providecommand \bibitemNoStop [0]{.\EOS\space}%
\providecommand \EOS [0]{\spacefactor3000\relax}%
\providecommand \BibitemShut  [1]{\csname bibitem#1\endcsname}%
\let\auto@bib@innerbib\@empty
%</preamble>
\bibitem [{\citenamefont {Zheng}\ \emph {et~al.}(2017)\citenamefont {Zheng},
  \citenamefont {Chung}, \citenamefont {Corboz}, \citenamefont {Ehlers},
  \citenamefont {Qin}, \citenamefont {Noack}, \citenamefont {Shi},
  \citenamefont {White}, \citenamefont {Zhang},\ and\ \citenamefont
  {Chan}}]{zheng2017}%
  \BibitemOpen
  \bibfield  {author} {\bibinfo {author} {\bibfnamefont {B.-X.}\ \bibnamefont
  {Zheng}}, \bibinfo {author} {\bibfnamefont {C.-M.}\ \bibnamefont {Chung}},
  \bibinfo {author} {\bibfnamefont {P.}~\bibnamefont {Corboz}}, \bibinfo
  {author} {\bibfnamefont {G.}~\bibnamefont {Ehlers}}, \bibinfo {author}
  {\bibfnamefont {M.-P.}\ \bibnamefont {Qin}}, \bibinfo {author} {\bibfnamefont
  {R.~M.}\ \bibnamefont {Noack}}, \bibinfo {author} {\bibfnamefont
  {H.}~\bibnamefont {Shi}}, \bibinfo {author} {\bibfnamefont {S.~R.}\
  \bibnamefont {White}}, \bibinfo {author} {\bibfnamefont {S.}~\bibnamefont
  {Zhang}},\ and\ \bibinfo {author} {\bibfnamefont {G.~K.-L.}\ \bibnamefont
  {Chan}},\ }\bibfield  {title} {\bibinfo {title} {Stripe order in the
  underdoped region of the two-dimensional hubbard model},\ }\href@noop {}
  {\bibfield  {journal} {\bibinfo  {journal} {Science}\ }\textbf {\bibinfo
  {volume} {358}},\ \bibinfo {pages} {1155} (\bibinfo {year}
  {2017})}\BibitemShut {NoStop}%
\bibitem [{\citenamefont {Qin}\ \emph {et~al.}(2020)\citenamefont {Qin},
  \citenamefont {Chung}, \citenamefont {Shi}, \citenamefont {Vitali},
  \citenamefont {Hubig}, \citenamefont {Schollw\"ock}, \citenamefont {White},\
  and\ \citenamefont {Zhang}}]{absence-qin}%
  \BibitemOpen
  \bibfield  {author} {\bibinfo {author} {\bibfnamefont {M.}~\bibnamefont
  {Qin}}, \bibinfo {author} {\bibfnamefont {C.-M.}\ \bibnamefont {Chung}},
  \bibinfo {author} {\bibfnamefont {H.}~\bibnamefont {Shi}}, \bibinfo {author}
  {\bibfnamefont {E.}~\bibnamefont {Vitali}}, \bibinfo {author} {\bibfnamefont
  {C.}~\bibnamefont {Hubig}}, \bibinfo {author} {\bibfnamefont
  {U.}~\bibnamefont {Schollw\"ock}}, \bibinfo {author} {\bibfnamefont {S.~R.}\
  \bibnamefont {White}},\ and\ \bibinfo {author} {\bibfnamefont
  {S.}~\bibnamefont {Zhang}} (\bibinfo {collaboration} {Simons Collaboration on
  the Many-Electron Problem}),\ }\bibfield  {title} {\bibinfo {title} {Absence
  of superconductivity in the pure two-dimensional hubbard model},\ }\href
  {https://doi.org/10.1103/PhysRevX.10.031016} {\bibfield  {journal} {\bibinfo
  {journal} {Phys. Rev. X}\ }\textbf {\bibinfo {volume} {10}},\ \bibinfo
  {pages} {031016} (\bibinfo {year} {2020})}\BibitemShut {NoStop}%
\bibitem [{\citenamefont {Arovas}\ \emph {et~al.}(2021)\citenamefont {Arovas},
  \citenamefont {Berg}, \citenamefont {Kivelson},\ and\ \citenamefont
  {Raghu}}]{hubreview-kivleson}%
  \BibitemOpen
  \bibfield  {author} {\bibinfo {author} {\bibfnamefont {D.~P.}\ \bibnamefont
  {Arovas}}, \bibinfo {author} {\bibfnamefont {E.}~\bibnamefont {Berg}},
  \bibinfo {author} {\bibfnamefont {S.}~\bibnamefont {Kivelson}},\ and\
  \bibinfo {author} {\bibfnamefont {S.}~\bibnamefont {Raghu}},\ }\href@noop {}
  {\bibinfo {title} {The hubbard model}} (\bibinfo {year} {2021}),\ \Eprint
  {https://arxiv.org/abs/2103.12097} {arXiv:2103.12097 [cond-mat.str-el]}
  \BibitemShut {NoStop}%
\bibitem [{\citenamefont {Qin}\ \emph {et~al.}(2021)\citenamefont {Qin},
  \citenamefont {Schafer}, \citenamefont {Andergassen}, \citenamefont
  {Corboz},\ and\ \citenamefont {Gull}}]{hubreview-qin}%
  \BibitemOpen
  \bibfield  {author} {\bibinfo {author} {\bibfnamefont {M.}~\bibnamefont
  {Qin}}, \bibinfo {author} {\bibfnamefont {T.}~\bibnamefont {Schafer}},
  \bibinfo {author} {\bibfnamefont {S.}~\bibnamefont {Andergassen}}, \bibinfo
  {author} {\bibfnamefont {P.}~\bibnamefont {Corboz}},\ and\ \bibinfo {author}
  {\bibfnamefont {E.}~\bibnamefont {Gull}},\ }\href@noop {} {\bibinfo {title}
  {The hubbard model: A computational perspective}} (\bibinfo {year} {2021}),\
  \Eprint {https://arxiv.org/abs/2104.00064} {arXiv:2104.00064
  [cond-mat.str-el]} \BibitemShut {NoStop}%
\bibitem [{\citenamefont {Jiang}\ and\ \citenamefont
  {Devereaux}(2019)}]{4leghub-hcjiang}%
  \BibitemOpen
  \bibfield  {author} {\bibinfo {author} {\bibfnamefont {H.-C.}\ \bibnamefont
  {Jiang}}\ and\ \bibinfo {author} {\bibfnamefont {T.~P.}\ \bibnamefont
  {Devereaux}},\ }\bibfield  {title} {\bibinfo {title} {Superconductivity in
  the doped hubbard model and its interplay with next-nearest hopping $t'$},\
  }\href@noop {} {\bibfield  {journal} {\bibinfo  {journal} {Science}\ }\textbf
  {\bibinfo {volume} {365}},\ \bibinfo {pages} {1424} (\bibinfo {year}
  {2019})}\BibitemShut {NoStop}%
\bibitem [{\citenamefont {Jiang}\ \emph {et~al.}(2020)\citenamefont {Jiang},
  \citenamefont {Zaanen}, \citenamefont {Devereaux},\ and\ \citenamefont
  {Jiang}}]{4leghub-yfjiang}%
  \BibitemOpen
  \bibfield  {author} {\bibinfo {author} {\bibfnamefont {Y.-F.}\ \bibnamefont
  {Jiang}}, \bibinfo {author} {\bibfnamefont {J.}~\bibnamefont {Zaanen}},
  \bibinfo {author} {\bibfnamefont {T.~P.}\ \bibnamefont {Devereaux}},\ and\
  \bibinfo {author} {\bibfnamefont {H.-C.}\ \bibnamefont {Jiang}},\ }\bibfield
  {title} {\bibinfo {title} {Ground state phase diagram of the doped hubbard
  model on the four-leg cylinder},\ }\href
  {https://doi.org/10.1103/PhysRevResearch.2.033073} {\bibfield  {journal}
  {\bibinfo  {journal} {Phys. Rev. Research}\ }\textbf {\bibinfo {volume}
  {2}},\ \bibinfo {pages} {033073} (\bibinfo {year} {2020})}\BibitemShut
  {NoStop}%
\bibitem [{\citenamefont {Huang}\ \emph {et~al.}(2018)\citenamefont {Huang},
  \citenamefont {Mendl}, \citenamefont {Jiang}, \citenamefont {Moritz},\ and\
  \citenamefont {Devereaux}}]{4leghub-huang}%
  \BibitemOpen
  \bibfield  {author} {\bibinfo {author} {\bibfnamefont {E.~W.}\ \bibnamefont
  {Huang}}, \bibinfo {author} {\bibfnamefont {C.~B.}\ \bibnamefont {Mendl}},
  \bibinfo {author} {\bibfnamefont {H.-C.}\ \bibnamefont {Jiang}}, \bibinfo
  {author} {\bibfnamefont {B.}~\bibnamefont {Moritz}},\ and\ \bibinfo {author}
  {\bibfnamefont {T.~P.}\ \bibnamefont {Devereaux}},\ }\bibfield  {title}
  {\bibinfo {title} {Stripe order from the perspective of the hubbard model},\
  }\href@noop {} {\bibfield  {journal} {\bibinfo  {journal} {npj Quantum
  Materials}\ }\textbf {\bibinfo {volume} {3}},\ \bibinfo {pages} {1} (\bibinfo
  {year} {2018})}\BibitemShut {NoStop}%
\bibitem [{\citenamefont {Dodaro}\ \emph {et~al.}(2017)\citenamefont {Dodaro},
  \citenamefont {Jiang},\ and\ \citenamefont {Kivelson}}]{4legtj-dodaro}%
  \BibitemOpen
  \bibfield  {author} {\bibinfo {author} {\bibfnamefont {J.~F.}\ \bibnamefont
  {Dodaro}}, \bibinfo {author} {\bibfnamefont {H.-C.}\ \bibnamefont {Jiang}},\
  and\ \bibinfo {author} {\bibfnamefont {S.~A.}\ \bibnamefont {Kivelson}},\
  }\bibfield  {title} {\bibinfo {title} {Intertwined order in a frustrated
  four-leg $t\ensuremath{-}j$ cylinder},\ }\href
  {https://doi.org/10.1103/PhysRevB.95.155116} {\bibfield  {journal} {\bibinfo
  {journal} {Phys. Rev. B}\ }\textbf {\bibinfo {volume} {95}},\ \bibinfo
  {pages} {155116} (\bibinfo {year} {2017})}\BibitemShut {NoStop}%
\bibitem [{\citenamefont {Chung}\ \emph {et~al.}(2020)\citenamefont {Chung},
  \citenamefont {Qin}, \citenamefont {Zhang}, \citenamefont {Schollw\"ock},\
  and\ \citenamefont {White}}]{4leghub-chung}%
  \BibitemOpen
  \bibfield  {author} {\bibinfo {author} {\bibfnamefont {C.-M.}\ \bibnamefont
  {Chung}}, \bibinfo {author} {\bibfnamefont {M.}~\bibnamefont {Qin}}, \bibinfo
  {author} {\bibfnamefont {S.}~\bibnamefont {Zhang}}, \bibinfo {author}
  {\bibfnamefont {U.}~\bibnamefont {Schollw\"ock}},\ and\ \bibinfo {author}
  {\bibfnamefont {S.~R.}\ \bibnamefont {White}} (\bibinfo {collaboration} {The
  Simons Collaboration on the Many-Electron Problem}),\ }\bibfield  {title}
  {\bibinfo {title} {Plaquette versus ordinary $d$-wave pairing in the
  ${t'}$-hubbard model on a width-4 cylinder},\ }\href
  {https://doi.org/10.1103/PhysRevB.102.041106} {\bibfield  {journal} {\bibinfo
   {journal} {Phys. Rev. B}\ }\textbf {\bibinfo {volume} {102}},\ \bibinfo
  {pages} {041106(R)} (\bibinfo {year} {2020})}\BibitemShut {NoStop}%
\bibitem [{\citenamefont {Jiang}\ \emph {et~al.}(2018)\citenamefont {Jiang},
  \citenamefont {Weng},\ and\ \citenamefont {Kivelson}}]{4legtj-hcjiang}%
  \BibitemOpen
  \bibfield  {author} {\bibinfo {author} {\bibfnamefont {H.-C.}\ \bibnamefont
  {Jiang}}, \bibinfo {author} {\bibfnamefont {Z.-Y.}\ \bibnamefont {Weng}},\
  and\ \bibinfo {author} {\bibfnamefont {S.~A.}\ \bibnamefont {Kivelson}},\
  }\bibfield  {title} {\bibinfo {title} {Superconductivity in the doped
  $\mathit{t}\ensuremath{-}\mathit{J}$ model: Results for four-leg cylinders},\
  }\href {https://doi.org/10.1103/PhysRevB.98.140505} {\bibfield  {journal}
  {\bibinfo  {journal} {Phys. Rev. B}\ }\textbf {\bibinfo {volume} {98}},\
  \bibinfo {pages} {140505(R)} (\bibinfo {year} {2018})}\BibitemShut {NoStop}%
\bibitem [{\citenamefont {Ido}\ \emph {et~al.}(2018{\natexlab{a}})\citenamefont
  {Ido}, \citenamefont {Ohgoe},\ and\ \citenamefont {Imada}}]{stripe-vmc1}%
  \BibitemOpen
  \bibfield  {author} {\bibinfo {author} {\bibfnamefont {K.}~\bibnamefont
  {Ido}}, \bibinfo {author} {\bibfnamefont {T.}~\bibnamefont {Ohgoe}},\ and\
  \bibinfo {author} {\bibfnamefont {M.}~\bibnamefont {Imada}},\ }\bibfield
  {title} {\bibinfo {title} {Competition among various charge-inhomogeneous
  states and $d$-wave superconducting state in hubbard models on square
  lattices},\ }\href {https://doi.org/10.1103/PhysRevB.97.045138} {\bibfield
  {journal} {\bibinfo  {journal} {Phys. Rev. B}\ }\textbf {\bibinfo {volume}
  {97}},\ \bibinfo {pages} {045138} (\bibinfo {year}
  {2018}{\natexlab{a}})}\BibitemShut {NoStop}%
\bibitem [{\citenamefont {Tocchio}\ \emph {et~al.}(2016)\citenamefont
  {Tocchio}, \citenamefont {Becca},\ and\ \citenamefont
  {Sorella}}]{stripe-vmc2}%
  \BibitemOpen
  \bibfield  {author} {\bibinfo {author} {\bibfnamefont {L.~F.}\ \bibnamefont
  {Tocchio}}, \bibinfo {author} {\bibfnamefont {F.}~\bibnamefont {Becca}},\
  and\ \bibinfo {author} {\bibfnamefont {S.}~\bibnamefont {Sorella}},\
  }\bibfield  {title} {\bibinfo {title} {Hidden mott transition and large-$u$
  superconductivity in the two-dimensional hubbard model},\ }\href
  {https://doi.org/10.1103/PhysRevB.94.195126} {\bibfield  {journal} {\bibinfo
  {journal} {Phys. Rev. B}\ }\textbf {\bibinfo {volume} {94}},\ \bibinfo
  {pages} {195126} (\bibinfo {year} {2016})}\BibitemShut {NoStop}%
\bibitem [{\citenamefont {Corboz}\ \emph {et~al.}(2014)\citenamefont {Corboz},
  \citenamefont {Rice},\ and\ \citenamefont {Troyer}}]{Corboz2014}%
  \BibitemOpen
  \bibfield  {author} {\bibinfo {author} {\bibfnamefont {P.}~\bibnamefont
  {Corboz}}, \bibinfo {author} {\bibfnamefont {T.~M.}\ \bibnamefont {Rice}},\
  and\ \bibinfo {author} {\bibfnamefont {M.}~\bibnamefont {Troyer}},\
  }\bibfield  {title} {\bibinfo {title} {Competing states in the $t$-$j$ model:
  Uniform $d$-wave state versus stripe state},\ }\href
  {https://doi.org/10.1103/PhysRevLett.113.046402} {\bibfield  {journal}
  {\bibinfo  {journal} {Phys. Rev. Lett.}\ }\textbf {\bibinfo {volume} {113}},\
  \bibinfo {pages} {046402} (\bibinfo {year} {2014})}\BibitemShut {NoStop}%
\bibitem [{\citenamefont {Corboz}\ \emph {et~al.}(2011)\citenamefont {Corboz},
  \citenamefont {White}, \citenamefont {Vidal},\ and\ \citenamefont
  {Troyer}}]{corboz2011stripes}%
  \BibitemOpen
  \bibfield  {author} {\bibinfo {author} {\bibfnamefont {P.}~\bibnamefont
  {Corboz}}, \bibinfo {author} {\bibfnamefont {S.~R.}\ \bibnamefont {White}},
  \bibinfo {author} {\bibfnamefont {G.}~\bibnamefont {Vidal}},\ and\ \bibinfo
  {author} {\bibfnamefont {M.}~\bibnamefont {Troyer}},\ }\bibfield  {title}
  {\bibinfo {title} {Stripes in the two-dimensional t-j model with infinite
  projected entangled-pair states},\ }\href@noop {} {\bibfield  {journal}
  {\bibinfo  {journal} {Physical Review B}\ }\textbf {\bibinfo {volume} {84}},\
  \bibinfo {pages} {041108(R)} (\bibinfo {year} {2011})}\BibitemShut {NoStop}%
\bibitem [{\citenamefont {LeBlanc}\ \emph {et~al.}(2015)\citenamefont
  {LeBlanc}, \citenamefont {Antipov}, \citenamefont {Becca}, \citenamefont
  {Bulik}, \citenamefont {Chan}, \citenamefont {Chung}, \citenamefont {Deng},
  \citenamefont {Ferrero}, \citenamefont {Henderson}, \citenamefont
  {Jim\'enez-Hoyos}, \citenamefont {Kozik}, \citenamefont {Liu}, \citenamefont
  {Millis}, \citenamefont {Prokof'ev}, \citenamefont {Qin}, \citenamefont
  {Scuseria}, \citenamefont {Shi}, \citenamefont {Svistunov}, \citenamefont
  {Tocchio}, \citenamefont {Tupitsyn}, \citenamefont {White}, \citenamefont
  {Zhang}, \citenamefont {Zheng}, \citenamefont {Zhu},\ and\ \citenamefont
  {Gull}}]{hub-bench2015}%
  \BibitemOpen
  \bibfield  {author} {\bibinfo {author} {\bibfnamefont {J.~P.~F.}\
  \bibnamefont {LeBlanc}}, \bibinfo {author} {\bibfnamefont {A.~E.}\
  \bibnamefont {Antipov}}, \bibinfo {author} {\bibfnamefont {F.}~\bibnamefont
  {Becca}}, \bibinfo {author} {\bibfnamefont {I.~W.}\ \bibnamefont {Bulik}},
  \bibinfo {author} {\bibfnamefont {G.~K.-L.}\ \bibnamefont {Chan}}, \bibinfo
  {author} {\bibfnamefont {C.-M.}\ \bibnamefont {Chung}}, \bibinfo {author}
  {\bibfnamefont {Y.}~\bibnamefont {Deng}}, \bibinfo {author} {\bibfnamefont
  {M.}~\bibnamefont {Ferrero}}, \bibinfo {author} {\bibfnamefont {T.~M.}\
  \bibnamefont {Henderson}}, \bibinfo {author} {\bibfnamefont {C.~A.}\
  \bibnamefont {Jim\'enez-Hoyos}}, \bibinfo {author} {\bibfnamefont
  {E.}~\bibnamefont {Kozik}}, \bibinfo {author} {\bibfnamefont {X.-W.}\
  \bibnamefont {Liu}}, \bibinfo {author} {\bibfnamefont {A.~J.}\ \bibnamefont
  {Millis}}, \bibinfo {author} {\bibfnamefont {N.~V.}\ \bibnamefont
  {Prokof'ev}}, \bibinfo {author} {\bibfnamefont {M.}~\bibnamefont {Qin}},
  \bibinfo {author} {\bibfnamefont {G.~E.}\ \bibnamefont {Scuseria}}, \bibinfo
  {author} {\bibfnamefont {H.}~\bibnamefont {Shi}}, \bibinfo {author}
  {\bibfnamefont {B.~V.}\ \bibnamefont {Svistunov}}, \bibinfo {author}
  {\bibfnamefont {L.~F.}\ \bibnamefont {Tocchio}}, \bibinfo {author}
  {\bibfnamefont {I.~S.}\ \bibnamefont {Tupitsyn}}, \bibinfo {author}
  {\bibfnamefont {S.~R.}\ \bibnamefont {White}}, \bibinfo {author}
  {\bibfnamefont {S.}~\bibnamefont {Zhang}}, \bibinfo {author} {\bibfnamefont
  {B.-X.}\ \bibnamefont {Zheng}}, \bibinfo {author} {\bibfnamefont
  {Z.}~\bibnamefont {Zhu}},\ and\ \bibinfo {author} {\bibfnamefont
  {E.}~\bibnamefont {Gull}} (\bibinfo {collaboration} {Simons Collaboration on
  the Many-Electron Problem}),\ }\bibfield  {title} {\bibinfo {title}
  {Solutions of the two-dimensional hubbard model: Benchmarks and results from
  a wide range of numerical algorithms},\ }\href
  {https://doi.org/10.1103/PhysRevX.5.041041} {\bibfield  {journal} {\bibinfo
  {journal} {Phys. Rev. X}\ }\textbf {\bibinfo {volume} {5}},\ \bibinfo {pages}
  {041041} (\bibinfo {year} {2015})}\BibitemShut {NoStop}%
\bibitem [{\citenamefont {White}\ and\ \citenamefont
  {Scalapino}(2009)}]{WS2009}%
  \BibitemOpen
  \bibfield  {author} {\bibinfo {author} {\bibfnamefont {S.~R.}\ \bibnamefont
  {White}}\ and\ \bibinfo {author} {\bibfnamefont {D.~J.}\ \bibnamefont
  {Scalapino}},\ }\bibfield  {title} {\bibinfo {title} {Pairing on striped t-
  t'- j lattices},\ }\href@noop {} {\bibfield  {journal} {\bibinfo  {journal}
  {Physical Review B}\ }\textbf {\bibinfo {volume} {79}},\ \bibinfo {pages}
  {220504(R)} (\bibinfo {year} {2009})}\BibitemShut {NoStop}%
\bibitem [{\citenamefont {Ponsioen}\ \emph {et~al.}(2019)\citenamefont
  {Ponsioen}, \citenamefont {Chung},\ and\ \citenamefont
  {Corboz}}]{hubw4stp-corboz2019}%
  \BibitemOpen
  \bibfield  {author} {\bibinfo {author} {\bibfnamefont {B.}~\bibnamefont
  {Ponsioen}}, \bibinfo {author} {\bibfnamefont {S.~S.}\ \bibnamefont
  {Chung}},\ and\ \bibinfo {author} {\bibfnamefont {P.}~\bibnamefont
  {Corboz}},\ }\bibfield  {title} {\bibinfo {title} {Period 4 stripe in the
  extended two-dimensional hubbard model},\ }\href
  {https://doi.org/10.1103/PhysRevB.100.195141} {\bibfield  {journal} {\bibinfo
   {journal} {Phys. Rev. B}\ }\textbf {\bibinfo {volume} {100}},\ \bibinfo
  {pages} {195141} (\bibinfo {year} {2019})}\BibitemShut {NoStop}%
\bibitem [{\citenamefont {Himeda}\ \emph {et~al.}(2002)\citenamefont {Himeda},
  \citenamefont {Kato},\ and\ \citenamefont {Ogata}}]{himeda2002stripe}%
  \BibitemOpen
  \bibfield  {author} {\bibinfo {author} {\bibfnamefont {A.}~\bibnamefont
  {Himeda}}, \bibinfo {author} {\bibfnamefont {T.}~\bibnamefont {Kato}},\ and\
  \bibinfo {author} {\bibfnamefont {M.}~\bibnamefont {Ogata}},\ }\bibfield
  {title} {\bibinfo {title} {Stripe states with spatially oscillating d-wave
  superconductivity in the two-dimensional t- t'- j model},\ }\href@noop {}
  {\bibfield  {journal} {\bibinfo  {journal} {Physical review letters}\
  }\textbf {\bibinfo {volume} {88}},\ \bibinfo {pages} {117001} (\bibinfo
  {year} {2002})}\BibitemShut {NoStop}%
\bibitem [{\citenamefont {Xu}\ and\ \citenamefont
  {Grover}(2020)}]{xu2020competing}%
  \BibitemOpen
  \bibfield  {author} {\bibinfo {author} {\bibfnamefont {X.~Y.}\ \bibnamefont
  {Xu}}\ and\ \bibinfo {author} {\bibfnamefont {T.}~\bibnamefont {Grover}},\
  }\href@noop {} {\bibinfo {title} {Competing nodal d-wave superconductivity
  and antiferromagnetism: a quantum monte carlo study}} (\bibinfo {year}
  {2020}),\ \Eprint {https://arxiv.org/abs/2009.06644} {arXiv:2009.06644
  [cond-mat.str-el]} \BibitemShut {NoStop}%
\bibitem [{\citenamefont {Jiang}\ and\ \citenamefont
  {Kivelson}(2021)}]{jiang2021-hightcinsl}%
  \BibitemOpen
  \bibfield  {author} {\bibinfo {author} {\bibfnamefont {H.-C.}\ \bibnamefont
  {Jiang}}\ and\ \bibinfo {author} {\bibfnamefont {S.~A.}\ \bibnamefont
  {Kivelson}},\ }\href@noop {} {\bibinfo {title} {High temperature
  superconductivity in a lightly doped quantum spin liquid}} (\bibinfo {year}
  {2021}),\ \Eprint {https://arxiv.org/abs/2104.01485} {arXiv:2104.01485
  [cond-mat.str-el]} \BibitemShut {NoStop}%
\bibitem [{\citenamefont {Tohyama}\ \emph {et~al.}(2018)\citenamefont
  {Tohyama}, \citenamefont {Mori},\ and\ \citenamefont {Sota}}]{tohyama}%
  \BibitemOpen
  \bibfield  {author} {\bibinfo {author} {\bibfnamefont {T.}~\bibnamefont
  {Tohyama}}, \bibinfo {author} {\bibfnamefont {M.}~\bibnamefont {Mori}},\ and\
  \bibinfo {author} {\bibfnamefont {S.}~\bibnamefont {Sota}},\ }\bibfield
  {title} {\bibinfo {title} {Dynamical density matrix renormalization group
  study of spin and charge excitations in the four-leg
  $t\text{\ensuremath{-}}{t}^{\ensuremath{'}}\text{\ensuremath{-}}j$ ladder},\
  }\href {https://doi.org/10.1103/PhysRevB.97.235137} {\bibfield  {journal}
  {\bibinfo  {journal} {Phys. Rev. B}\ }\textbf {\bibinfo {volume} {97}},\
  \bibinfo {pages} {235137} (\bibinfo {year} {2018})}\BibitemShut {NoStop}%
\bibitem [{\citenamefont {Zaanen}\ and\ \citenamefont
  {Gunnarsson}(1989)}]{stripe-Zaanen}%
  \BibitemOpen
  \bibfield  {author} {\bibinfo {author} {\bibfnamefont {J.}~\bibnamefont
  {Zaanen}}\ and\ \bibinfo {author} {\bibfnamefont {O.}~\bibnamefont
  {Gunnarsson}},\ }\bibfield  {title} {\bibinfo {title} {Charged magnetic
  domain lines and the magnetism of high-${T}_{c}$ oxides},\ }\href
  {https://doi.org/10.1103/PhysRevB.40.7391} {\bibfield  {journal} {\bibinfo
  {journal} {Phys. Rev. B}\ }\textbf {\bibinfo {volume} {40}},\ \bibinfo
  {pages} {7391} (\bibinfo {year} {1989})}\BibitemShut {NoStop}%
\bibitem [{\citenamefont {Poilblanc}\ and\ \citenamefont
  {Rice}(1989)}]{hatreefock-rice}%
  \BibitemOpen
  \bibfield  {author} {\bibinfo {author} {\bibfnamefont {D.}~\bibnamefont
  {Poilblanc}}\ and\ \bibinfo {author} {\bibfnamefont {T.~M.}\ \bibnamefont
  {Rice}},\ }\bibfield  {title} {\bibinfo {title} {Charged solitons in the
  hartree-fock approximation to the large-u hubbard model},\ }\href
  {https://doi.org/10.1103/PhysRevB.39.9749} {\bibfield  {journal} {\bibinfo
  {journal} {Phys. Rev. B}\ }\textbf {\bibinfo {volume} {39}},\ \bibinfo
  {pages} {9749} (\bibinfo {year} {1989})}\BibitemShut {NoStop}%
\bibitem [{\citenamefont {Machida}(1989)}]{machida1}%
  \BibitemOpen
  \bibfield  {author} {\bibinfo {author} {\bibfnamefont {K.}~\bibnamefont
  {Machida}},\ }\bibfield  {title} {\bibinfo {title} {Magnetism in la2cuo4
  based compounds},\ }\href@noop {} {\bibfield  {journal} {\bibinfo  {journal}
  {Physica C: Superconductivity}\ }\textbf {\bibinfo {volume} {158}},\ \bibinfo
  {pages} {192} (\bibinfo {year} {1989})}\BibitemShut {NoStop}%
\bibitem [{\citenamefont {Kato}\ \emph {et~al.}(1990)\citenamefont {Kato},
  \citenamefont {Machida}, \citenamefont {Nakanishi},\ and\ \citenamefont
  {Fujita}}]{machida2}%
  \BibitemOpen
  \bibfield  {author} {\bibinfo {author} {\bibfnamefont {M.}~\bibnamefont
  {Kato}}, \bibinfo {author} {\bibfnamefont {K.}~\bibnamefont {Machida}},
  \bibinfo {author} {\bibfnamefont {H.}~\bibnamefont {Nakanishi}},\ and\
  \bibinfo {author} {\bibfnamefont {M.}~\bibnamefont {Fujita}},\ }\bibfield
  {title} {\bibinfo {title} {Soliton lattice modulation of incommensurate spin
  density wave in two dimensional hubbard model-a mean field study},\
  }\href@noop {} {\bibfield  {journal} {\bibinfo  {journal} {Journal of the
  Physical Society of Japan}\ }\textbf {\bibinfo {volume} {59}},\ \bibinfo
  {pages} {1047} (\bibinfo {year} {1990})}\BibitemShut {NoStop}%
\bibitem [{\citenamefont {White}(1992)}]{white1992dmrg}%
  \BibitemOpen
  \bibfield  {author} {\bibinfo {author} {\bibfnamefont {S.~R.}\ \bibnamefont
  {White}},\ }\bibfield  {title} {\bibinfo {title} {Density matrix formulation
  for quantum renormalization groups},\ }\href@noop {} {\bibfield  {journal}
  {\bibinfo  {journal} {Physical review letters}\ }\textbf {\bibinfo {volume}
  {69}},\ \bibinfo {pages} {2863} (\bibinfo {year} {1992})}\BibitemShut
  {NoStop}%
\bibitem [{\citenamefont {White}(1993)}]{white1993dmrg2}%
  \BibitemOpen
  \bibfield  {author} {\bibinfo {author} {\bibfnamefont {S.~R.}\ \bibnamefont
  {White}},\ }\bibfield  {title} {\bibinfo {title} {Density-matrix algorithms
  for quantum renormalization groups},\ }\href@noop {} {\bibfield  {journal}
  {\bibinfo  {journal} {Physical Review B}\ }\textbf {\bibinfo {volume} {48}},\
  \bibinfo {pages} {10345} (\bibinfo {year} {1993})}\BibitemShut {NoStop}%
\bibitem [{\citenamefont {White}\ and\ \citenamefont
  {Scalapino}(1998)}]{WS1998}%
  \BibitemOpen
  \bibfield  {author} {\bibinfo {author} {\bibfnamefont {S.~R.}\ \bibnamefont
  {White}}\ and\ \bibinfo {author} {\bibfnamefont {D.~J.}\ \bibnamefont
  {Scalapino}},\ }\bibfield  {title} {\bibinfo {title} {Density matrix
  renormalization group study of the striped phase in the 2d
  $\mathit{t}\ensuremath{-}\mathit{J}$ model},\ }\href
  {https://doi.org/10.1103/PhysRevLett.80.1272} {\bibfield  {journal} {\bibinfo
   {journal} {Phys. Rev. Lett.}\ }\textbf {\bibinfo {volume} {80}},\ \bibinfo
  {pages} {1272} (\bibinfo {year} {1998})}\BibitemShut {NoStop}%
\bibitem [{\citenamefont {Becca}\ \emph {et~al.}(2001)\citenamefont {Becca},
  \citenamefont {Capriotti},\ and\ \citenamefont {Sorella}}]{vmcnostripe1}%
  \BibitemOpen
  \bibfield  {author} {\bibinfo {author} {\bibfnamefont {F.}~\bibnamefont
  {Becca}}, \bibinfo {author} {\bibfnamefont {L.}~\bibnamefont {Capriotti}},\
  and\ \bibinfo {author} {\bibfnamefont {S.}~\bibnamefont {Sorella}},\
  }\bibfield  {title} {\bibinfo {title} {Stripes and spin incommensurabilities
  are favored by lattice anisotropies},\ }\href
  {https://doi.org/10.1103/PhysRevLett.87.167005} {\bibfield  {journal}
  {\bibinfo  {journal} {Phys. Rev. Lett.}\ }\textbf {\bibinfo {volume} {87}},\
  \bibinfo {pages} {167005} (\bibinfo {year} {2001})}\BibitemShut {NoStop}%
\bibitem [{\citenamefont {Sorella}\ \emph {et~al.}(2002)\citenamefont
  {Sorella}, \citenamefont {Martins}, \citenamefont {Becca}, \citenamefont
  {Gazza}, \citenamefont {Capriotti}, \citenamefont {Parola},\ and\
  \citenamefont {Dagotto}}]{vmcnostripe2}%
  \BibitemOpen
  \bibfield  {author} {\bibinfo {author} {\bibfnamefont {S.}~\bibnamefont
  {Sorella}}, \bibinfo {author} {\bibfnamefont {G.~B.}\ \bibnamefont
  {Martins}}, \bibinfo {author} {\bibfnamefont {F.}~\bibnamefont {Becca}},
  \bibinfo {author} {\bibfnamefont {C.}~\bibnamefont {Gazza}}, \bibinfo
  {author} {\bibfnamefont {L.}~\bibnamefont {Capriotti}}, \bibinfo {author}
  {\bibfnamefont {A.}~\bibnamefont {Parola}},\ and\ \bibinfo {author}
  {\bibfnamefont {E.}~\bibnamefont {Dagotto}},\ }\bibfield  {title} {\bibinfo
  {title} {Superconductivity in the two-dimensional
  $\mathit{t}\ensuremath{-}\mathit{J}$ model},\ }\href
  {https://doi.org/10.1103/PhysRevLett.88.117002} {\bibfield  {journal}
  {\bibinfo  {journal} {Phys. Rev. Lett.}\ }\textbf {\bibinfo {volume} {88}},\
  \bibinfo {pages} {117002} (\bibinfo {year} {2002})}\BibitemShut {NoStop}%
\bibitem [{\citenamefont {Chou}\ and\ \citenamefont {Lee}(2010)}]{vmctjstripe}%
  \BibitemOpen
  \bibfield  {author} {\bibinfo {author} {\bibfnamefont {C.-P.}\ \bibnamefont
  {Chou}}\ and\ \bibinfo {author} {\bibfnamefont {T.-K.}\ \bibnamefont {Lee}},\
  }\bibfield  {title} {\bibinfo {title} {Mechanism of formation of half-doped
  stripes in underdoped cuprates},\ }\href
  {https://doi.org/10.1103/PhysRevB.81.060503} {\bibfield  {journal} {\bibinfo
  {journal} {Phys. Rev. B}\ }\textbf {\bibinfo {volume} {81}},\ \bibinfo
  {pages} {060503(R)} (\bibinfo {year} {2010})}\BibitemShut {NoStop}%
\bibitem [{\citenamefont {Ido}\ \emph {et~al.}(2018{\natexlab{b}})\citenamefont
  {Ido}, \citenamefont {Ohgoe},\ and\ \citenamefont {Imada}}]{hubstp-Ido}%
  \BibitemOpen
  \bibfield  {author} {\bibinfo {author} {\bibfnamefont {K.}~\bibnamefont
  {Ido}}, \bibinfo {author} {\bibfnamefont {T.}~\bibnamefont {Ohgoe}},\ and\
  \bibinfo {author} {\bibfnamefont {M.}~\bibnamefont {Imada}},\ }\bibfield
  {title} {\bibinfo {title} {Competition among various charge-inhomogeneous
  states and $d$-wave superconducting state in hubbard models on square
  lattices},\ }\href {https://doi.org/10.1103/PhysRevB.97.045138} {\bibfield
  {journal} {\bibinfo  {journal} {Phys. Rev. B}\ }\textbf {\bibinfo {volume}
  {97}},\ \bibinfo {pages} {045138} (\bibinfo {year}
  {2018}{\natexlab{b}})}\BibitemShut {NoStop}%
\bibitem [{\citenamefont {Tocchio}\ \emph {et~al.}(2019)\citenamefont
  {Tocchio}, \citenamefont {Montorsi},\ and\ \citenamefont
  {Becca}}]{hubstp-tocchio}%
  \BibitemOpen
  \bibfield  {author} {\bibinfo {author} {\bibfnamefont {L.~F.}\ \bibnamefont
  {Tocchio}}, \bibinfo {author} {\bibfnamefont {A.}~\bibnamefont {Montorsi}},\
  and\ \bibinfo {author} {\bibfnamefont {F.}~\bibnamefont {Becca}},\ }\bibfield
   {title} {\bibinfo {title} {{Metallic and insulating stripes and their
  relation with superconductivity in the doped Hubbard model}},\ }\href
  {https://doi.org/10.21468/SciPostPhys.7.2.021} {\bibfield  {journal}
  {\bibinfo  {journal} {SciPost Phys.}\ }\textbf {\bibinfo {volume} {7}},\
  \bibinfo {pages} {21} (\bibinfo {year} {2019})}\BibitemShut {NoStop}%
\bibitem [{\citenamefont {White}\ and\ \citenamefont
  {Scalapino}(1999)}]{tt'j-ws1999}%
  \BibitemOpen
  \bibfield  {author} {\bibinfo {author} {\bibfnamefont {S.~R.}\ \bibnamefont
  {White}}\ and\ \bibinfo {author} {\bibfnamefont {D.~J.}\ \bibnamefont
  {Scalapino}},\ }\bibfield  {title} {\bibinfo {title} {Competition between
  stripes and pairing in a ${t\ensuremath{-}t}^{\ensuremath{'}}\ensuremath{-}j$
  model},\ }\href {https://doi.org/10.1103/PhysRevB.60.R753} {\bibfield
  {journal} {\bibinfo  {journal} {Phys. Rev. B}\ }\textbf {\bibinfo {volume}
  {60}},\ \bibinfo {pages} {R753} (\bibinfo {year} {1999})}\BibitemShut
  {NoStop}%
\bibitem [{\citenamefont {Psaltakis}\ and\ \citenamefont
  {Fenton}(1983)}]{psaltakis1983}%
  \BibitemOpen
  \bibfield  {author} {\bibinfo {author} {\bibfnamefont {G.}~\bibnamefont
  {Psaltakis}}\ and\ \bibinfo {author} {\bibfnamefont {E.}~\bibnamefont
  {Fenton}},\ }\bibfield  {title} {\bibinfo {title} {Superconductivity and
  spin-density waves: organic superconductors},\ }\href@noop {} {\bibfield
  {journal} {\bibinfo  {journal} {Journal of Physics C: Solid State Physics}\
  }\textbf {\bibinfo {volume} {16}},\ \bibinfo {pages} {3913} (\bibinfo {year}
  {1983})}\BibitemShut {NoStop}%
\bibitem [{\citenamefont {Zhang}(1997)}]{zhang1997so5}%
  \BibitemOpen
  \bibfield  {author} {\bibinfo {author} {\bibfnamefont {S.-C.}\ \bibnamefont
  {Zhang}},\ }\bibfield  {title} {\bibinfo {title} {A unified theory based on
  so (5) symmetry of superconductivity and antiferromagnetism},\ }\href@noop {}
  {\bibfield  {journal} {\bibinfo  {journal} {Science}\ }\textbf {\bibinfo
  {volume} {275}},\ \bibinfo {pages} {1089} (\bibinfo {year}
  {1997})}\BibitemShut {NoStop}%
\bibitem [{\citenamefont {Aperis}\ \emph {et~al.}(2008)\citenamefont {Aperis},
  \citenamefont {Varelogiannis}, \citenamefont {Littlewood},\ and\
  \citenamefont {Simons}}]{aperis2008}%
  \BibitemOpen
  \bibfield  {author} {\bibinfo {author} {\bibfnamefont {A.}~\bibnamefont
  {Aperis}}, \bibinfo {author} {\bibfnamefont {G.}~\bibnamefont
  {Varelogiannis}}, \bibinfo {author} {\bibfnamefont {P.}~\bibnamefont
  {Littlewood}},\ and\ \bibinfo {author} {\bibfnamefont {B.}~\bibnamefont
  {Simons}},\ }\bibfield  {title} {\bibinfo {title} {Coexistence of spin
  density wave, d-wave singlet and staggered $\pi$-triplet superconductivity},\
  }\href@noop {} {\bibfield  {journal} {\bibinfo  {journal} {Journal of
  Physics: Condensed Matter}\ }\textbf {\bibinfo {volume} {20}},\ \bibinfo
  {pages} {434235} (\bibinfo {year} {2008})}\BibitemShut {NoStop}%
\bibitem [{\citenamefont {Rowe}\ \emph {et~al.}(2012)\citenamefont {Rowe},
  \citenamefont {Knolle}, \citenamefont {Eremin},\ and\ \citenamefont
  {Hirschfeld}}]{rowe2012spin}%
  \BibitemOpen
  \bibfield  {author} {\bibinfo {author} {\bibfnamefont {W.}~\bibnamefont
  {Rowe}}, \bibinfo {author} {\bibfnamefont {J.}~\bibnamefont {Knolle}},
  \bibinfo {author} {\bibfnamefont {I.}~\bibnamefont {Eremin}},\ and\ \bibinfo
  {author} {\bibfnamefont {P.~J.}\ \bibnamefont {Hirschfeld}},\ }\bibfield
  {title} {\bibinfo {title} {Spin excitations in layered antiferromagnetic
  metals and superconductors},\ }\href@noop {} {\bibfield  {journal} {\bibinfo
  {journal} {Physical Review B}\ }\textbf {\bibinfo {volume} {86}},\ \bibinfo
  {pages} {134513} (\bibinfo {year} {2012})}\BibitemShut {NoStop}%
\bibitem [{\citenamefont {Almeida}\ \emph {et~al.}(2017)\citenamefont
  {Almeida}, \citenamefont {Fernandes},\ and\ \citenamefont
  {Miranda}}]{almeida2017induced}%
  \BibitemOpen
  \bibfield  {author} {\bibinfo {author} {\bibfnamefont {D.~E.}\ \bibnamefont
  {Almeida}}, \bibinfo {author} {\bibfnamefont {R.~M.}\ \bibnamefont
  {Fernandes}},\ and\ \bibinfo {author} {\bibfnamefont {E.}~\bibnamefont
  {Miranda}},\ }\bibfield  {title} {\bibinfo {title} {Induced spin-triplet
  pairing in the coexistence state of antiferromagnetism and singlet
  superconductivity: Collective modes and microscopic properties},\ }\href@noop
  {} {\bibfield  {journal} {\bibinfo  {journal} {Physical Review B}\ }\textbf
  {\bibinfo {volume} {96}},\ \bibinfo {pages} {014514} (\bibinfo {year}
  {2017})}\BibitemShut {NoStop}%
\bibitem [{\citenamefont {Fishman}\ \emph {et~al.}(2020)\citenamefont
  {Fishman}, \citenamefont {White},\ and\ \citenamefont
  {Stoudenmire}}]{itensor}%
  \BibitemOpen
  \bibfield  {author} {\bibinfo {author} {\bibfnamefont {M.}~\bibnamefont
  {Fishman}}, \bibinfo {author} {\bibfnamefont {S.~R.}\ \bibnamefont {White}},\
  and\ \bibinfo {author} {\bibfnamefont {E.~M.}\ \bibnamefont {Stoudenmire}},\
  }\href@noop {} {\bibinfo {title} {The \mbox{ITensor} software library for
  tensor network calculations}} (\bibinfo {year} {2020}),\ \Eprint
  {https://arxiv.org/abs/2007.14822} {arXiv:2007.14822} \BibitemShut {NoStop}%
\bibitem [{\citenamefont {Scalapino}(1995)}]{dwavesc}%
  \BibitemOpen
  \bibfield  {author} {\bibinfo {author} {\bibfnamefont {D.~J.}\ \bibnamefont
  {Scalapino}},\ }\bibfield  {title} {\bibinfo {title} {The case for $d_{x^2-
  y^2}$ pairing in the cuprate superconductors},\ }\href@noop {} {\bibfield
  {journal} {\bibinfo  {journal} {Physics Reports}\ }\textbf {\bibinfo {volume}
  {250}},\ \bibinfo {pages} {329} (\bibinfo {year} {1995})}\BibitemShut
  {NoStop}%
\bibitem [{\citenamefont {White}\ \emph {et~al.}(1994)\citenamefont {White},
  \citenamefont {Noack},\ and\ \citenamefont
  {Scalapino}}]{white1994resonating}%
  \BibitemOpen
  \bibfield  {author} {\bibinfo {author} {\bibfnamefont {S.~R.}\ \bibnamefont
  {White}}, \bibinfo {author} {\bibfnamefont {R.~M.}\ \bibnamefont {Noack}},\
  and\ \bibinfo {author} {\bibfnamefont {D.~J.}\ \bibnamefont {Scalapino}},\
  }\bibfield  {title} {\bibinfo {title} {Resonating valence bond theory of
  coupled heisenberg chains},\ }\href@noop {} {\bibfield  {journal} {\bibinfo
  {journal} {Physical review letters}\ }\textbf {\bibinfo {volume} {73}},\
  \bibinfo {pages} {886} (\bibinfo {year} {1994})}\BibitemShut {NoStop}%
\bibitem [{\citenamefont {Xiang}\ and\ \citenamefont
  {d'Ambrumenil}(1992)}]{xiang92}%
  \BibitemOpen
  \bibfield  {author} {\bibinfo {author} {\bibfnamefont {T.}~\bibnamefont
  {Xiang}}\ and\ \bibinfo {author} {\bibfnamefont {N.}~\bibnamefont
  {d'Ambrumenil}},\ }\bibfield  {title} {\bibinfo {title} {Charge-spin
  separation and the ground-state wave function of the one-dimensional t-j
  model},\ }\href {https://doi.org/10.1103/PhysRevB.45.8150} {\bibfield
  {journal} {\bibinfo  {journal} {Phys. Rev. B}\ }\textbf {\bibinfo {volume}
  {45}},\ \bibinfo {pages} {8150} (\bibinfo {year} {1992})}\BibitemShut
  {NoStop}%
\bibitem [{\citenamefont {S\'en\'echal}\ \emph {et~al.}(2005)\citenamefont
  {S\'en\'echal}, \citenamefont {Lavertu}, \citenamefont {Marois},\ and\
  \citenamefont {Tremblay}}]{senechal2005}%
  \BibitemOpen
  \bibfield  {author} {\bibinfo {author} {\bibfnamefont {D.}~\bibnamefont
  {S\'en\'echal}}, \bibinfo {author} {\bibfnamefont {P.-L.}\ \bibnamefont
  {Lavertu}}, \bibinfo {author} {\bibfnamefont {M.-A.}\ \bibnamefont
  {Marois}},\ and\ \bibinfo {author} {\bibfnamefont {A.-M.~S.}\ \bibnamefont
  {Tremblay}},\ }\bibfield  {title} {\bibinfo {title} {Competition between
  antiferromagnetism and superconductivity in high-${T}_{c}$ cuprates},\ }\href
  {https://doi.org/10.1103/PhysRevLett.94.156404} {\bibfield  {journal}
  {\bibinfo  {journal} {Phys. Rev. Lett.}\ }\textbf {\bibinfo {volume} {94}},\
  \bibinfo {pages} {156404} (\bibinfo {year} {2005})}\BibitemShut {NoStop}%
\bibitem [{\citenamefont {Foley}\ \emph {et~al.}(2019)\citenamefont {Foley},
  \citenamefont {Verret}, \citenamefont {Tremblay},\ and\ \citenamefont
  {S\'en\'echal}}]{foley2019coexistence}%
  \BibitemOpen
  \bibfield  {author} {\bibinfo {author} {\bibfnamefont {A.}~\bibnamefont
  {Foley}}, \bibinfo {author} {\bibfnamefont {S.}~\bibnamefont {Verret}},
  \bibinfo {author} {\bibfnamefont {A.-M.~S.}\ \bibnamefont {Tremblay}},\ and\
  \bibinfo {author} {\bibfnamefont {D.}~\bibnamefont {S\'en\'echal}},\
  }\bibfield  {title} {\bibinfo {title} {Coexistence of superconductivity and
  antiferromagnetism in the hubbard model for cuprates},\ }\href
  {https://doi.org/10.1103/PhysRevB.99.184510} {\bibfield  {journal} {\bibinfo
  {journal} {Phys. Rev. B}\ }\textbf {\bibinfo {volume} {99}},\ \bibinfo
  {pages} {184510} (\bibinfo {year} {2019})}\BibitemShut {NoStop}%
\bibitem [{\citenamefont {Kyung}(2000)}]{kyung2000}%
  \BibitemOpen
  \bibfield  {author} {\bibinfo {author} {\bibfnamefont {B.}~\bibnamefont
  {Kyung}},\ }\bibfield  {title} {\bibinfo {title} {Mean-field study of the
  interplay between antiferromagnetism and d-wave superconductivity},\
  }\href@noop {} {\bibfield  {journal} {\bibinfo  {journal} {Physical Review
  B}\ }\textbf {\bibinfo {volume} {62}},\ \bibinfo {pages} {9083} (\bibinfo
  {year} {2000})}\BibitemShut {NoStop}%
\bibitem [{\citenamefont {Kato}\ and\ \citenamefont
  {Machida}(1988)}]{pitriplet-kato}%
  \BibitemOpen
  \bibfield  {author} {\bibinfo {author} {\bibfnamefont {M.}~\bibnamefont
  {Kato}}\ and\ \bibinfo {author} {\bibfnamefont {K.}~\bibnamefont {Machida}},\
  }\bibfield  {title} {\bibinfo {title} {Superconductivity and spin-density
  waves: Application to heavy-fermion materials},\ }\href@noop {} {\bibfield
  {journal} {\bibinfo  {journal} {Physical Review B}\ }\textbf {\bibinfo
  {volume} {37}},\ \bibinfo {pages} {1510} (\bibinfo {year}
  {1988})}\BibitemShut {NoStop}%
\bibitem [{\citenamefont {Machida}\ \emph {et~al.}(1980)\citenamefont
  {Machida}, \citenamefont {Nokura},\ and\ \citenamefont
  {Matsubara}}]{pitriplet-machida}%
  \BibitemOpen
  \bibfield  {author} {\bibinfo {author} {\bibfnamefont {K.}~\bibnamefont
  {Machida}}, \bibinfo {author} {\bibfnamefont {K.}~\bibnamefont {Nokura}},\
  and\ \bibinfo {author} {\bibfnamefont {T.}~\bibnamefont {Matsubara}},\
  }\bibfield  {title} {\bibinfo {title} {Theory of antiferromagnetic
  superconductors},\ }\href {https://doi.org/10.1103/PhysRevB.22.2307}
  {\bibfield  {journal} {\bibinfo  {journal} {Phys. Rev. B}\ }\textbf {\bibinfo
  {volume} {22}},\ \bibinfo {pages} {2307} (\bibinfo {year}
  {1980})}\BibitemShut {NoStop}%
\bibitem [{\citenamefont {Murakami}\ and\ \citenamefont
  {Fukuyama}(1998)}]{pitriplet-murakami}%
  \BibitemOpen
  \bibfield  {author} {\bibinfo {author} {\bibfnamefont {M.}~\bibnamefont
  {Murakami}}\ and\ \bibinfo {author} {\bibfnamefont {H.}~\bibnamefont
  {Fukuyama}},\ }\bibfield  {title} {\bibinfo {title} {Backward scattering and
  coexistent state in two-dimensional electron system},\ }\href@noop {}
  {\bibfield  {journal} {\bibinfo  {journal} {Journal of the Physical Society
  of Japan}\ }\textbf {\bibinfo {volume} {67}},\ \bibinfo {pages} {2784}
  (\bibinfo {year} {1998})}\BibitemShut {NoStop}%
\bibitem [{\citenamefont {Jeckelmann}\ \emph {et~al.}(1998)\citenamefont
  {Jeckelmann}, \citenamefont {Scalapino},\ and\ \citenamefont
  {White}}]{jeckelmann1998}%
  \BibitemOpen
  \bibfield  {author} {\bibinfo {author} {\bibfnamefont {E.}~\bibnamefont
  {Jeckelmann}}, \bibinfo {author} {\bibfnamefont {D.~J.}\ \bibnamefont
  {Scalapino}},\ and\ \bibinfo {author} {\bibfnamefont {S.~R.}\ \bibnamefont
  {White}},\ }\bibfield  {title} {\bibinfo {title} {Comparison of different
  ladder models},\ }\href@noop {} {\bibfield  {journal} {\bibinfo  {journal}
  {Physical Review B}\ }\textbf {\bibinfo {volume} {58}},\ \bibinfo {pages}
  {9492} (\bibinfo {year} {1998})}\BibitemShut {NoStop}%
\bibitem [{\citenamefont {White}\ and\ \citenamefont
  {Scalapino}(1997)}]{white1997hole}%
  \BibitemOpen
  \bibfield  {author} {\bibinfo {author} {\bibfnamefont {S.~R.}\ \bibnamefont
  {White}}\ and\ \bibinfo {author} {\bibfnamefont {D.~J.}\ \bibnamefont
  {Scalapino}},\ }\bibfield  {title} {\bibinfo {title} {Hole and pair
  structures in the tj model},\ }\href@noop {} {\bibfield  {journal} {\bibinfo
  {journal} {Physical Review B}\ }\textbf {\bibinfo {volume} {55}},\ \bibinfo
  {pages} {6504} (\bibinfo {year} {1997})}\BibitemShut {NoStop}%
\bibitem [{Note1()}]{Note1}%
  \BibitemOpen
  \bibinfo {note} {The chemical potential is of form $\mu (l_x)=\mu
  _0+a\protect \sqrt {1+(b|2l_x-L_x|/L_x)^2}$ with $a$ and $b$ to be adjusted
  and different for left and right half. This form of chemical potential varies
  slower and connects smoothly in the middle.}\BibitemShut {Stop}%
\bibitem [{\citenamefont {White}\ and\ \citenamefont
  {Chernyshev}(2007)}]{White_and_Chernyshev}%
  \BibitemOpen
  \bibfield  {author} {\bibinfo {author} {\bibfnamefont {S.~R.}\ \bibnamefont
  {White}}\ and\ \bibinfo {author} {\bibfnamefont {A.~L.}\ \bibnamefont
  {Chernyshev}},\ }\bibfield  {title} {\bibinfo {title} {Ne\'el order in square
  and triangular lattice heisenberg models},\ }\href
  {https://doi.org/10.1103/PhysRevLett.99.127004} {\bibfield  {journal}
  {\bibinfo  {journal} {Phys. Rev. Lett.}\ }\textbf {\bibinfo {volume} {99}},\
  \bibinfo {pages} {127004} (\bibinfo {year} {2007})}\BibitemShut {NoStop}%
\bibitem [{\citenamefont {Stoudenmire}\ and\ \citenamefont
  {White}(2012)}]{2ddmrg-miles}%
  \BibitemOpen
  \bibfield  {author} {\bibinfo {author} {\bibfnamefont {E.~M.}\ \bibnamefont
  {Stoudenmire}}\ and\ \bibinfo {author} {\bibfnamefont {S.~R.}\ \bibnamefont
  {White}},\ }\bibfield  {title} {\bibinfo {title} {Studying two-dimensional
  systems with the density matrix renormalization group},\ }\href@noop {}
  {\bibfield  {journal} {\bibinfo  {journal} {Annu. Rev. Condens. Matter
  Phys.}\ }\textbf {\bibinfo {volume} {3}},\ \bibinfo {pages} {111} (\bibinfo
  {year} {2012})}\BibitemShut {NoStop}%
\bibitem [{\citenamefont {da~Silva~Neto}\ \emph {et~al.}(2015)\citenamefont
  {da~Silva~Neto}, \citenamefont {Comin}, \citenamefont {He}, \citenamefont
  {Sutarto}, \citenamefont {Jiang}, \citenamefont {Greene}, \citenamefont
  {Sawatzky},\ and\ \citenamefont {Damascelli}}]{da2015charge}%
  \BibitemOpen
  \bibfield  {author} {\bibinfo {author} {\bibfnamefont {E.~H.}\ \bibnamefont
  {da~Silva~Neto}}, \bibinfo {author} {\bibfnamefont {R.}~\bibnamefont
  {Comin}}, \bibinfo {author} {\bibfnamefont {F.}~\bibnamefont {He}}, \bibinfo
  {author} {\bibfnamefont {R.}~\bibnamefont {Sutarto}}, \bibinfo {author}
  {\bibfnamefont {Y.}~\bibnamefont {Jiang}}, \bibinfo {author} {\bibfnamefont
  {R.~L.}\ \bibnamefont {Greene}}, \bibinfo {author} {\bibfnamefont {G.~A.}\
  \bibnamefont {Sawatzky}},\ and\ \bibinfo {author} {\bibfnamefont
  {A.}~\bibnamefont {Damascelli}},\ }\bibfield  {title} {\bibinfo {title}
  {Charge ordering in the electron-doped superconductor nd2--xcexcuo4},\
  }\href@noop {} {\bibfield  {journal} {\bibinfo  {journal} {Science}\ }\textbf
  {\bibinfo {volume} {347}},\ \bibinfo {pages} {282} (\bibinfo {year}
  {2015})}\BibitemShut {NoStop}%
\bibitem [{\citenamefont {Belinicher}\ \emph {et~al.}(1996)\citenamefont
  {Belinicher}, \citenamefont {Chernyshev},\ and\ \citenamefont
  {Shubin}}]{tt'j-sasha}%
  \BibitemOpen
  \bibfield  {author} {\bibinfo {author} {\bibfnamefont {V.~I.}\ \bibnamefont
  {Belinicher}}, \bibinfo {author} {\bibfnamefont {A.~L.}\ \bibnamefont
  {Chernyshev}},\ and\ \bibinfo {author} {\bibfnamefont {V.~A.}\ \bibnamefont
  {Shubin}},\ }\bibfield  {title} {\bibinfo {title} {Generalized
  t-t\ensuremath{'}-j model: Parameters and single-particle spectrum for
  electrons and holes in copper oxides},\ }\href
  {https://doi.org/10.1103/PhysRevB.53.335} {\bibfield  {journal} {\bibinfo
  {journal} {Phys. Rev. B}\ }\textbf {\bibinfo {volume} {53}},\ \bibinfo
  {pages} {335} (\bibinfo {year} {1996})}\BibitemShut {NoStop}%
\bibitem [{\citenamefont {Gong}\ \emph {et~al.}(2021)\citenamefont {Gong},
  \citenamefont {Zhu},\ and\ \citenamefont {Sheng}}]{gong2021robust}%
  \BibitemOpen
  \bibfield  {author} {\bibinfo {author} {\bibfnamefont {S.}~\bibnamefont
  {Gong}}, \bibinfo {author} {\bibfnamefont {W.}~\bibnamefont {Zhu}},\ and\
  \bibinfo {author} {\bibfnamefont {D.~N.}\ \bibnamefont {Sheng}},\ }\href@noop
  {} {\bibinfo {title} {Robust d-wave superconductivity in the square-lattice
  $t$-$j$ model}} (\bibinfo {year} {2021}),\ \Eprint
  {https://arxiv.org/abs/2104.03758} {arXiv:2104.03758 [cond-mat.str-el]}
  \BibitemShut {NoStop}%
\bibitem [{\citenamefont {Lieb}\ and\ \citenamefont {Mattis}(1962)}]{lieb1962}%
  \BibitemOpen
  \bibfield  {author} {\bibinfo {author} {\bibfnamefont {E.}~\bibnamefont
  {Lieb}}\ and\ \bibinfo {author} {\bibfnamefont {D.}~\bibnamefont {Mattis}},\
  }\bibfield  {title} {\bibinfo {title} {Ordering energy levels of interacting
  spin systems},\ }\href@noop {} {\bibfield  {journal} {\bibinfo  {journal}
  {Journal of Mathematical Physics}\ }\textbf {\bibinfo {volume} {3}},\
  \bibinfo {pages} {749} (\bibinfo {year} {1962})}\BibitemShut {NoStop}%
\bibitem [{\citenamefont {Sandvik}(1997)}]{sandvik97}%
  \BibitemOpen
  \bibfield  {author} {\bibinfo {author} {\bibfnamefont {A.~W.}\ \bibnamefont
  {Sandvik}},\ }\bibfield  {title} {\bibinfo {title} {Finite-size scaling of
  the ground-state parameters of the two-dimensional heisenberg model},\
  }\href@noop {} {\bibfield  {journal} {\bibinfo  {journal} {Physical Review
  B}\ }\textbf {\bibinfo {volume} {56}},\ \bibinfo {pages} {11678} (\bibinfo
  {year} {1997})}\BibitemShut {NoStop}%
\end{thebibliography}%

\end{document}